\documentclass[
	reprint,
	superscriptaddress,
	nofootinbib,
	showkeys,
	aps,
	pra,
	longbibliography
]{revtex4-2}

\usepackage[utf8]{inputenc}

\usepackage[hyperref]{xcolor}
	\definecolor{goethe-blau}{cmyk}{1.0,0.2,0.0,0.4}
	\definecolor{hellgrau}{cmyk}{0.04,0.04,0.05,0.02}
	\definecolor{sandgrau}{cmyk}{0.12,0.09,0.13,0.0}
	\definecolor{dunkelgrau}{cmyk}{0.25,0.25,0.30,0.75}
	\definecolor{purple}{cmyk}{0.08,1.0,0.3,0.36}
	\definecolor{emo-rot}{cmyk}{0.04,1.0,0.8,0.07}
	\definecolor{senfgelb}{cmyk}{0.01,0.25,1.0,0.05}
	\definecolor{gruen}{cmyk}{0.62,0.4,0.87,0.09}
	\definecolor{magenta}{cmyk}{0.08,0.86,0.12,0.12}
	\definecolor{orange}{cmyk}{0.0,0.7,1.0,0.04}
	\definecolor{sonnengelb}{cmyk}{0.0,0.12,0.95,0.0}
	\definecolor{helles-gruen}{cmyk}{0.4,0.17,0.81,0.07}
	\definecolor{lichtblau}{cmyk}{0.8,0.0,0.06,0.04}

\usepackage[
	colorlinks,
	pdfpagelabels,
	breaklinks,
	pdfstartview=FitH,
	bookmarksopen=true,
	bookmarksnumbered=true,
	bookmarksopenlevel=2,
	plainpages=false,
	hypertexnames=false,
	citecolor=emo-rot,
	linkcolor=goethe-blau,
	urlcolor=purple,
	pdftitle={Bound-state formation and thermalization within the Lindblad approach},		
	pdfauthor={Jan Rais, Tim Neidig, Hendrik van Hees, Carsten Greiner},	
	pdfkeywords={}					
]{hyperref}

\usepackage{graphicx}
\usepackage{dcolumn}
\usepackage{braket}
\usepackage[tbtags]{amsmath}
\usepackage{amssymb}
\usepackage{epstopdf}
\usepackage{color}
\usepackage{xspace}
\usepackage[capitalize]{cleveref}
\usepackage{bm}
\usepackage{tabularx}
\usepackage{lipsum}
\usepackage{multirow}
\usepackage[bottom]{footmisc}
\usepackage{orcidlink}
\usepackage{upgreek}
\usepackage{comment}

\usepackage[acronym]{glossaries-extra}
\glssetcategoryattribute{acronym}{nohyperfirst}{true}
\setabbreviationstyle[acronym]{long-short}

\DeclareMathOperator{\Tr}{Tr}
\newacronym{kt}{KT}{Kurganov-Tadmor}
\newacronym{fv}{FV}{Finite Volume}
\newacronym{muscl}{MUSCL}{Monotonic Upstream-centered Scheme for Conservation Laws}
\newacronym{pde}{PDE}{partial differential equation}
\newacronym{cfl}{CFL}{Courant-Friedrichs-Lewy}
\newacronym{clme}{CLME}{Caldeira-Leggett-master equation}

\newacronym{lhs}{l.h.s.}{left hand side}
\newacronym{rhs}{r.h.s.}{right hand side}

\newcommand{\reff}{Ref.~}
\newcommand{\reffs}{Refs.~}

\newcommand{\vdistance}{\vphantom{\bigg(\bigg)}}

\newcommand{\dd}{{\rm d}}

\newcommand{\ee}{{\rm e}}

\newcommand{\kBoltzmann}{k_{\mathrm{B}}}

\newcommand{\REM}[1]{}

\definecolor{magenta}{cmyk}{0,1,0,0}

\allowdisplaybreaks

\begin{document}
	
\title{Bound-state formation and thermalization within the Lindblad approach}
	
\author{Jan Rais \orcidlink{0000-0001-8691-6930}}
\email{rais@itp.uni-frankfurt.de} \affiliation{Institut f\"ur
  Theoretische Physik, Johann Wolfgang Goethe-Universit\"at,
  Max-von-Laue-Strasse 1, 60438 Frankfurt am Main, Germany}
	
	
\author{Hendrik van Hees \orcidlink{0000-0003-0729-2117}}
\email{hees@itp.uni-frankfurt.de} \affiliation{Institut f\"ur
  Theoretische Physik, Johann Wolfgang Goethe-Universit\"at,
  Max-von-Laue-Strasse 1, 60438 Frankfurt am Main, Germany}%

\author{Carsten Greiner \orcidlink{0000-0001-8933-1321}}
\email{carsten.greiner@itp.uni-frankfurt.de} \affiliation{Institut f\"ur
  Theoretische Physik, Johann Wolfgang Goethe-Universit\"at,
  Max-von-Laue-Strasse 1, 60438 Frankfurt am Main, Germany}


\date{\today}
	
\begin{abstract}
  The Lindblad equation, as one approach to open quantum systems, describes the density matrix of a particle or a chain of interacting particles, which are in contact with a thermal bath.  
  Still, it is not fully understood yet, how arbitrary systems evolve towards a stationary distribution, which guarantees thermalization in a thermodynamical context, and how to systematically incorporate the variety of assumptions that are made in this approach in order to preserve thermal Gibbs states. 
 Despite these shortcomings, Lindblad dynamics was successfully employed in heavy-ion physics (quarkonia) and  also became of interest in quantum-computer applications.

  In this paper, we consider a problem borrowed from heavy-ion collisions, namely the formation of bound states, as for example the deuteron, in the non-relativistic regime by using the already well understood techniques of Lindblad dynamics.
  However, only recently, we were able to extend this toolbox by showing, that the position-space Lindblad equation can be reformulated in terms of a diffusion-advection equation with sources and therefore
  provides a hydrodynamical formulation of a dissipative quantum master equation.
  Making use of this advanced machinery and insights, we describe the possible formation of a bound state, which is realized by a  P\"oschl-Teller-like potential, of a particle in  interaction with a heat bath in a 1-dim setting.
  We analyse the possibility of a thermalization and the time-scale of the formation, population and depopulation of the bound state.  
  Finally, we also show an example of a much deeper potential, where we allow for three bound states, just in the spirit of quarkonia.
  Besides this, we discuss general aspects of open quantum systems, like decoherence, entropy production etc. 
\end{abstract}
	
\keywords{Lindblad equation, open quantum systems, computational fluid dynamics, bound state formation, formation time, thermalization, decoherence}

\maketitle

\section{Introduction}

Open quantum systems are understood as quantum systems which are not
(perfectly) isolated from their environment.
Isolating a system, and therefore excluding any interaction with a (thermal) environment is a highly challenging task of experimental physics, but has major relevance.  
The most prominent example of how important it is to minimize the system-bath interaction is currently given by the field of quantum computing \cite{Nielsen:2012yss}.  
In contrast to the question, how to isolate the system, one can also ask how a quantum system can be influenced in a systematized way by a thermal environment \reffs \cite{gardiner00,BRE02}.

\subsection{Open quantum systems in heavy-ion collisions}\label{sec:motivation}

A particularly interesting field of research for open quantum systems is nuclear and high-energy physics.
 Both are investigated in heavy-ion collisions \cite{Brambilla:2016wgg,Blaizot:2017ypk,Blaizot:2018oev,Andronic:2024oxz,Brambilla:2022ynh,Brambilla:2017zei,Akamatsu:2020ypb}:  
in heavy-ion collisions, energies far above the freeze-out temperature, which is given by approximately $60- 150$ MeV (depending on the collision energies probed in large systems like Au+Au or Pb+Pb)\cite{Braun-Munzinger:2008szb,Becattini:2009fv}, are reached. 
Colliders, such as the LHC at CERN or the RHIC at BNL focus on the measurement of the matter which is produced at the collision energy regime of $\sqrt{s}=2.4$ GeV $-13000$ GeV.  
Therefore, while cooling -- even at high energies -- bound states such as heavy quarkonia ($J/\psi$, $\Upsilon$ and excited states), but also the (anti-)deuteron are probes for the medium of strongly interacting matter under extreme conditions \cite{Andronic:2017pug}.  
The typical deuteron yields at collision energies of $\sqrt{s_{NN}} = 2.76$ TeV of  Pb-Pb collisions with 0-10\% centrality are $\dd N/\dd y \approx 10^{-1}$ \cite{Andronic:2017pug,ALICE:2015wav} and therefore a phenomenologically remarkable finding: the typical binding energy of the deuteron is $\sim 2.3$ MeV, which is orders of magnitude below the hadronic chemical freeze-out temperature of $\sim 155$ MeV. 
In the literature, several attempts have been made to describe this phenomenon as a collective effect which appears either by coalescence \cite{Nagle:1996vp,Mattiello:1995xg,Nagle:1994wj,Gyulassy:1982pe,Liu:2019nii,Hillmann:2021zgj,Oliinychenko:2018ugs} or may be caused by $n$-body collisions \cite{Neidig:2021bal,Staudenmaier:2021lrg,Danielewicz:1991dh}.

Also we have tried to provide a better description and understanding of this effect:	
In \reff \cite{Rais:2022gfg}, we have introduced a very basic potential, which mimics a bound state in one dimension.  
This potential was chosen because it represents the most simple potential which causes bound states, in contrast to the very elaborated approaches,  which are assumed to be relevant in heavy-ion collisions, such as for example the BONN potential \cite{Machleidt:1982xp}. 
We used this toy model to investigate the probability of the population or
depopulation of a particle obeying a potential bound state which interacts with a single pulse or a
chain of pulses.
From our results we concluded, that the bound state is
formed immediately during the interaction, but the energy spectrum pf the particle is
dependent on the strength and range of the interacting potential.
	
In fact, this interaction rate can not be simulated by one pulse or a chain of pulses properly, because the system interacts all the time with its environment, and the bath has to follow more advanced constraints, which are usually adjusted to a Ohmic frequency spectrum.
However, there are various descriptions for system-bath interactions, depending on the strengths of these interactions, their spectra, their temperatures and their energy regimes.
	
One way to realize a temporally continuous interaction with an environment is to couple the system of interest, whether it is a single particle or a molecule, to an (infinite) number of surrounding particles, representing an Ohmic heat bath. 
Since these particles are typically assumed to be lighter, one can approximate their interaction with the system in terms of ``classical springs". 
This model has been introduced by Feynman and Vernon \cite{Feynman:1963fq} and later exemplified by Caldeira and Leggett \cite{Caldeira:1982iu}.  
	
In this work, we will use the so-called Caldeira-Leggett model, in form of a Lindblad equation, to describe this setup and discuss the formation of bound states, assuming a system particle in a P\"oschl-Teller potential with parameters motivated from nuclear physics.

\subsection{Goal of this work}

In this work we investigate the formation of quantum bound states with energies which are typical for heavy-ion collisions. 
As motivated in \cref{sec:motivation}, a physical system, which itself is part of a many body system is for example a deuteron with a binding energy much lower than the energy of the surrounding system.

Therefore, we introduce a potential, which is a modified version of the one, we have introduced in \reff \cite{Rais:2022gfg}, namely a P\"oschl-Teller potential within a square-well potential. 
This can be understood as follows: the P\"oschl-Teller part of the potential is used to mimic a bound state and can be adjusted such, that either only one bound state is generated, or a higher number of bound states.
In our case, we adjust the potential to the binding energy of the deuteron (we are also going to discuss a scenario with more bound states).
The square-well potential is introduced to identify discrete and normalizable energy eigenstates, and therefore a full set of eigenvectors in a (truncated) Hilbert space and can be interpreted as the size of the fireball in a heavy-ion collision. 

We employ this potential, to use several formulations of the Lindblad equation and to investigate the formation or destruction time of a bound state and to treat thermalization in terms of Boltzmann-like statistical distributions. 
We find, that for various regimes and various parameters thermalization is reached (up to small numerical discrepancy) and the equilibration time is usually dependent on the temperature of the heat bath and the damping coefficient, as well as the cut-off frequency of the Ohmic heat bath spectrum. 

We employ a new method \cite{rais2024}, known from hydrodynamics, to compute the reduced density matrix $\rho(x,y,t)$ at arbitrary times towards equilibration, following Lindblad dynamics.
As a consequence, we can directly tackle the questions of thermalization and thermalization time. 
To use these (numerical) tools, it is necessary to reformulate the Lindblad equation into a conservative form and separate the real part of the reduced density matrix from the imaginary part.

Furthermore, we use $\rho(x,-x,t=t_\text{eq})$ and $\rho_{nn}(t)$\footnote{
	The matrix coefficients $\rho_{mn}(t)$ are calculated via projection on the initial wave functions $\psi_n(x)$ of the considered system,
	\begin{align}\label{eq:coeff}
		\rho_{nm}(t) = \int_{-L}^{L}\dd x \int_{-L}^{L} \dd y \, \rho(x,y,t) \braket{\psi_n\vert x}\braket{y\vert \psi_m }\, .
\end{align}}, the corresponding occupation number, in order to clarify the question about thermalization and calculate the entropy and purity of the system to know, at which time the system thermalizes.\footnote{In this context the terminus thermalization means, that the stationary case is a Gibbs state and is described by a Boltzmann-distribution,\begin{align*}
		p_i = \exp\left[- \frac{E_i- \mu}{T} \right] \, .
\end{align*}}
This analysis is worked out for various heat bath temperatures $T$, damping coefficients $\gamma$ and initial conditions, as well as for different types and formulations of the Lindblad equation.

In this context, the entropy,
\begin{align*}
	S(t) = - \text{Tr}[\rho(t) \ln \rho(t)]\,,
\end{align*}
 and purity,
 \begin{align*}
 	P(t) = \text{Tr}[\rho^2(t)]\, ,
 \end{align*}
  of the system are obtained by diagonalizing the reduced density matrix.
At the same time, this allows us to discuss the effective eigenfuntions of the system-plus-reservoir Hamiltonian to answer the question of how a previously bound state is modified after interacting with the environment. 

Studying the entropy of the full system and the occupation numbers of the states enables us to compare the time scales of the formation or destruction of the bound state to the equilibration of the full system, which will turn out to be different depending on the initial condition.
Among other results, it turns out, that higher states tend to equilibrate faster than energetically lower states.\footnote{In this work, the equilibration of a certain state is the minimal time, where $\partial_t \rho_{nn} = 0$ and the equilibration time of the full system is the minimal time, where $\partial_t S(t) =0$. It turns out, that both time scales can differ.}
Furthermore, \reffs \cite{Bernad2018,Homa2019} argued, that the interaction with the heat bath can lead to a modification of the wave functions and as a consequence the energy eigenvalues under certain conditions. 
This calls for clarification: Are there mode shifts, and are they of sufficient strength to lead to a unbinding of a previously bound state?
We answer this question by calculating the wave functions of the thermalized effective Hamiltonian.

\subsection{Structure}

We start in \cref{sec:Lindblad_Diff_adv} with a recapitulation of our findings from \reff \cite{rais2024} - the fluid-dynamical reformulation of the Lindblad equation.
This helps for a proper understanding of the applicability of the Lindblad equation, and the mathematical meaning of each of the diffusion coefficients. 
In \cref{sec:numerics} we briefly recapitulate the corresponding new numerical method for solving Lindblad equations.
In \cref{sec:bound_state} we discuss the mathematical framework, which is implemented into the scheme to finally tackle the question, how bound states can be described within the Lindblad approach.  
This makes up the main part of this work and treats the question of thermalization, formation time and dependence on the initial condition, damping constant, diffusion parameter, and the heat bath temperature.  
We also briefly tackle the question of how a system with more than one bound state thermalizes.
In \cref{sec:bound_state} we summarize our work and present future directions of research.

\subsection{Conventions}
\label{sec:conventions}

We work on nuclear scales and set the mass to the
reduced mass of a deuteron, $m = m_{\text{red,d}} = 470$ MeV,  and $\hbar = \kBoltzmann = 1$. The
computational domain is set to $L \times L = 40 \times 40$
fm$^2$, which is, if not explicitly mentioned divided into
$300\times 300$ cells to discretize the system for numerical
evaluations.\footnote{Even though we are considering a one-dimensional quantum
	system, cf.\ \reff\cite{Rais:2022gfg}, the density matrix formalism
	leads to a two-dimensional problem in coordinate space,
	$\rho(x,y,t)$. }

\section{The Lindblad equation} 

In the previous section, we introduced our basic motivation to
investigate Lindblad dynamics using the Caldeira-Leggett model. In
Ref.~\cite{rais2024} we have given a phenomenological introduction of
the basic nature and the assumptions, that are made to obtain the
Lindblad equation from the basic Hamiltonian, which has been introduced
in \reff \cite{Feynman:1963fq} to describe a system interacting with a heat
bath,
\begin{align}\label{eq:hamiltonian}
  \hat{H} =  \hat{H}_\text{S} + \hat{H}_\text{B} + \hat{H}_{\text{SB}}\, .
\end{align}
Here, S denotes the system of interest, B the environment (thermal
heat bath), which remains in thermal equilibrium during the described
process, and SB the interaction between the system and the heat bath.
Systems, which include dissipation can not be described by a Schr\"odinger
or a von-Neumann equation, because these equations are symmetric in
time. 
Therefore, a possibility to describe such systems is provided by
the application of phenomenological equations such as Fokker-Planck or
Langevin equations, which explicitly break the time symmetry \cite{Lindenberg:1984zz}.  
One possible observable is provided by the density matrix, which for closed
quantum systems fulfils the von-Neumann equation,
\begin{align}
  \dot{\rho} = -\text{i}[\hat{H}, \rho ]\, .
\end{align}	
To extend this equation to take system-bath interactions
into account, the first assumption is that the bath and the system are
uncorrelated at some initial time $t_0$.  
This allows to describe the
entire framework at $t=0$ by the product state
\begin{align}
  \rho_\text{T} = \rho \otimes \rho_\text{B},	
\end{align}	 	
where $\rho_\text{B} = \ee^{-\beta \hat{H}_\text{B}}/Z$ with
$Z=\Tr \ee^{-\beta \hat{H}_\text{B}}$, and $\rho_\text{T}$ is the total density
matrix, which describes the dynamics of both, system and environment,
which together form a closed quantum system.  
We consider the Caldeira-Leggett model, where
\begin{align}\label{hamiltonian}
  &	\hat{H}_\text{B} + \hat{H}_{\text{SB}} =\\
  &= \sum_{\alpha = 1}^{N} \left[\frac{\hat{p}_\alpha^2}{2 M_\alpha} + \frac{1}{2} M_\alpha \Omega_\alpha^2 \left(\hat{x}_\alpha - \frac{c_\alpha}{M_\alpha \Omega_\alpha^2} \hat{x}\right)^2\right]\nonumber,\\
  \hat{H}_\text{S} &= \frac{\hat{p}^2}{2M} + V(\hat{x})\, ,
\end{align}
where we choose a coordinate-space representation as introduced in \reffs
\cite{Caldeira:1982iu,BRE02}.
Thereby, the $\hat{x}_{\alpha}$ refer to the bath degrees of
freedom, while $\hat{x}$ belongs to the system.  
Here, $N$ with $N´ \rightarrow \infty$ denotes the number of surrounding particles, which are coupled via ``springs" to the system, while not interacting among each other.
Assuming, that the damping is weak \cite{Thingna_2012}, and the process is Markovian \cite{BRE02,Caldeira:1982iu}, Caldeira and Leggett showed that
\begin{align}\label{CLME}
  \dot{\hat{\rho}} =& - \text{i} \left[\hat{H}_S, \hat{\rho}_S\right] - \text{i} \gamma \left[ \hat{x}, \left\{ \hat{p}, \hat{\rho}_S(t)\right\}\right] +\\
  &- 2mT\gamma \left[\hat{x},\left[\hat{x}, \hat{\rho}_S(t)\right]\right]\, \nonumber,
\end{align} 
which is known as the Caldeira-Leggett master equation (CLME).
It is assumed, that $\gamma= \eta/2m$ is the characteristic damping rate
of the system, while $\eta$ is the friction coefficient \cite{Caldeira:1982iu}.  

In the case of the harmonic oscillator with frequency $\omega$, the heat bath temperature is $T \gg \omega$ and  the coherence length pertaining to the state $\rho_{nn}$ has to be greater then $\lambda_{\text{dB}} = \frac{1}{\sqrt{4MT}}$, the de-Broglie wave length.  
If this is not the case, the CLME is known to violate the positivity of $\rho$, especially, because the CLME in this form is only valid at high temperatures \cite{Homa2019,Bernad2018}.  

Meanwhile, the Lindblad equation, which is a structure-wise similar master equation, preserves the conservation of the norm and obeys the positivity condition per construction \cite{Lindblad:1975ef}.  
The general form of the Lindblad-Gorini-Kossakowski-Sudarshan equation,
\begin{align}\label{eq:Lindblad}
  &\mathcal{L}\left[\hat{\rho}_S\right] =\\
  &= -\text{i} \left[\hat{\tilde{H}}, \hat{\rho}_S\right] +\sum_{i,j=1} \left(\hat{L}_i \hat{\rho}_S  \hat{L}_j -  \frac{1}{2}  \left\{\hat{L}_i^\dagger \hat{L}_j, \hat{\rho}_S\right\}\right),\nonumber
\end{align}
contains $\hat{L}_i$ Lindblad operators, whose choice is ``guided by intuition" \cite{Gao:1997} and possibly can be as diverse as transitions are allowed in the systems´ \cite{Koide:2023awf}.  
On the other hand, there are recent discussions about the systematic derivation of these operators, cf. Refs. \cite{Tupkary:2021bcd,Manzano:2020yyw}. 

Regarding \cref{eq:Lindblad}, we have to point out that there is a large variety of master equations in the literature, which obey Lindblad form and therefore can be called Lindblad equation.   
\cref{eq:Lindblad} is simply the most general form of a Lindblad equation \cite{BRE02}. 
However,  the Lindblad operators $L_i$ can in general be different for every system under consideration, time dependent, and also dependent on the eigenstate to which the $i$-th operator is related to \cite{Koide:2023awf}.  
 
The most general CLME in Lindblad form reads
\begin{align}\label{eq:Lindblad_general}
		\dot{\hat{\rho}}_S =& - \text{i} \left[\hat{H}_S, \hat{\rho}_S\right] - \text{i} \gamma \left[\hat{ x}, \left\{ \hat{p}, \hat{\rho}_S(t)\right\}\right] + \\
		& - D_{pp} \left[\hat{x},\left[\hat{x}, \hat{\rho}_S(t)\right]\right] + 
		+ 2D_{px} \left[\hat{x},\left[\hat{p}, \hat{\rho}_S(t)\right]\right] + \nonumber\\
		&- D_{xx} \left[\hat{p},\left[\hat{p}, \hat{\rho}_S(t)\right]\right],\nonumber
\end{align} 
with (hypothetically time dependent) coefficients $D_{pp}$, $D_{xp}=D_{px}$ and
$D_{xx}$, which are derived in different limits and for the temperature
of the bath in various ways. 
However, in our work we will consider these coefficients to be constant. 

The position-space representation of \cref{eq:Lindblad_general} is given by
\begin{align}\label{eq:Lindblad_space}
	\text{i} \frac{\partial}{\partial t} &\rho(x,y,t) =\\
	&= \left[\frac{1}{2m} \left( \frac{\partial^2}{\partial y^2} -    \frac{\partial^2}{\partial x^2}\right) + V(x) - V(y) +  \right. \nonumber \\
	&	\left. -\text{i}D_{pp} (x-y)^2
	- \text{i} \gamma (x-y)\left(\frac{\partial}{\partial x} - \frac{ \partial}{\partial y}\right) + \right. \nonumber \\
	&	\left.- 2D_{px} (x-y) \left(\frac{\partial}{\partial x} + \frac{ \partial}{\partial y}\right) + \right.\nonumber\\
	&\left. + \text{i} D_{xx} \left(\frac{\partial}{\partial x} + \frac{ \partial}{\partial y}\right)  \right] \rho(x,y,t)\, ,\nonumber
\end{align}
that we use for this work to describe the Lindblad dynamics of our systems of interest, cf. \cite{Gao:1997,BRE02,Homa2019,Bernad2018,rais2024}.  

In particular, we use the following damping coefficients $D_{pp}, D_{px}$ and $D_{xx} = 0$.
\begin{align}
	D_{pp} &= 2m\gamma T\\
	D_{px} &= \begin{cases}
		-\frac{ \gamma T}{\Omega}\,\qquad \text{\cite{BRE02}}\,, \\ 
		\frac{\Omega \gamma}{6\pi T}\,\qquad  \text{\cite{DIOSI1993517}}\, ,\\
	\end{cases}\\
	D_{xx} &= 0\,\qquad  \text{\cite{rais2024}} .	
\end{align}
Here, $\Omega$ is the cut-off frequency of the Ohmic heat bath, $T$ the temperature of the heat bath, $m$ the (reduced) mass of the system particle and $\gamma$ the damping coefficient.\footnote{This parametrization was derived in \cite{BRE02,Caldeira:1982iu,DEKKER19811}.
Furthermore, in \cite{Rais2025} we have shown, that in order to satisfy norm conservation, the spatial diffusion of the density matrix in a finite sized spatial domain has to be zero. This is also feasible, because this term was not introduced by the original work of Caldeira and Leggett.}

In order to study this position space version of \cref{eq:Lindblad}
We have implemented a numerical method, which, to our knowledge has not been used to describe Lindblad dynamics before, but turns out to be a highly reliable and efficient tool. This method has been successfully tested and is discussed in detail in Ref.~\cite{rais2024} and is based on the code presented in \cite{Zorbach:2024rre}.

Hence, next we present the reformulated form of the Lindblad equation as an advection-diffusion-type equation and then comment on the specific numerical implementation in  \cref{sec:numerics}.

\subsection{The Lindblad equation as an advection-diffusion equation}
\label{sec:Lindblad_Diff_adv}

In \reff \cite{rais2024}, we have rewritten the Lindblad equation \cref{eq:Lindblad_space} into an advection-diffusion equation in conservative form. 
 Usually, the conservative form refers to a conservation low, which is satisfied by these types of equations.
 For us, this is the norm of the density matrix, $\Tr \hat{\rho}=\int \dd x \, \rho(x,x,t)=1$ that is conserved by construction for Lindblad equations.  
 Splitting the density matrix into real and imaginary parts, rearranging the terms and performing integrations by parts, cf. \reff \cite{rais2024}, we can rewrite \cref{eq:Lindblad_space} into
\begin{align}\label{eq:dgl}
	& \partial_t \vec{u} + \partial_x \vec{f}^{\, x} [ \vec{x}, \vec{u} \, ] + \partial_y \vec{f}^{\, y} [ \vec{x}, \vec{u} \, ] =	\vdistance
	\\
	= \, & \partial_x \vec{Q}^{\, x} [ \partial_x \vec{u}, \partial_y \vec{u} \, ] + \partial_y \vec{Q}^{\, y} [ \partial_x \vec{u}, \partial_y \vec{u} \, ] + \vec{S} [ t, \vec{x}, \vec{u} \, ] \, .	\vdistance	\nonumber
\end{align} 
Here,
$\vec{u} = \vec{u} ( \vec{x}, t ) = ( \rho_I ( x, y, t ), \rho_R ( x, y,
t ) )^T$ is a vector that contains the imaginary- and real part of the
density matrix and plays the role of a two-component fluid field.
The so-clled advection and diffusion fluxes $\vec{f}^{x,y}$, $\vec{Q}^{x,y}$ and the source term $\vec{S}$ are
given by
\begin{align}\label{eq:fqs1}
	&\vec{f}^x[\vec{x}, \vec{u}] =\\
	&=\begin{pmatrix}
		- 2 D_{p x} \, ( x - y ) \, \rho_R + \gamma \, ( x - y ) \, \rho_I
		\\
		+ 2 D_{p x} \, ( x - y ) \, \rho_I + \gamma \, ( x - y ) \, \rho_R
	\end{pmatrix}\, ,\nonumber\\
	\label{eq:fqs2}
	&\vec{f}^y[\vec{x}, \vec{u}] =\\
	&=\begin{pmatrix}
		- 2 D_{p x} \, ( x - y ) \, \rho_R - \gamma \, ( x - y ) \, \rho_I
		\\
		+ 2 D_{p x} \, ( x - y ) \, \rho_I - \gamma \, ( x - y ) \, \rho_R
	\end{pmatrix}\, ,\nonumber\\
	\label{eq:fqs3}
	&\vec{Q}^x[\partial_x \vec{u}, \partial_y \vec{u}] =\\
	&=	\begin{pmatrix}
		\frac{\partial}{\partial x} \big[ \frac{1}{2 m} \, \rho_R + D_{x x} \, \rho_I \big] + D_{x x} \, \frac{\partial}{\partial y} \, \rho_I
		\\
		\frac{\partial}{\partial x} \, \big[ - \frac{1}{2 m} \, \rho_I + D_{x x} \, \rho_R \big] + D_{x x} \, \frac{\partial}{\partial y} \, \rho_R
	\end{pmatrix}\, ,\nonumber\\
	\label{eq:fqs4}
	&\vec{Q}^y[\partial_x \vec{u}, \partial_y \vec{u}]=\\
	& =	\begin{pmatrix}
		\frac{\partial}{\partial y} \big[ - \frac{1}{2 m} \, \rho_R + D_{x x} \, \rho_I \big] + D_{x x} \, \frac{\partial}{\partial x} \, \rho_I
		\\
		\frac{\partial}{\partial y} \, \big[ \frac{1}{2 m} \,  \rho_I + D_{x x} \, \rho_R \big] + D_{x x} \, \frac{\partial}{\partial x} \, \rho_R
	\end{pmatrix} \, ,\nonumber\\
	\label{eq:fqs5}
	&\vec{S}[t, \vec{x},\vec{u}] =\\
	&=	\begin{pmatrix}
		( V ( y ) - V ( x ) ) \, \rho_R + \big[ 2 \gamma - D_{p p} \, ( x - y )^2 \big] \,\rho_I
		\\
		( V ( x ) - V ( y ) ) \, \rho_I + \big[ 2 \gamma - D_{p p} \, ( x - y )^2 \big]\rho_R
	\end{pmatrix}.\nonumber
\end{align}
Note that the equation is now formulated entirely in terms of real
quantities, which is conceptually similar to the hydrodynamical formulation of
quantum mechanics, such as it was investigated by the Madelung equations for
the wave function \cite{Madelung:1927ksh} and later widely discussed in
various literature \cite{Bohm:1951xw,Bohm:1951xx}. 
Even today it is a topic of interest in theoretical physics \cite{Schuch:2018fvh,Schuch:2023nlm,Schuch:2023sfx}.
A detailed comparison of our formulation to these works is however postponed to future work.

In \reff \cite{rais2024}, we have shown, that this form indeed allows a
mathematical interpretation of the Lindblad equation in terms of the
general expression of an advection-diffusion equation with sources and
sinks,
\begin{align}\label{eq:diff-adv}
	\tfrac{\partial}{\partial t} \, \xi = \vec{\nabla} \cdot \big( D \, \vec{\nabla} \, \xi - \vec{v} \, \xi \big) + S ( \xi ) \, .
\end{align}
Here, $\xi$ is usually some concentration, temperature or fluid density, $D$ the
diffusion coefficients, $\vec{v}$ the field velocity, and $S$ is called
the source term related to sources or sinks of $\xi$.  In our case,
$\xi = \vec{u}$, the vector of the real and imaginary parts of the
probability density (matrix).  Note here, that the ``source term'' is
also dependent on the density matrix itself, which makes it unclear to
interpret this term as being a pure source or sink.  Introducing such a
new perspective on a dissipative quantum master equation, we have to
briefly discuss  the density matrix $\rho(x,y,t)$ as a ``hydrodynamical quantity''. 
 Therefore, it is crucial to comment on the meaning of the real and imaginary parts of the density matrix, as two coupled quantities in \cref{eq:diff-adv}.

For this purpose it is better to rewrite \cref{eq:dgl} in terms of centre-of-mass and relative
coordinates, $r \equiv\frac{1}{2}(x-y)$ and $q\equiv \frac{1}{2} (x+y)$, with
density matrix
$\tilde{\vec{u}} = ( \tilde{\rho}_R(r,q,t) , 
\tilde{\rho}_I(r,q,t))$.  Then the
Lindblad equation in the separated form reads
\begin{align}\label{eq:Lindblad_rel}
	&\partial_t \tilde{\vec{u}} =\\
	&=\left[
	\begin{pmatrix}
		 D_{x x} & 0
		\\
		0 & D_{x x}
	\end{pmatrix} \partial_q^2
	 + \begin{pmatrix}
		\gamma& 0
		\\
		0 & \gamma
	\end{pmatrix} \partial_r \, + 
		 \right.\nonumber\\	
	&\left.
	-4\begin{pmatrix}
		D_{pp}& 0
		\\
		0& D_{pp}
	\end{pmatrix} r^2
	 \right] \tilde{\vec{u}} \, + \nonumber\\
	 &+ \left[
	  \begin{pmatrix}
	 	0& \frac{1}{2m}
	 	\\
	 	-\frac{1}{2m} & 0
	 \end{pmatrix}\partial_q\partial_r
	 +4 \begin{pmatrix}
	 	0& D_{px}
	 	\\
	 	-D_{px} & 0
	 \end{pmatrix} r\partial_q \, +
	\right.\nonumber\\
	 &\left.+ \begin{pmatrix}
	 	0& V(r-q) - V(r+q)
	 	\\
	 	V(r+q) -V(r-q) & 0
	 \end{pmatrix} 
	  \right]\tilde{\vec{u}}\, , \nonumber
\end{align}
We divided the Lindblad equation into two parts, separated by the two
square brackets, a symmetric and an antisymmetric part.  The second one
is the von-Neumann part, with an additional term proportional to $D_{px}$.  
This antisymmetric part has the structure of a
von-Neumann or Schr\"odinger equation, because it mixes the real and
imaginary parts in the differential equation.  It depends
on the relative coordinate $r$ as well as the derivative with respect to the centre-of-mass
coordinate $q$.\footnote{At this point it is appealing to relate \cref{eq:Lindblad_rel} to space-momentum correlations, 
and therefore, to rewrite \cref{eq:Lindblad_rel} in terms of the Wigner transform. This might provide deeper a insight into the physical meaning of the density matrix in terms of the probability distribution in the phase space \cite{Schleich}. }

In the first parentheses, the symmetric part, we have separated the terms proportional to the diffusion coefficients
$D_{x x}$, $D_{p p}$, and the damping $\gamma$.
Note, that only the $D_{p p}$
term has been originally derived by Caldeira and Leggett
\cite{Caldeira:1982iu}.  This term can indeed be interpreted as being
proportional to momentum diffusion, because it depends on the
relative coordinate, and therefore, the smaller $r$ gets, the more
momentum diffuses.  If one solely studies this term, it has the solution
\begin{align}
	\tilde{\vec{u}}(r,t) = \tilde{\vec{u}}_0 \exp(-4D_{pp}r^2t)\, ,
\end{align}
which shows, that for higher values of the relative coordinate the
suppression rate at these points is larger, which corresponds to
momentum diffusion and therefore leads to faster diagonalization of the
density matrix (i.e. decoherence).

The first term is an ordinary diffusion equation, which leads to spatial diffusion
along the diagonal of the density matrix ($q$-direction).

The $\gamma-$term depends on the derivative of the
relative coordinate $r$. It does not mix the real and imaginary
parts of $\tilde{\vec{u}}$, because the matrix entries are only
diagonal.
 This part can be solved for example with the method of characteristics, cf. \reff \cite{couranthilbert}, and is a plain advection orthogonal to the diagonal. 

To summarize, we demonstrated, that reformulating the Lindblad
equation, as given in \cref{eq:Lindblad_space}, we are enabled to
provide a better interpretation of its structure in terms of both the real and imaginary part of the density matrix.
We have seen, that the Lindblad equation can be separated in symmetric
and antisymmetric parts, where the antisymmetric part is mixing the
real and imaginary parts of the density matrix, while the symmetric parts
do not. 
Next, we make use of this formulation to apply modern numerical methods from the field of computational fluid dynamics.
\begin{figure}
	\begin{center}
		\hspace*{\fill}%
		\includegraphics[width=1.0\columnwidth,clip=true]{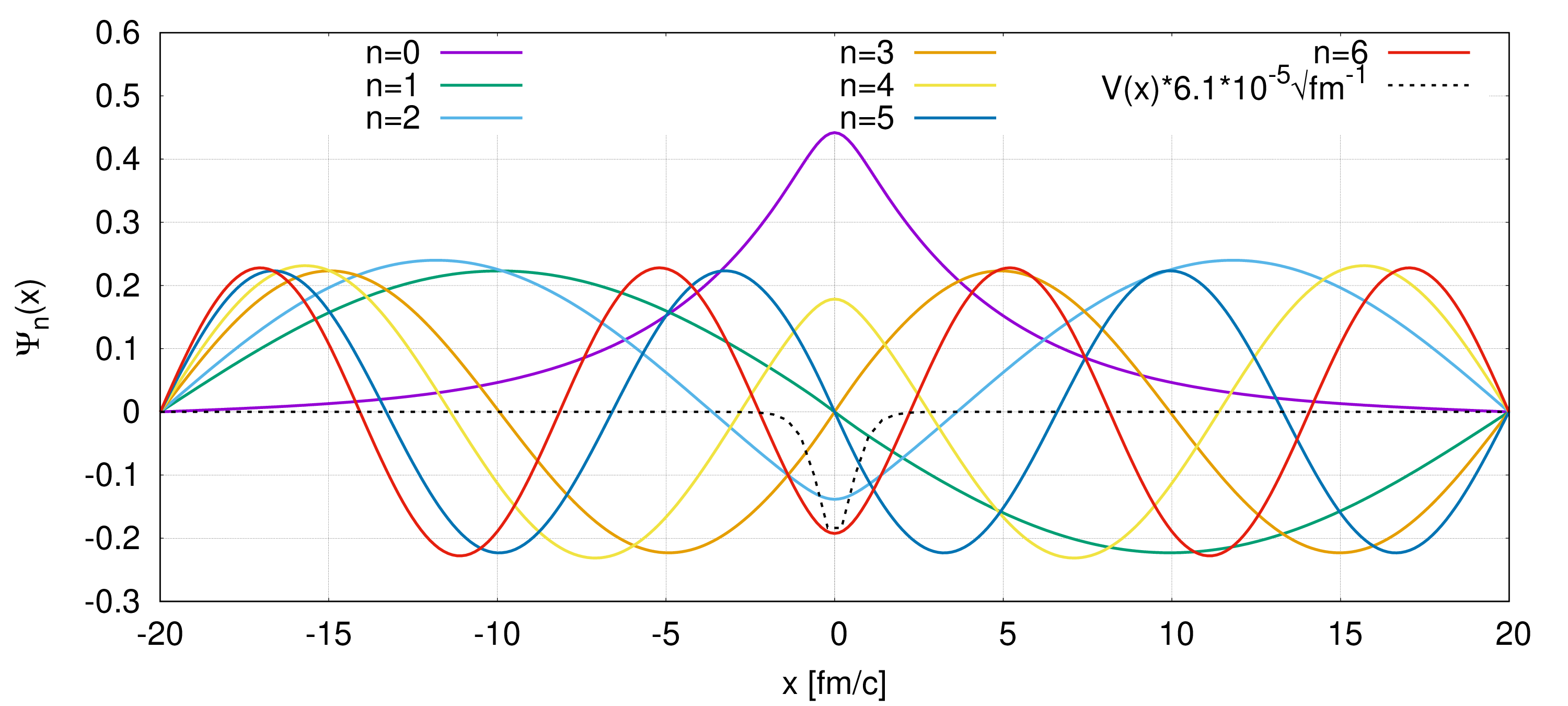}
		\hspace*{\fill}%
		
		\caption{
			The  wave functions of a P\"oschl-Teller potential, \cref{eq:potential}, of the first seven energy eigenstates with parameters $\alpha = 1.45$ fm$^{-1}$ and $V_0 = 16.5$ MeV, calculated with a standard shooting method. The dashed black line refers to the potential, which is rescaled for better illustration.
		}
		\label{fig:waves}
	\end{center}
\end{figure}
\begin{figure}
	\begin{center}
		\hspace*{\fill}%
		\includegraphics[width=1.0\columnwidth,clip=true]{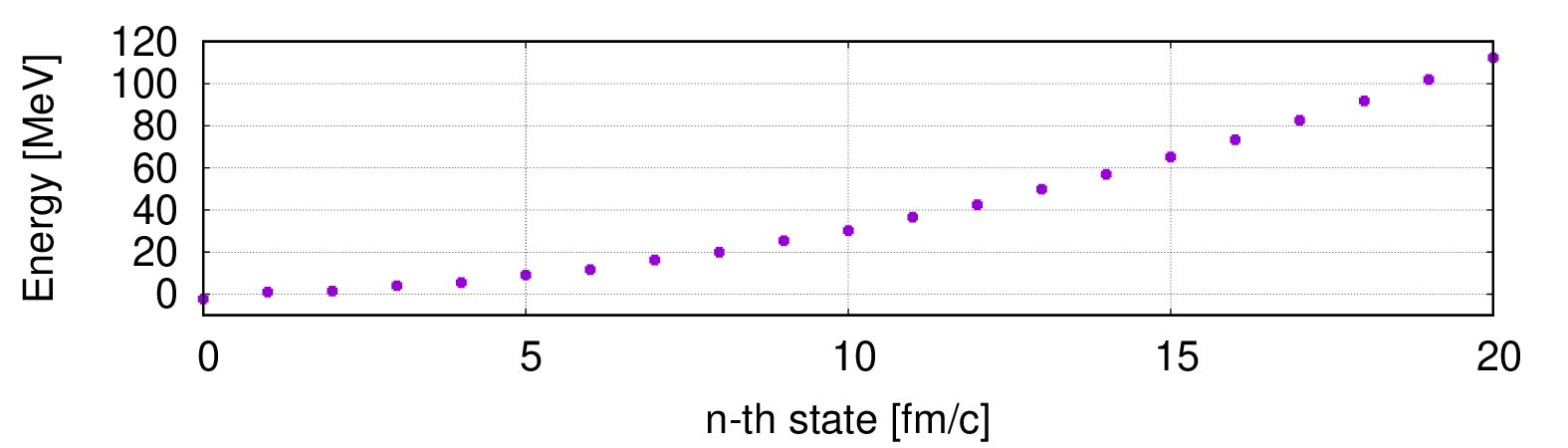}
		\hspace*{\fill}%
		
		\caption{
			The first 20 energy eigenvalues of the wave functions from \cref{fig:waves}, following the parameters also introduced in \cref{fig:waves}. Notice, that for $n=0$ the eigenvalue is negative, namely $-2.3$ MeV.
		}
		\label{fig:energies}
	\end{center}
\end{figure}
\begin{figure*}
	\begin{center}
		\hspace*{\fill}%
		\includegraphics[width=\linewidth]{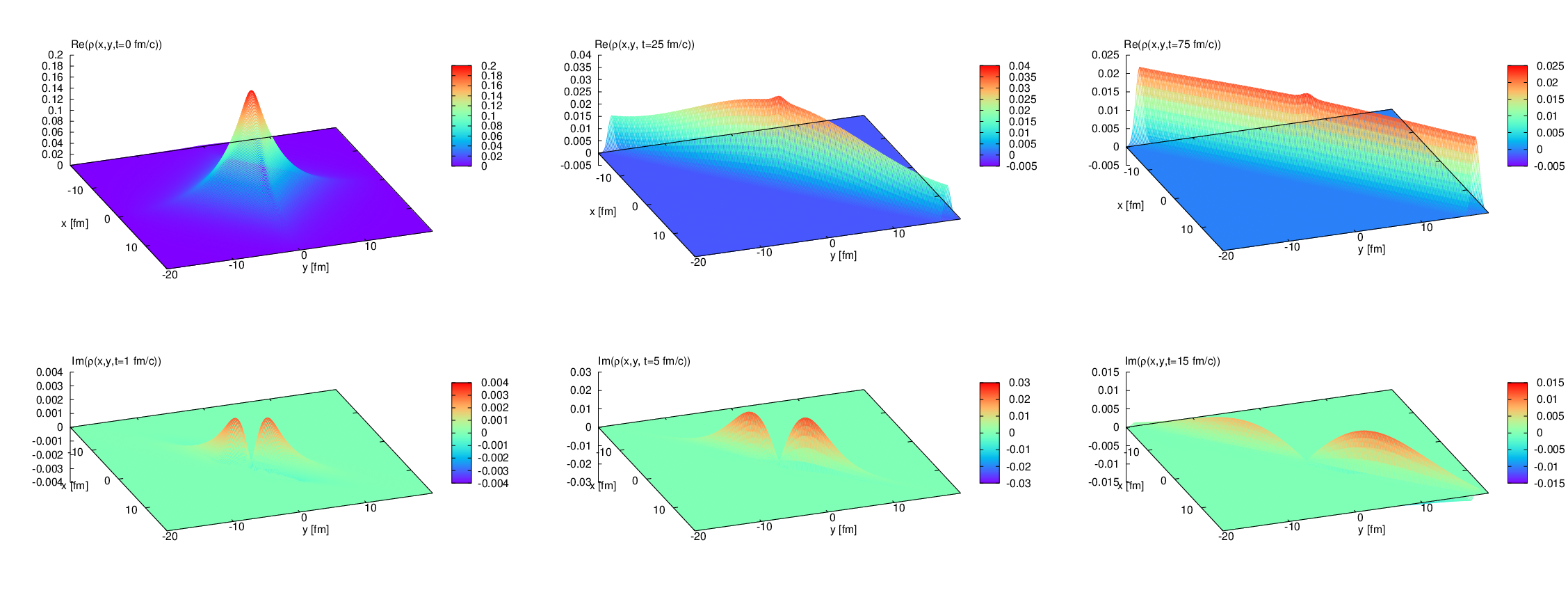}
		\hspace*{\fill}%
		
		\caption{
			The temporal evolution of the initially populated bound state towards thermal equilibrium at times $t=0$ fm/c, $t=25$ fm/c and $t=75$ fm/c for the real part (upper panel) and at times $t=1$ fm/c, 5 fm/c and 15 fm/c for the imaginary part (lower panel). 
			The parameters are $T=250$ MeV, $\Omega = 4T$, $\gamma = 0.1$ c/fm, and $D_{px} = - \gamma T/ \Omega$, with $V(x)$, given by \cref{eq:potential} and parameters from \cref{fig:waves}. 
		}
		\label{fig:3d_init1}
	\end{center}
\end{figure*}
\begin{figure*}
	\begin{center}
		\hspace*{\fill}%
		\includegraphics[width=\linewidth]{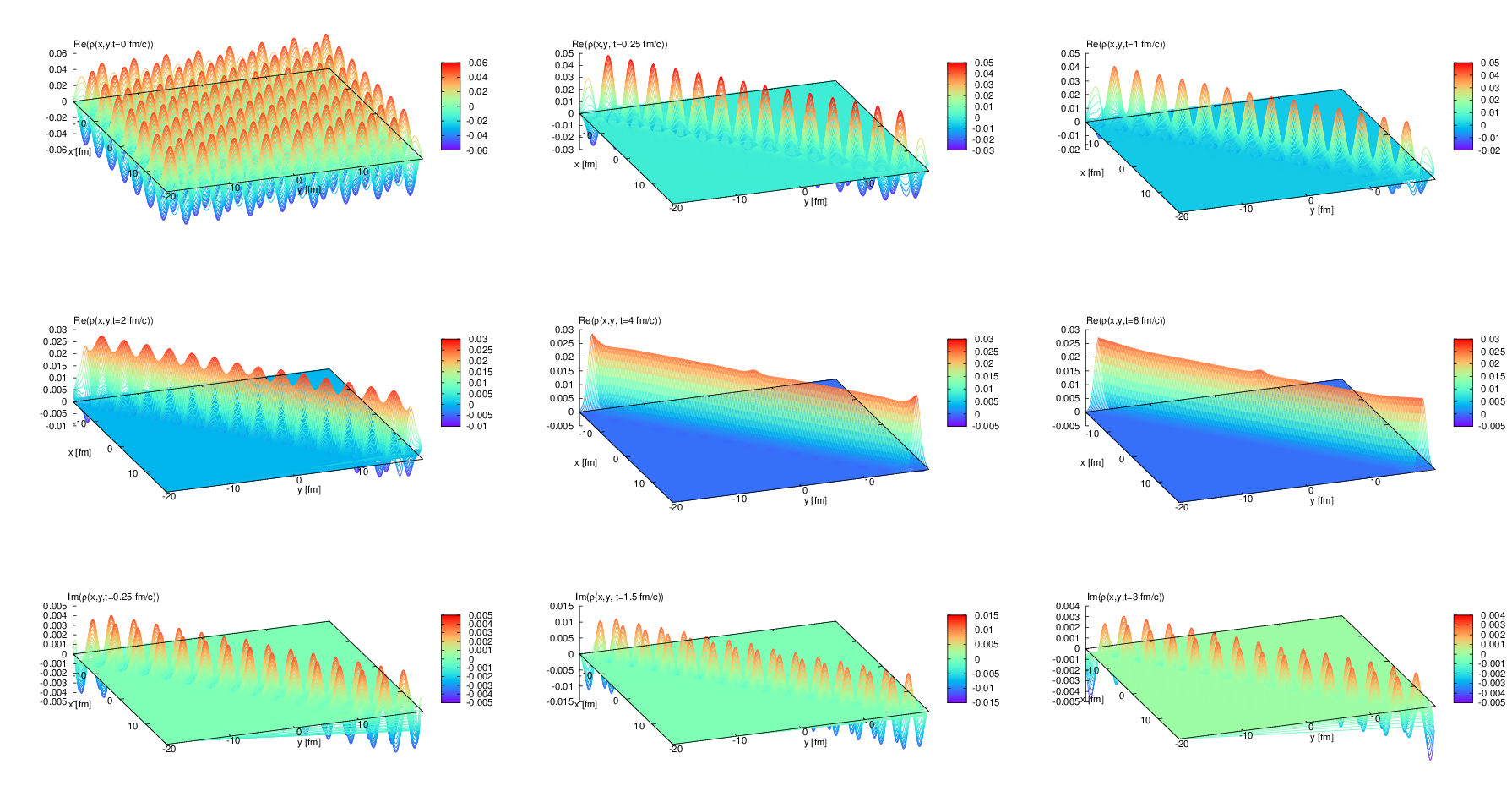}
		\hspace*{\fill}%
		
		\caption{
			The temporal evolution of the initially populated $16^{\text{th}}$ state towards thermal equilibrium at times $t=0$ fm/c, $t=0.25$ fm/c, $t=1$ fm/c, $t=2$ fm/c, $t=4$ fm/c, and $t=8$ fm/c for the real part (upper two panels) and at times $t=0.25$ fm/c, 1.5 fm/c and 3 fm/c for the imaginary part (lower panel). 
			The parameters are $T=250$ MeV, $\Omega = 4T$, $\gamma = 0.1$ c/fm and $D_{px} = - \gamma T/ \Omega$, with $V(x)$,  given by \cref{eq:potential} with parameters from \cref{fig:waves}. 
		}
		\label{fig:3d_init10}
	\end{center}
\end{figure*}

\begin{figure}
	\begin{center}
		\hspace*{\fill}%
		\includegraphics[width=.9\columnwidth,clip=true]{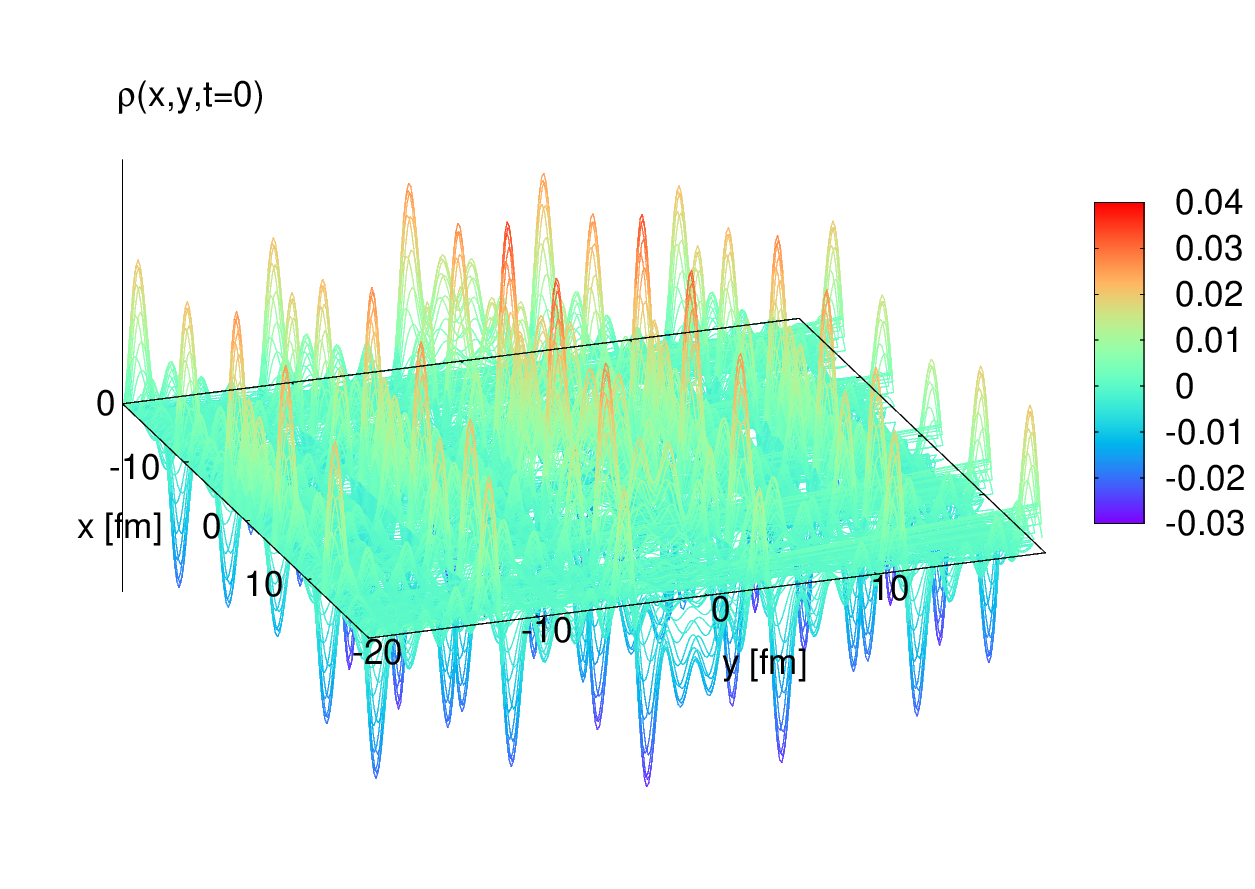}
		\hspace*{\fill}%
		
		\caption{
			The initial condition, $\rho(x,y,0)$ for a initially populated mixed state, \cref{eq:inhom}.
			The parameters are $T=200$ MeV, $\Omega = 4T$, $\gamma = 0.1$ c/fm, and $D_{px} = - \gamma T/ \Omega$, with $V(x)$, given by \cref{eq:potential} and parameters from \cref{fig:waves}. 
		}
		\label{fig:inhom3d}
	\end{center}
\end{figure}

\begin{figure*}
	\begin{center}
		\hspace*{\fill}%
		\includegraphics[width=2.0\columnwidth,clip=true]{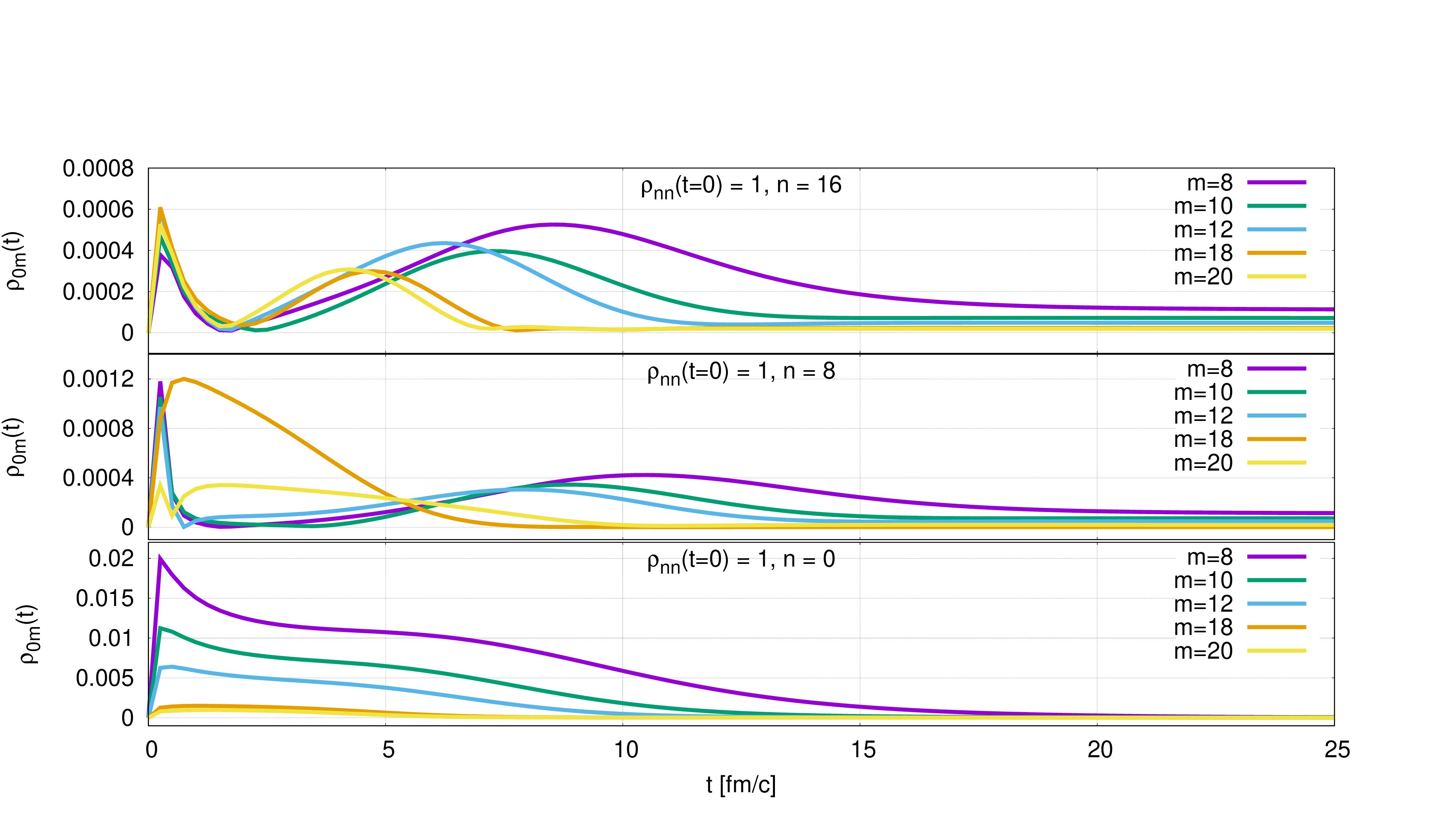}
		\hspace*{\fill}%
		
		\caption{
			First row of the off-diagonal matrix coefficients calculated with \cref{eq:coeff} for different initial conditions $n=0, 8, 16$ for the  density matrix $\rho(x,y,t)$. Here we show $\rho_{0,m}$, where $m = \{8,10,12,18,20\}$,  and the parameters are  $T=200$ MeV, $\Omega = 4T$, $\gamma = 0.1$ c/fm and $D_{px} = - \gamma T/ \Omega$.
		}
		\label{fig:decoherence}
	\end{center}
\end{figure*}

\begin{figure}
	\begin{center}
		\hspace*{\fill}%
		\includegraphics[width=1.0\columnwidth,clip=true]{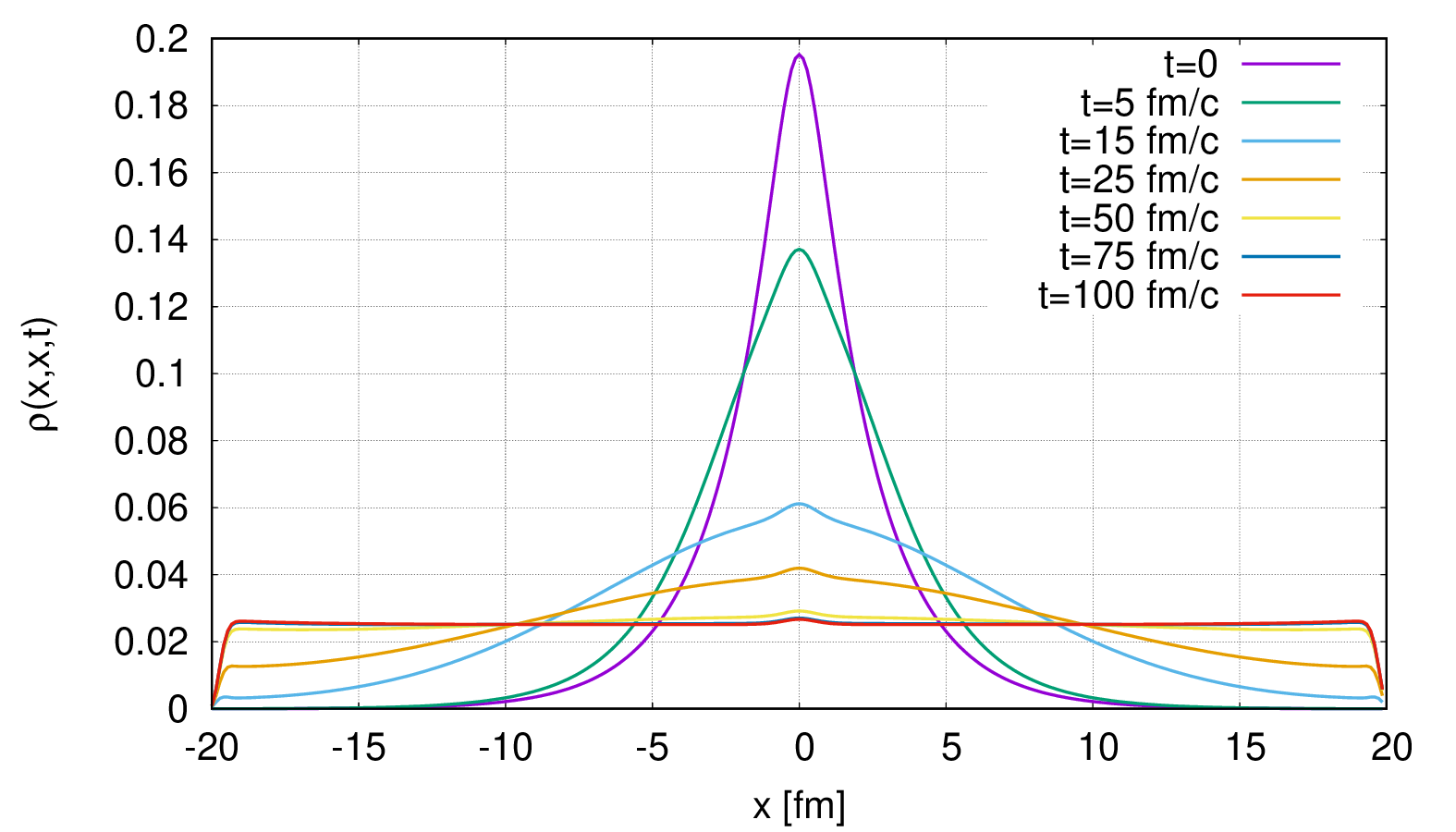}
		\hspace*{\fill}%
		
		\caption{
			$\rho(x,x,t)$ for different times $t$ for the parameters and setup illustrated in \cref{fig:3d_init1}.
		}
		\label{fig:rho_xx_init1}
	\end{center}
\end{figure}
\begin{figure}
	\begin{center}
		\hspace*{\fill}%
		\includegraphics[width=1.0\columnwidth,clip=true]{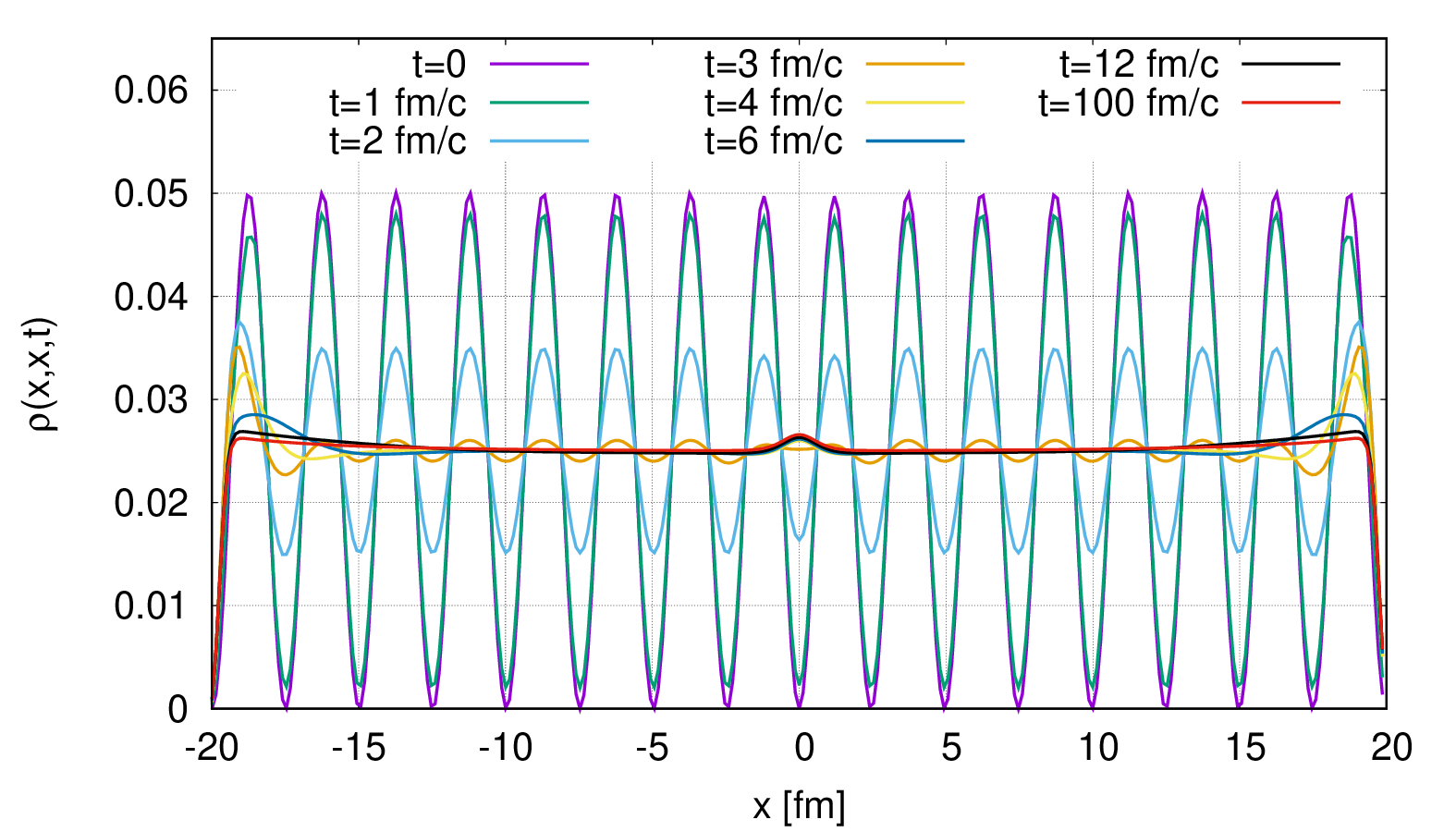}
		\hspace*{\fill}%
		
		\caption{
			$\rho(x,x,t)$ for different times $t$ for the parameters and setup illustrated in \cref{fig:3d_init10}.
		}
		\label{fig:rho_xx_init16}
	\end{center}
\end{figure}

\begin{figure*}
	\begin{center}
		\hspace*{\fill}%
		\includegraphics[width=\linewidth]{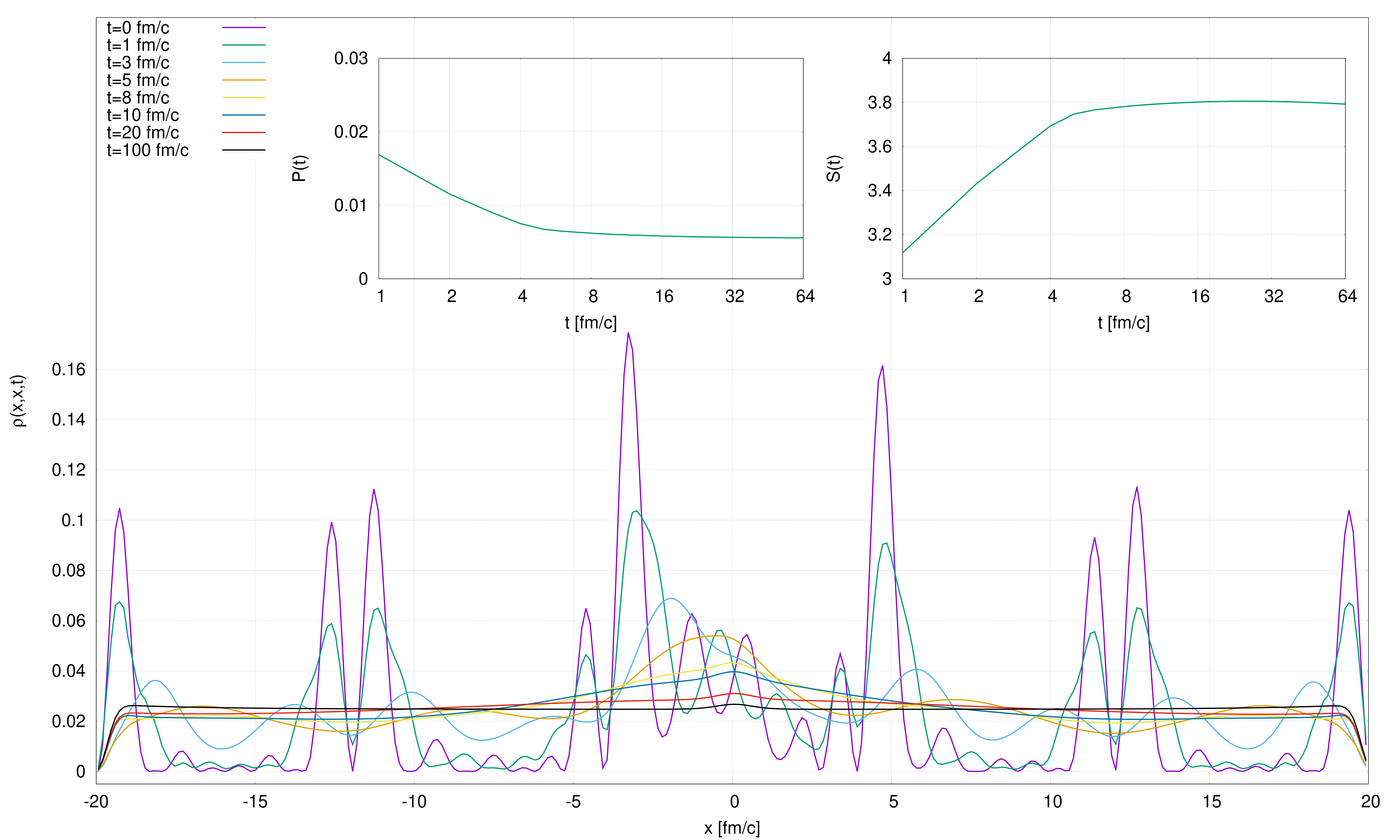}
		\hspace*{\fill}%
		
		\caption{
			$\rho(x,x,t)$, purity, and entropy for the initial condition, \cref{eq:inhom} illustrated in \cref{fig:inhom3d} for different times $t$. The parameters are given in \cref{fig:inhom3d}. The inner two  illustrations show the corresponding purity and entropy of the given initial conditions, which is discussed in \cref{sec:purity,sec:entropy}.	}
		\label{fig:inhom_xx}
	\end{center}
\end{figure*}

\begin{figure}
	\begin{center}
		\hspace*{\fill}%
		\includegraphics[width=1.0\columnwidth,clip=true]{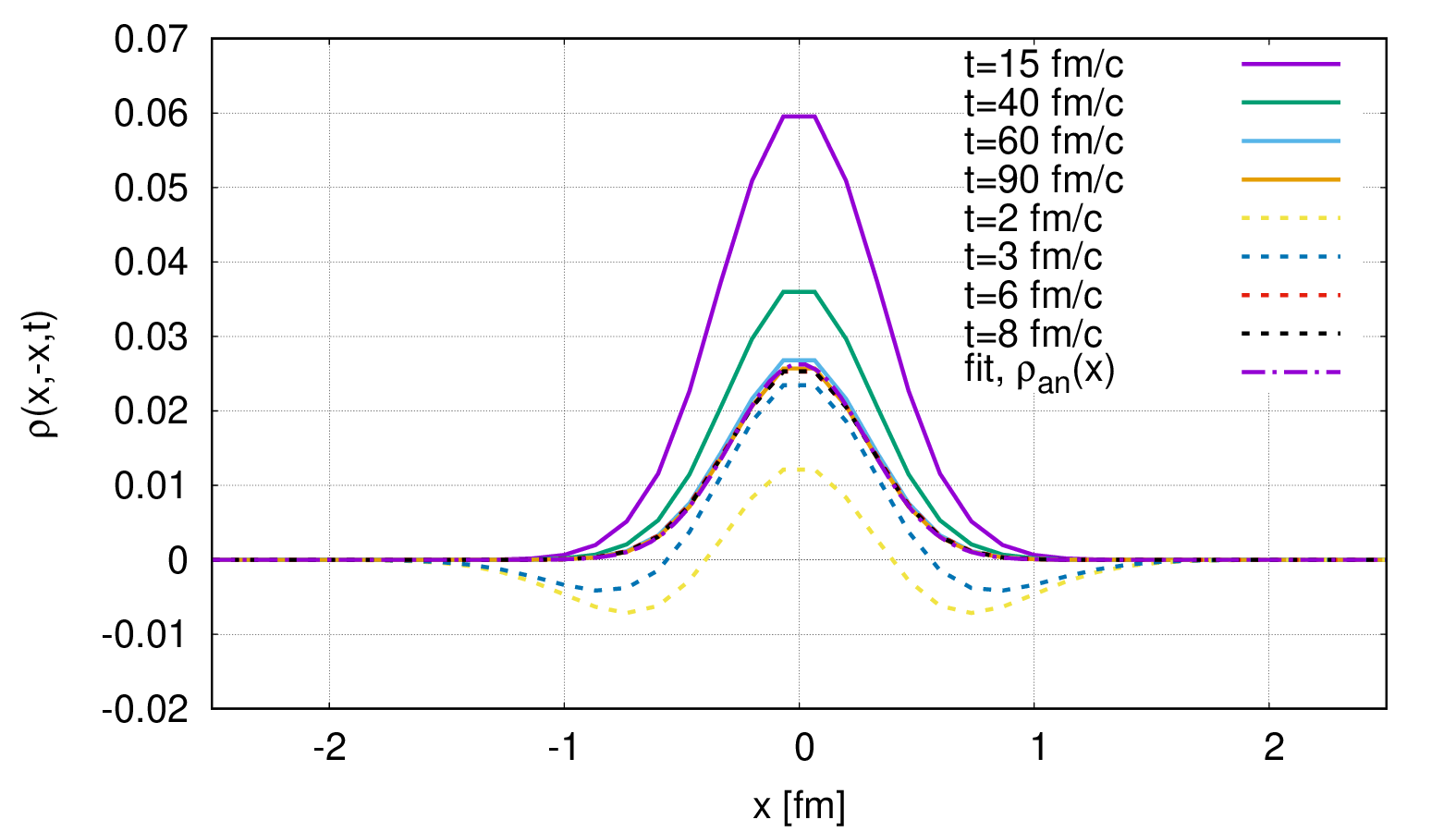}
		\hspace*{\fill}%
		
		\caption{
			$\rho(x,-x,t)$ for different times $t$ for the parameters and setup illustrated in \cref{fig:3d_init1,fig:3d_init10}. The dashed lines correspond to the initial condition, where $n=16$ and the solid lines correspond to the initial condition, where $n=0$. The dashed-dotted line corresponds to the approximated thermalized  result for $\rho(x,-x,t)$, \cref{eq:cross_diag}.
			The non-smooth shape of the curves for higher time  come from the smaller resolution, which we have generally used. For the initial condition where $n=16$ it was necessary to use a resolution of 1000x1000 grid-points.
		}
		\label{fig:rho_cross}
	\end{center}
\end{figure}


\section{Numerical method}
\label{sec:numerics}

In this section, we briefly introduce the numerical method, which we
apply in order to solve the Lindblad equation.  This is detailed in \reff
\cite{rais2024}, where also the boundary conditions and the
implementation of the Lindblad equation are discussed.  Let us mention
here, that the method we are using belongs to the large variety of
finite-volume methods, which is the underlying concept for the scheme
which we are using, the KT (Kurganov-Tadmor) central scheme, which is developed
and explicitly discussed in \reff \cite{KURGANOV2000241}.

To summarize the most important characteristics of this method, and the way it is implemented to solve the Lindblad equation, is the reformulation of  \cref{eq:Lindblad_space} into the one given by \cref{eq:dgl} as a conservative equation.

Since we are interested in the temporal evolution of $\vec{u}(t,\vec{x})$, from some initial time $t_0$ to $t_N>t_0$, we define the finite computational domain $\Omega = \mathcal{V} \times \left[t_0,t_N\right]$, where $\mathcal{V} \subset \mathbb{R}^2$  is the discretized spatial volume and $t_{0,N}$ the initial/final time.
The initial condition is therefore given by $\vec{u}(t_0,\vec{x})$, with  Dirichlet (Neumann) boundary condition specifying $(\partial_{x,y})\vec{u}(t,\vec{x})\vert_{x,y\in \partial\mathcal{V}}$.

Finite-volume methods discretize the computational domain into spatial control volumes $\mathcal{V}_i$, covering the spatial computational domain $\mathcal{V}$.
Therefore, let
\begin{align}
	\bar{\vec{u}}_i (t) \equiv \frac{1}{\mathcal{V}_i} \int_{\mathcal{V}_i} \dd \xi_x \dd \xi_y \vec{u}(t,\vec{\xi})
\end{align}
be the sliding cell average, where $\mathcal{V}_i = \left\{ \vec{\xi}: \vert \xi_x - x_i \vert \leq \frac{\Delta x}{2}, \vert \xi_y - y_i \vert \leq \frac{\Delta y}{2}\right\}$.
Then \cref{eq:dgl} can be integrated over the control volumes centred at $\vec{x}_i$, using the divergence theorem on the fluxes to obtain an ``integral form".

To evaluate these integrals to evolve the cell averages $\vec{\bar{u}}_j$ in the time from $t_n$ to $t_{n+1}$, one has to calculate the fluxes $\vec{f}^{x/y}$ and $\vec{Q}^{x/y}$ at the cell boundaries $x_{j\pm \frac{1}{2}} = x_j \pm \frac{\Delta x}{2}$ and $x_{k\pm \frac{1}{2}} = x_k \pm \frac{\Delta x}{2}$.
This requires some kind of reconstruction of $\vec{u}$ on the cell surfaces.
In general, the time step can be done in various ways, based on Riemann solvers (e.g. the Roe \cite{ROE1981357} or the HLLE solver \cite{Harten:1997,Einfeldt:1988}) or the one we are using from A. Kurganov and E. Tadmor \cite{KURGANOV2000241}.
This method is using a piecewise linear reconstruction of the cell averages and a slope limiter to avoid oscillations.
Furthermore, it has the advantage, that it has a well defined $\Delta t \rightarrow 0$ limit, while keeping the spatial directions discrete. 
This allows to use standard time integrators, which can be arbitrary and therefore do not require to fulfill the CFL condition  by hand \cite{CFL}, as it is the case in other solving methods, such as the widely used Crank-Nicholson method \cite{Crank_Nicolson_1947,10.1093/comjnl/9.1.110,10.1093/comjnl/7.2.163,8b370aba-ebed-340f-8ce2-87c0149f028b}.

For details of the implementation and the recombination method, which is a type of a Monotoc Upstream-centered Scheme for Conservation Laws  (MUSCL) \cite{Harten:1997}, we refer to the original work \cite{KURGANOV2000241} and our implementation and adoptions, cf. \reffs \cite{rais2024,Koenigstein:2021syz,Zorbach:2024rre}, where the details of the implementation are presented.
In \reff \cite{rais2024} there is also a comment on the boundary conditions and their implementation, which are chosen to be reflective for finite sized spatial domains, in order to avoid dissipation out of the computational domain.
Let us explicitly  mention, that the implementation, which we use in this work and in \reff \cite{rais2024} was originally produced for \reff \cite{Zorbach:2024rre}.

\section{Bound-States in Lindblad Approach}\label{sec:bound_state}

In this chapter, we discuss a physically motivated application, namely the P\"oschl-Teller potential, to mimic bound states and to treat the question of thermalization. 
We discuss the Lindblad dynamics of this bound state problem for various initial conditions, different parameters used in the Lindblad equation and also the dependence on the number of bound states, regarding the purity, entropy, decoherence, and population of the various states of the system. 

Furthermore, we construct an effective Hamiltonian, which describes the system-plus-bath interaction to investigate its eigenfunctions and answer the question, if a previously bound state can be shifted in its energy such, that it is no longer bound when being correlated with the thermal bath in equilibrium.

We discuss the themalization time scales of the full system and the bound state and close with a discussion, where we show the Lindblad dynamics of a system with three bound states.
This will point to the direction of describing more strongly bound particles (inspired by charmonia).

\subsection{Framework and general discussion}

In order to understand the formation of non-relativistic particles in heavy-ion collisions as it is motivated in \cref{sec:motivation}, we consider the following bound-state problem: we use a one-dimensional P\"oschl-Teller potential, which is embedded into a square-well potential.
The reason, why we have embedded the P\"oschl-Teller potential into a square well potential is to allow us full decomposability of the wave functions via $\psi(x) = \sum_n c_n \psi_n$.
This leads to normalizable states and allows the formulation in terms of matrix elements $\rho_{nm}(t)$, which are the coefficients of the projection of the spatial density matrix onto the energy eigenfunctions of this very system. 
The only constraint, this approach has to obey is, that the outer ``box" has to be large enough, such that the correlation length of the considered particle is smaller than the size of the outer walls of the potential. 
The potential is given by
\begin{align}\label{eq:potential}
	V(x) = \begin{cases}
		-V_0\,\frac{1}{\cosh^2(\alpha x)}\, , &\qquad\text{for } \vert x\vert \leq 20 \text{fm}\\
		\infty\, ,& \qquad\text{for } \vert x\vert  > 20 \text{fm},\\
	\end{cases}
\end{align}
for $\alpha > 0$. 
We follow the idea, introduced in \reff \cite{Rais:2022gfg}, but with a smoothened potential, which was already discussed in \reff \cite{DeBoni:2017ocl} from a different perspective. 
The P\"oschl-Teller-like part of the potential is indicated in \cref{fig:waves} as a black dashed line.. 
The parameter $\alpha$ is chosen such that the variance corresponds to the deuteron radius.
The outer square-well potential is located at a distance, which is much larger than the correlation length of a particle of the size of a deuteron and satisfies normalizable, discrete eigenstates, to allow an evaluation of the system in terms of decomposing the final density matrix.
However, the square-well potential forces us to compute the wave functions numerically, where we use a standard shooting method.
In our calculation, we include $N= 50$ states until we truncate the Hilbert space, such that the energy space is given by $E\in \left[-2.3 \text{ MeV},  640\text{ MeV}\right]$.
$V_{0}$ is the depth of the potential and allows us to control the number of bound states and also their binding energy.
In our case, $V_0 = 16.5$ MeV leads to the binding energy of the deuteron, namely,
\begin{align}
 E_{\text{bind}} \equiv E_0= -2.3 \text{ MeV}\, .
 \end{align}

In \cref{fig:waves,fig:energies} we plot the energy eigenfunctions of the potential \cref{eq:potential} and the corresponding energy eigenvalues, \cref{fig:energies}.
Notice the bound state in \cref{fig:energies}, which we label with $n=0$.

For the initial conditions, which we use to solve the Lindblad dynamical time evolution, given by \cref{eq:Lindblad_space}, we use
\begin{align}
	\rho(x,y,0) = \sum_{m,n = 0}^N c_{mn} \braket{x\vert \psi_m}\braket{ \psi_n\vert y}\, ,
\end{align}
where $N$ is the highest considered state, in our case $N=50$ ($E_{50}\approx 640$ MeV), and therefore the state, where the Hilbert space of the system particle is truncated. 
Generally, three different types of initial conditions are interesting for us: (1.) the case where the bound state is fully populated, $c_{mn} = \delta_{mn} \delta_{n0}$, (2.) a higher state is initially populated (we will consider the cases $c_{mn} = \delta_{mn} \delta_{nj}$, $j\in \{8,16\}$) and (3.) a mixture of states, for example $c_{mn} = \frac{1}{N}\delta_{mn}$.

\subsection{Qualitative discussion -- time evolution}
In \cref{fig:3d_init1,fig:3d_init10} we show the temporal evolution of the real and imaginary parts of the density matrix within the Lindblad framework for the initial conditions $c_{mn} = \delta_{mn} \delta_{n\,0}$ and $c_{mn} = \delta_{mn} \delta_{n\, 16} $.
In \cref{fig:inhom3d} we show the initial condition 
\begin{align}\label{eq:inhom}
	c_{mn} =\sum_{k \in \mathbb{K}} \frac{1}{5}\delta_{mn}\delta_{nk}\, ,
\end{align}
where $\mathbb{K} = \{0,9,19,29,39\}$.

The imaginary part of the density matrices are initially zero, which is the reason, why we show only plots for times $t>0$, cf. \cref{fig:3d_init1,fig:3d_init10}. 
As expected, related to decoherence, one can see, that the imaginary parts build up to a certain, maximal value and thereafter decrease until they approximately vanish up to some numerical uncertainties (note the scales on the respective plot axes).

However, decoherence can also be studied by inspecting $\sqrt{\rho_R(x,y,t)^2 + \rho_I(x,y,t)^2 }$, which is illustrated in \cref{fig:decoherence}. 
Here we show $\rho_{mn}(t)$, which can be calculated with \cref{eq:coeff}, where $L = 20$ fm and for the case, where $m=0$, so that we are regarding only correlations to the bound states.
At this point, this choice of the off-diagonal entries is completely arbitrary and is only done to provide an insight to the behaviour of the off-diagonal elements. 
Since the main feature of decoherence is that a many body system shows classical behaviour while thermalizing, we expect, that the off-diagonal elements vanish for some finite times.
 In \cref{fig:decoherence} this can be seen for the given initial conditions $n=0, 8$,´ and $16$. 
 Also for all initial conditions, the decoherence time, the time until the off-diagonal entries vanish or at least are constant (which is the case close to the diagonal) is similar for all cases.
 In the case, where $m=0$ this is approximately 20 fm/c and for an initial condition, where $n=0$ is lowest, but does not diverge notably also for $n=8, 16$.
 This shows, that decoherence length is not directly related to thermalization, because, as can be seen in \cref{fig:rho_xx_init1}, at $t=15$ fm/c the system is far from equilibrium, and also the thermalization times decrease significantly for cases, where the initial condition is composed of a higher occupied state.
 One can see, that the off-diagonal parts are populated immediately, and thereafter decrease.
However, for the cases, where $n = 8,16$ initially, there is a second peak in \cref{fig:decoherence} at some later time $t\approx 10$ fm/c.
This can be interpreted as an additional correlation between the states, because at this time, the imaginary part is already very small. 
However, transitions take place at larger time scales and apparently the second peak corresponds to the process, where correlations between the $m$-th state to the bound state are more probable.
Also boundary effects, such as the reflection at the boarders of the square well in the coordinate space have to be taken into account and may cause this second peak. 
In \reff \cite{rais2024}, we have shown, that in the case, where a system without dissipation is considered, the density matrix indeed shows interference patterns, which are caused by the boarders of the system.
Also in the case, where dissipation is included, for initial conditions, which cause a  notable population of the density matrix close to the boarders of the system, these boundary effects affect the system. 
At this point, and due to the fact, that this dynamics are highly non-trivial, we can only speculate about a reasonable description of the decoherence dynamics in \cref{fig:decoherence}. 
A deeper understanding of this might be obtained by discussing the decoherence of this process separately in phase space, regarding the Wigner transformation.
This is however beyond the scope of the present work and discussed elsewhere \cite{Rais2025}.

\cref{fig:rho_xx_init1,fig:rho_xx_init16} also demonstrate, that although the initial condition is completely different for both cases, the same stationary (and potentially thermal) equilibrium is reached.
The same is found in \cref{fig:inhom_xx} for the initial condition, where five states are initially populated. 
In this figure, we also show the purity and entropy of this system, which is discussed in detail in \cref{sec:entropy,sec:purity}.
Let us mention here, that for most of the time evolutions shown in  \cref{fig:inhom_xx} the pattern is not symmetric.
The reason is that the initial condition, \cref{eq:inhom} is a superposition of symmetric and antisymmetric wave functions. 
Remarkably, this does not influence the final result, which is almost perfectly symmetric, and the norm of the density matrix, $\int \dd x \rho(x,x,t)  = 1$, which is, during the calculation violated only up to 2\% after 100 fm/c, even for this highly oscillation setting.




\subsection{Thermalization of the bound state problem}\label{sec:thermalization}

\begin{figure*}
	\begin{center}
		\hspace*{\fill}%
		\includegraphics[width=2.0\columnwidth,clip=true]{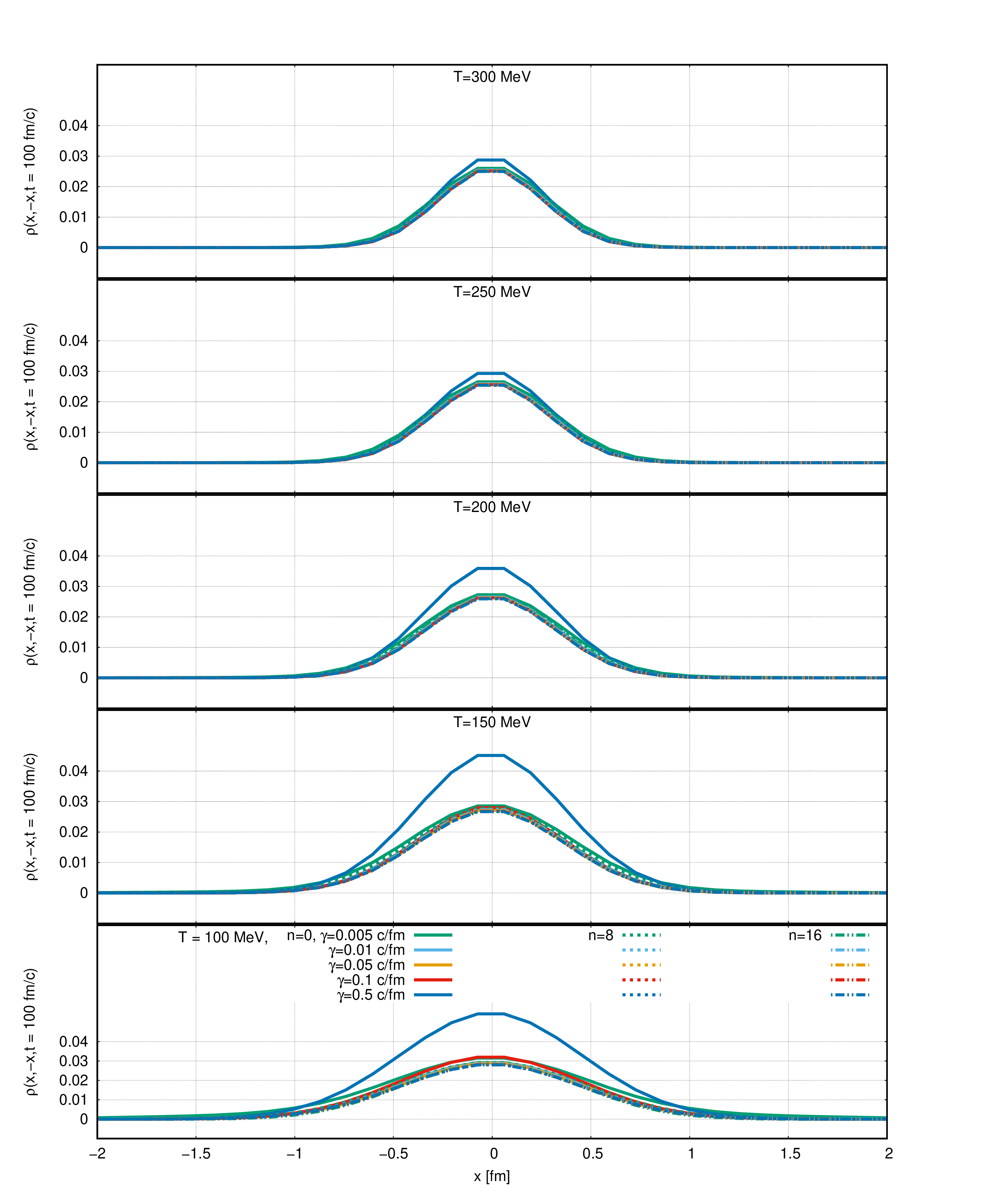}
		\hspace*{\fill}%
		
		\caption{
			$\rho(x,-x,t)$ for the pure Caldeira-Leggett master equation, with $D_{px} = 0$ and $\Omega = 4T$ for different bath temperatures $T$. 
			The different colors correspond to different damping coefficients $\gamma$ and the different line-types correspond to the different initial conditions $n=0, 8,16$.
		}
		\label{fig:cross_4T_0Dpx}
	\end{center}
\end{figure*}

\begin{figure}
	\begin{center}
		\hspace*{\fill}%
		\includegraphics[width=1.0\columnwidth,clip=true]{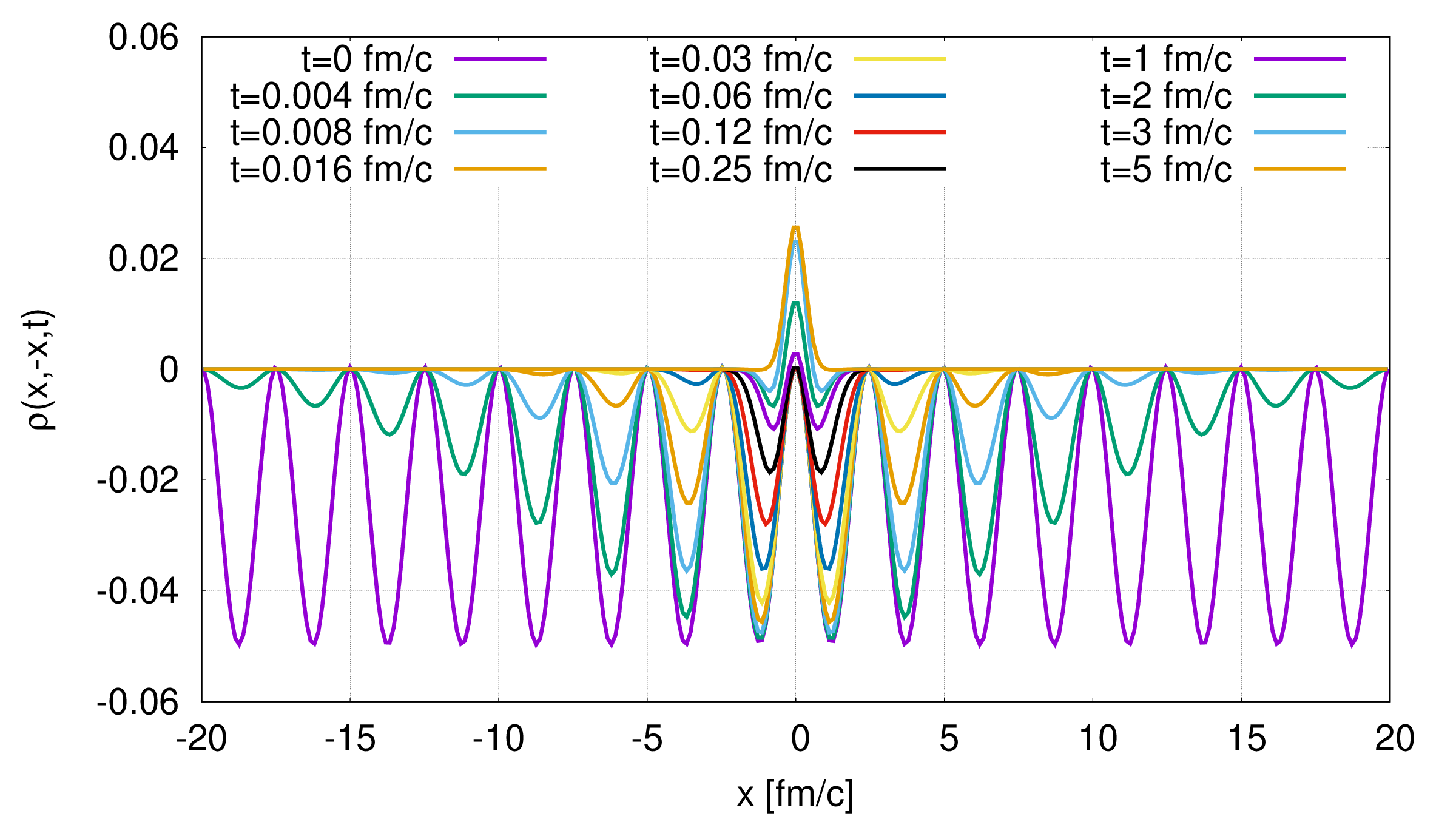}
		\hspace*{\fill}%
		
		\caption{
			$\rho(x,-x,t)$ at times right after bringing the system in contact with the bath,  where the state $n=16$ is originally populated. 
			Here, the temperature of the heat bath is $T=200$ MeV, $\gamma = 0.1$, $\Omega = 4T$ and $D_{px} = -\gamma T /\Omega$.
				}
		\label{fig:early_times}
	\end{center}
\end{figure}

\begin{figure}
	\begin{center}
		\hspace*{\fill}%
		\includegraphics[width=1.0\columnwidth,clip=true]{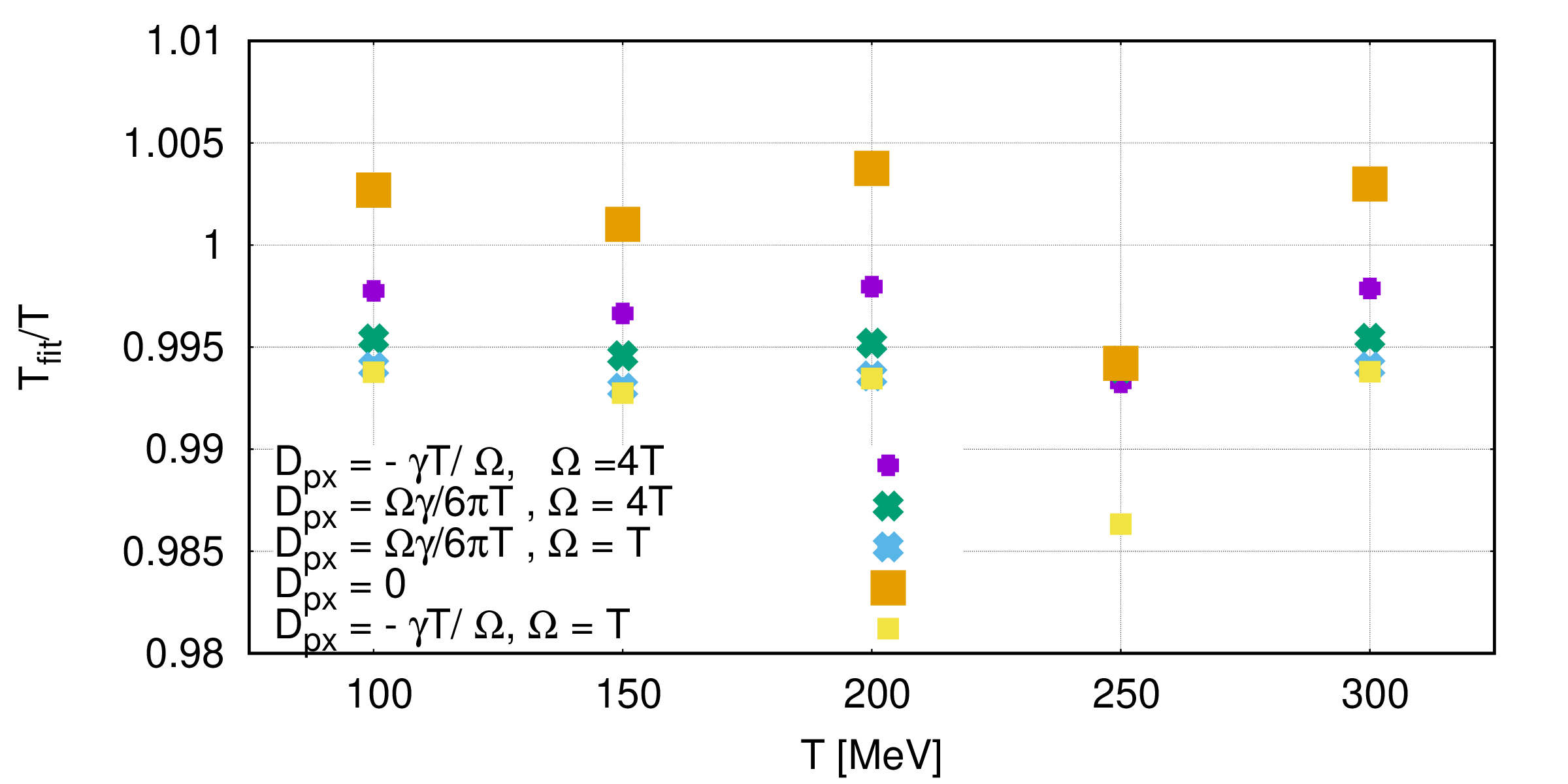}
		\hspace*{\fill}%
		
		\caption{
			The fit temperature of the function given in \cref{eq:cross_diag}, which is taken to be the fit function of the off-diagonal $\rho(x,-x,t)$ normalized to the heat bath temperature $T$ for different parameter combinations for $D_{px}$ and $\Omega$ and a fixed medium value for $\gamma = 0.1$ c/fm. At $T=250$ MeV all points except the yellow one over overlaying each other. 
		}
		\label{fig:T_fit}
	\end{center}
\end{figure}

\begin{figure}
	\begin{center}
		\hspace*{\fill}%
		\includegraphics[width=\linewidth]{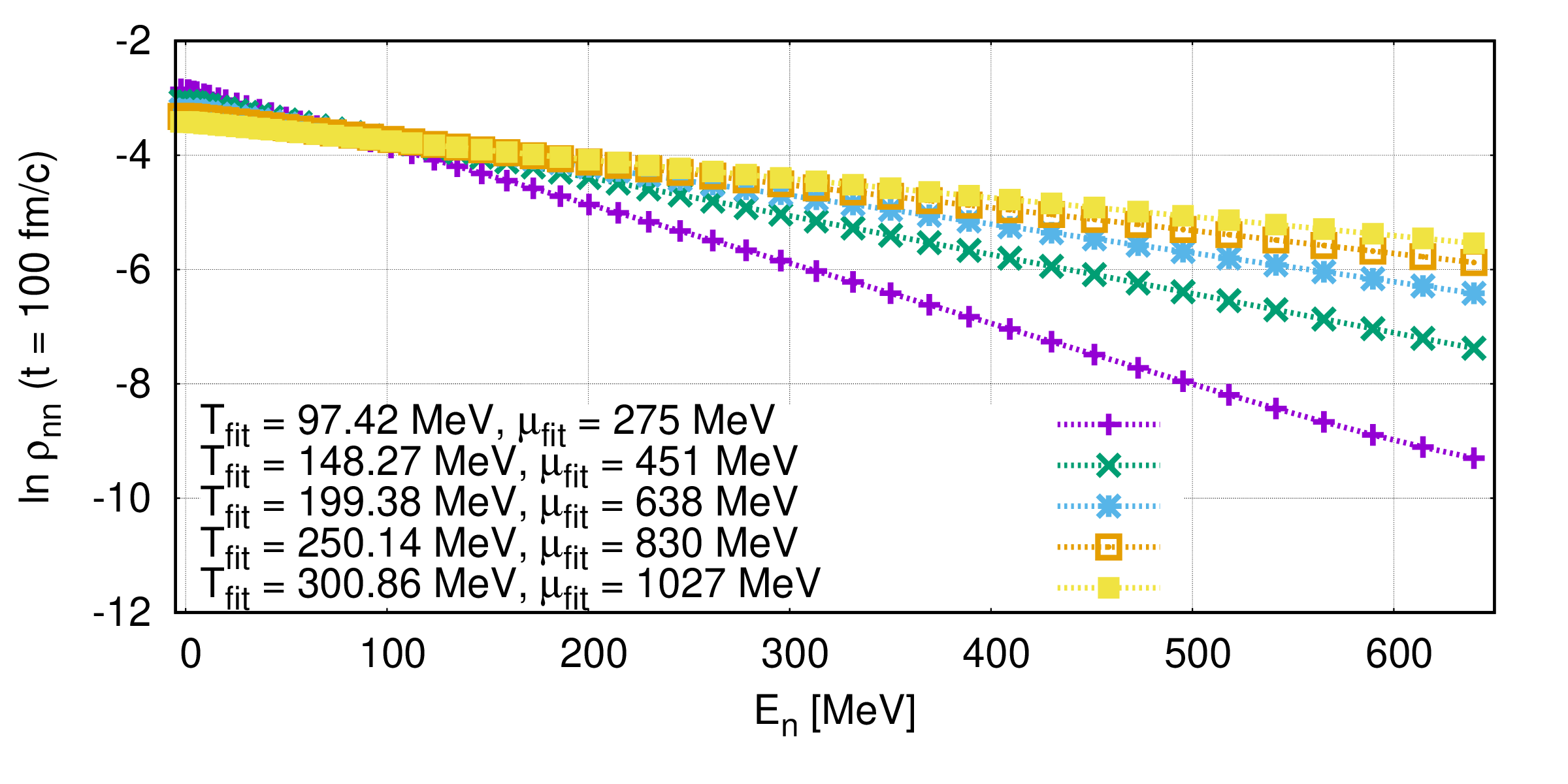}
		\hspace*{\fill}%
		\caption{
			Logarithm of the final distribution of $\rho_{nn}(t)$ at time $t_{\text{eq}} = t = 100$ fm/c for different heat bath temperatures $T$. 
			Here, $\gamma = 0.1$ c/fm, $\Omega = 4T$, $D_{px} = 0$ and the initial condition is $n=8$. 
			The dotted lines are the fitted distributions following \cref{eq:Boltzmann}, which is also  used to fit the temperature, which is described in \cref{tab:fit_temperatures} for all parameter sets.
		}
		\label{fig:Boltzmann_fit}
	\end{center}
\end{figure}

\begin{figure}
	\begin{center}
		\hspace*{\fill}%
		\includegraphics[width=1.0\columnwidth,clip=true]{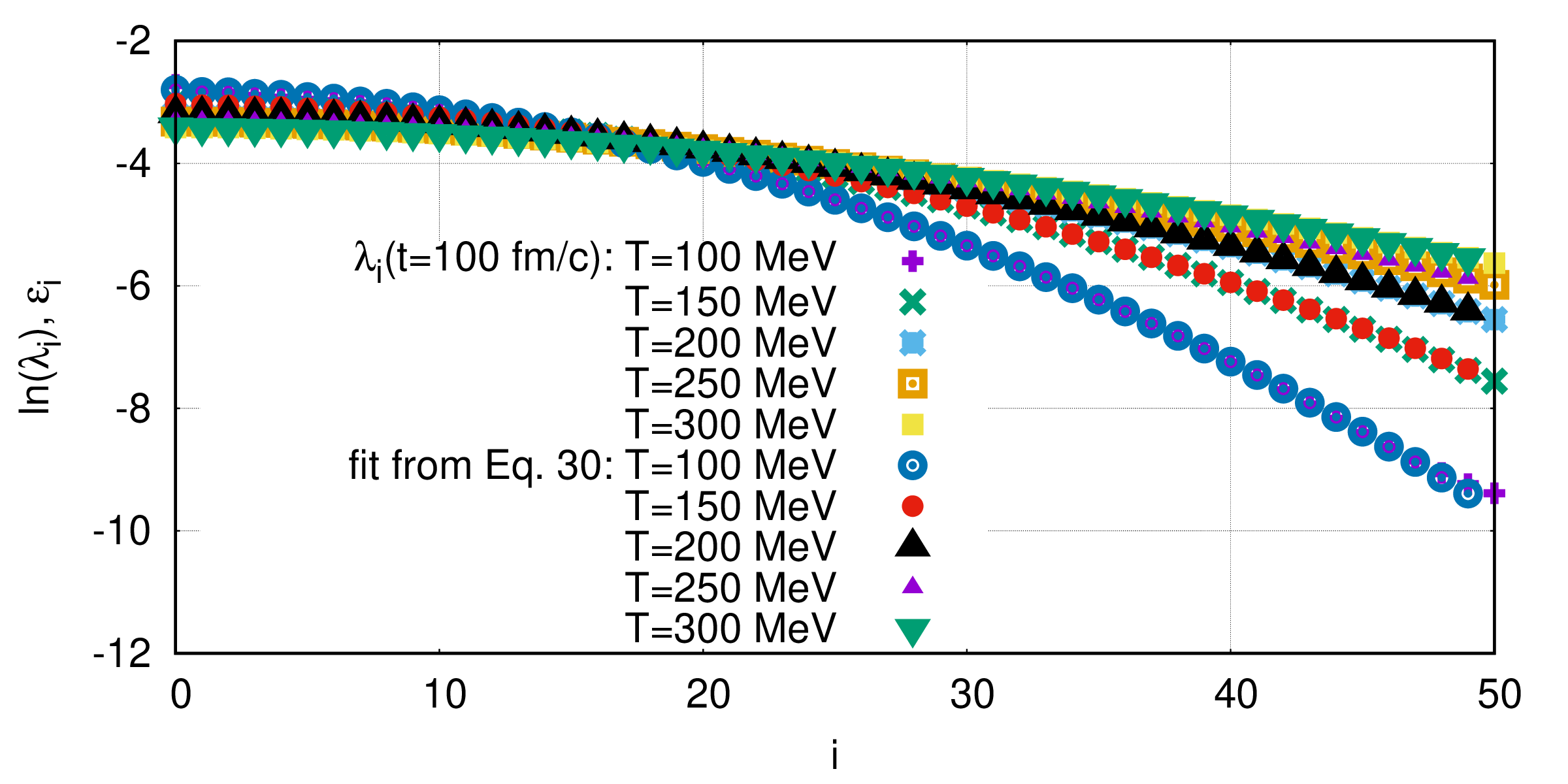}
		\hspace*{\fill}%
		
		\caption{
			Logarithm of the eigenvalues $\lambda_i$, \cref{eq:effectiveE}, compared to the argument of \cref{eq:Boltzmann}, \cref{eq:argument} for different heat bath temperatures $T$ and a damping $\gamma = 0.1$ c/fm, cutoff frequency $\Omega= 4T$ for the pure Caldeira-Leggett model, $D_{px} = 0$, at time $t=100$ fm/c. The parameters $T$ and $\mu$ of $\epsilon_i$ are the fitted values obtained also \cref{fig:Boltzmann_fit}.
		}
		\label{fig:eigenvalues}
	\end{center}
\end{figure}

\begin{figure*}
	\begin{center}
		\hspace*{\fill}%
		\includegraphics[width=\linewidth]{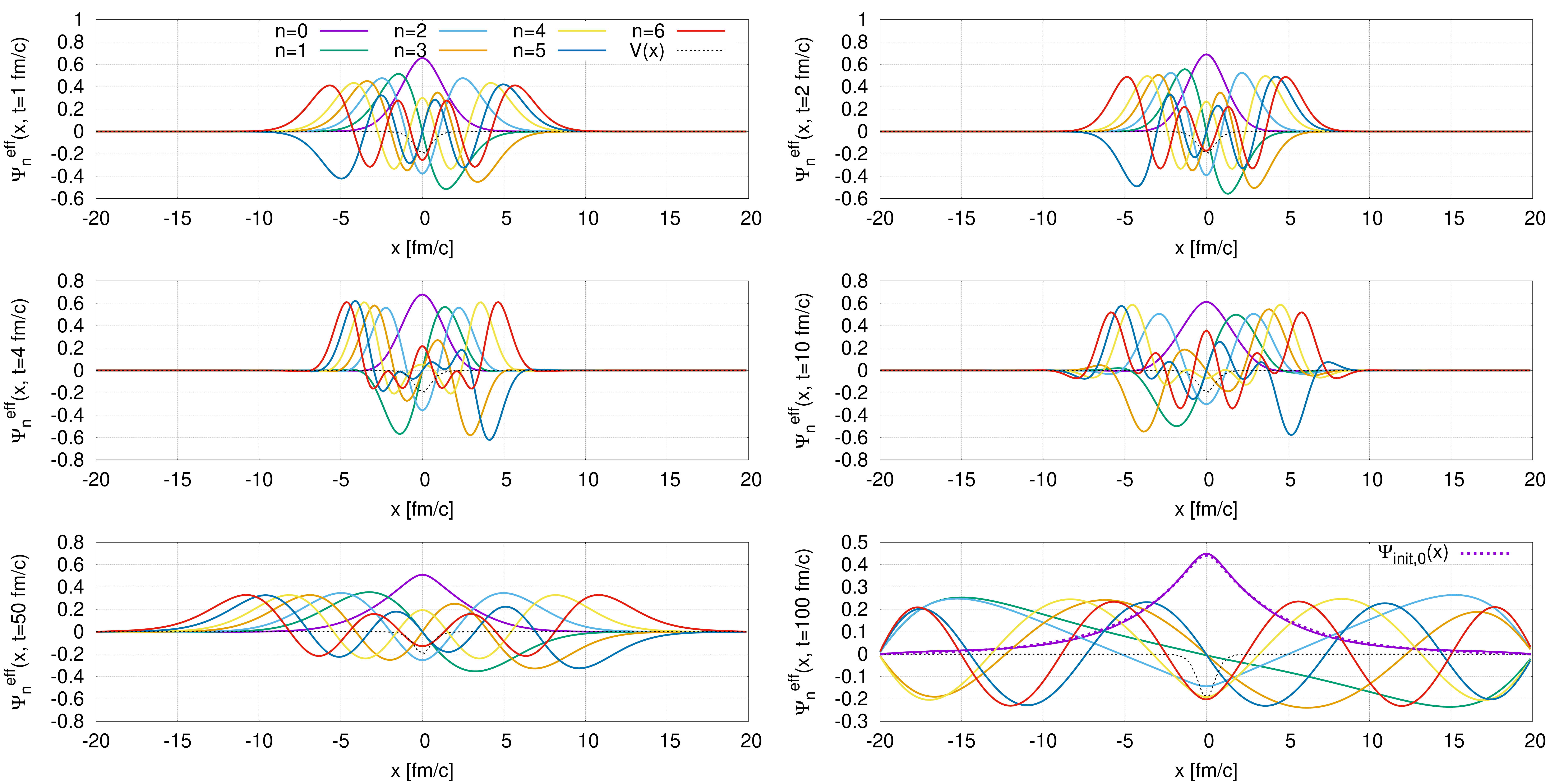}
		\hspace*{\fill}%
		
		\caption{
			Effective wave functions (system-plus-interaction) as the eigenvectors of the diagonalized density matrix $\rho(x,y,t)$ at different times $t$ for an initial condition, where $n=0$, the bound state is originally populated. The dashed  line illustrates the rescaled potential, cf. \cref{fig:waves} and  \cref{eq:potential}. Here, $T = 200$ MeV, $\gamma = 0.1$ c/fm, $\Omega = T$, and $D_{px} = - \gamma T /\Omega$.
		}
		\label{fig:wave_eff_init1}
	\end{center}
\end{figure*}

		

\begin{figure*}
	\begin{center}
		\hspace*{\fill}%
		\includegraphics[width=\linewidth]{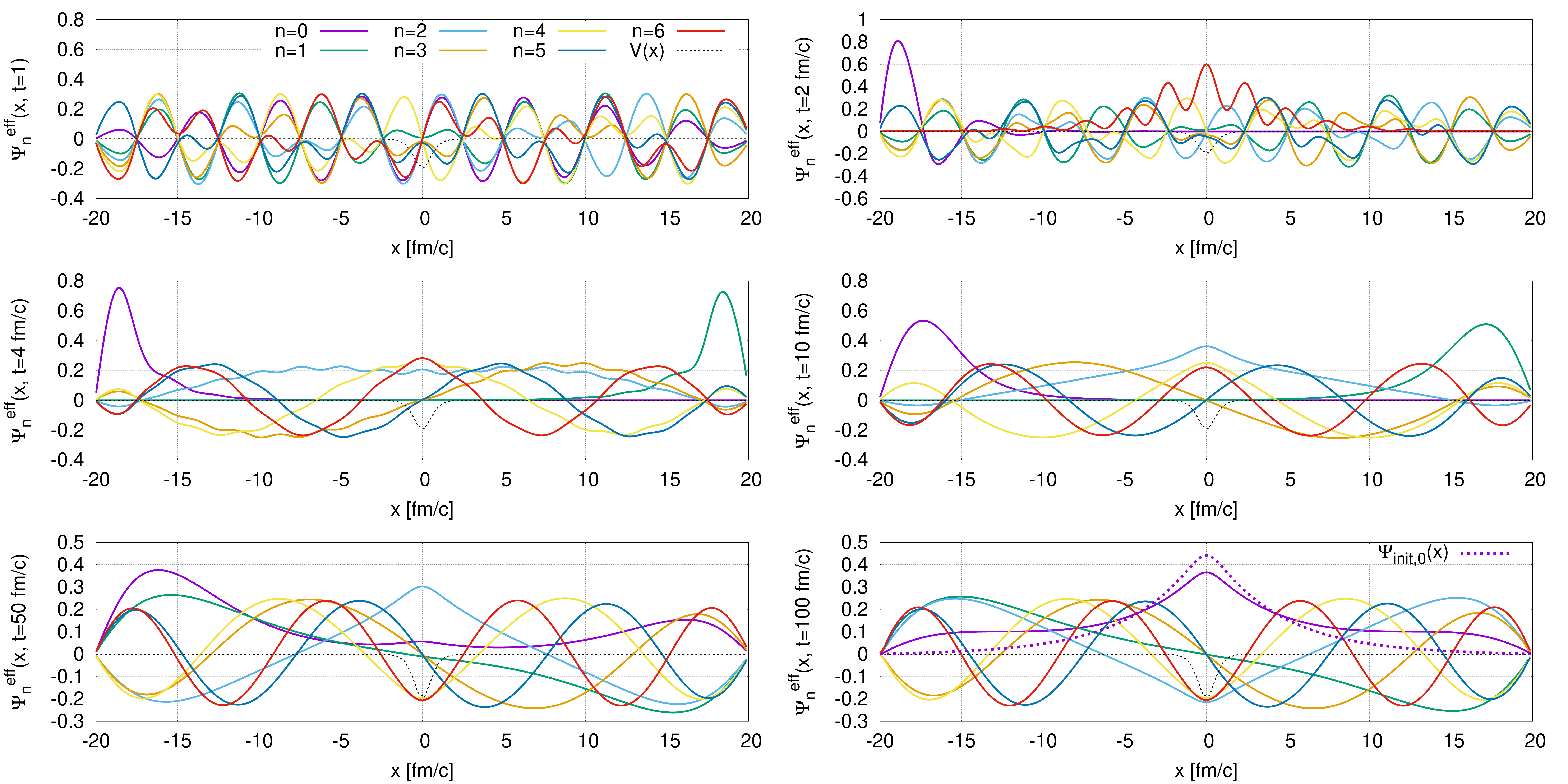}
		\hspace*{\fill}%
		
		\caption{
		Effective wave functions (system-plus-interaction) as the eigenvectors of the diagonalized density matrix $\rho(x,y,t)$ at different times $t$ for an initial condition, where $n=16$. The black dashed line illustrates the rescaled potential, cf. \cref{fig:waves} and  \cref{eq:potential}. The parameters are the same as in \cref{fig:wave_eff_init1}.
		}
		\label{fig:wave_eff_init16}
	\end{center}
\end{figure*}

		

Before we start discussing thermalization, let us mention, that we have several parameters, the diffusion constant $D_{px}$ and $\gamma$ at a given temperature $T$, which we are allowed to choose in order to solve the Lindblad equation properly.
In \reff \cite{DIOSI1993517} and \cite{DEKKER198467} it is derived that $D_{p x}$ can either be $D_{p x} = -\frac{\gamma T}{\Omega}$, high temperature limit, and  $D_{p x} = \Omega \gamma / 6\pi T$ in the medium temperature limit.
In the pure Caldeira-Leggett master equation $D_{px} = 0$ which brings up the question, if solving the pure Caldeira-Leggett master equation (which is not in Lindblad form) leads to thermalization, too.

Another important aspect is given by the ``right" choice of the cutoff-frequency $\Omega$, and its impact on the final thermalization. 
This is justified, because we have seen in \reff \cite{rais2024}, that for the case, where $V(x)$ is the harmonic potential, the deviation of the analytical result of the CLME from the barometric formula in the thermal distribution is a cutoff-effect, which is smaller, if the temperature is higher, or if the cutoff-frequency is lower.
Therefore, we additionally discuss the cases, where $\Omega = T$ and $\Omega = 4T$.
Exemplary, we show the evolution of  $\rho(x,-x,t)$ for different times, and \cref{fig:rho_cross}, and the final distribution of $\rho(x,-x,t)$ for the case, where $T = 4\Omega$ and $D_{px} = 0$, cf. \cref{fig:cross_4T_0Dpx}, for different temperatures and damping. 
As we can see in \cref{fig:cross_4T_0Dpx}, the distribution gets narrower for higher temperatures.
This is expected from \cref{eq:cross_diag} and also discussed below.

\begin{figure*}
	\begin{center}
		\hspace*{\fill}%
		\includegraphics[width=2.0\columnwidth,clip=true]{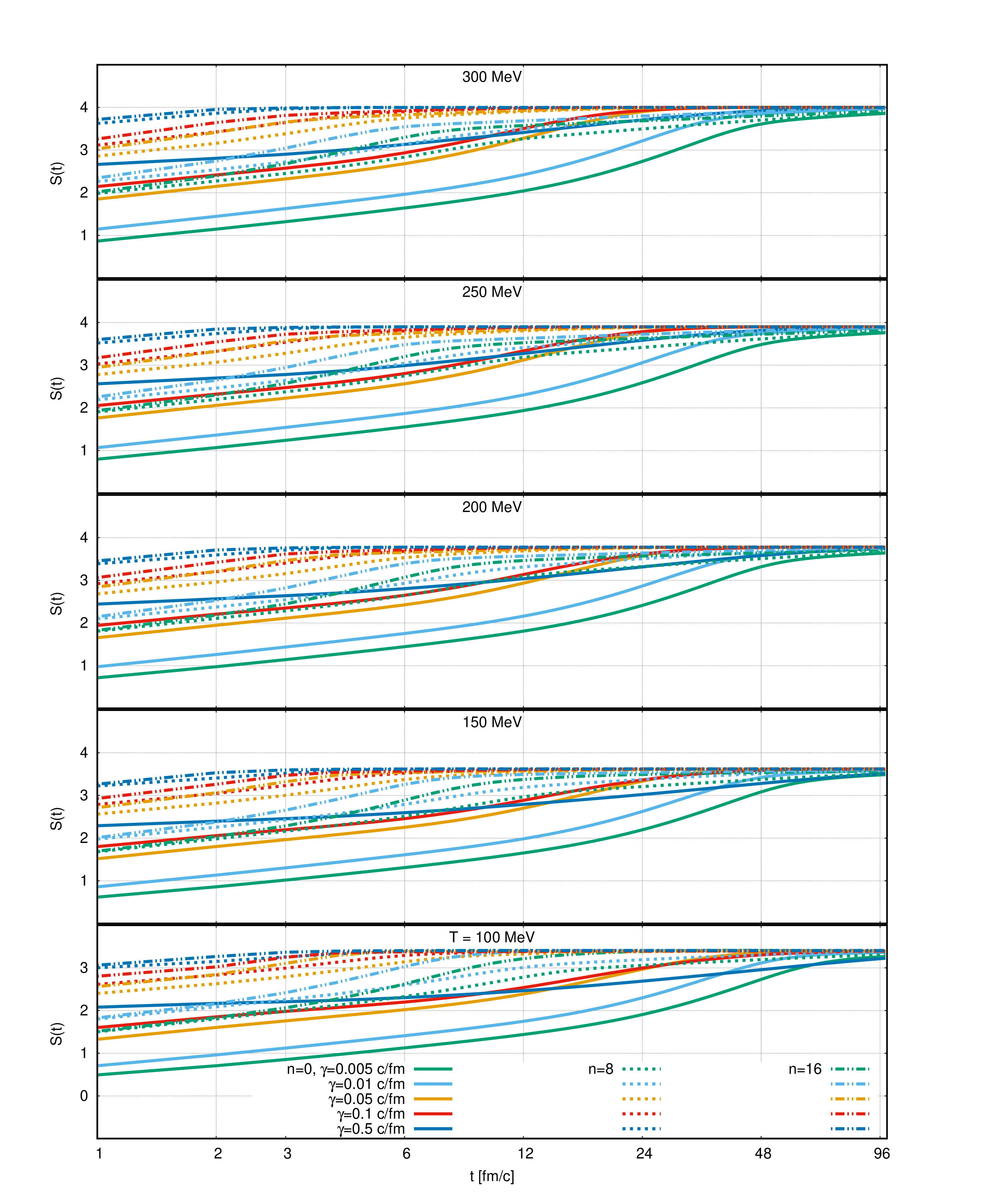}
		\hspace*{\fill}%
		
		\caption{
			Entropy $S(t)$ for different bath temperatures $T$ with $D_{px}=0$ and $\Omega = 4 T$. The different line colors correspond to different damping $\gamma$, while the different line-styles correspond to different initial conditions $n=0,8,16$.
		}
		\label{fig:entropy_4T_0Dpx}
	\end{center}
\end{figure*}

In the cases, where the temperature is lower (100-200 MeV), the case, where the bound state is initially populated, is not fully equilibrated, even after 100 fm/c, cf. \cref{fig:entropy_4T_0Dpx} and later discussion.
One can see, that especially in the case, where we expect the fastest equilibration of the bound state, $\gamma=0.5$ c/fm, it takes the system longer to equilibrate, than in all the other cases. 
This seems to be related to an over-damping of the system and will be detailed in \cref{sec:bound}.
Also the final distribution is not dependent on the damping $\gamma$.

To answer the question, which choice of $D_{px}$ and $\Omega$ is most suitable for our purposes, we have to compare all possible combinations for different bath temperatures $T$ and damping coefficients $\gamma$. 

Let us turn to the question of thermalization. 
Having discussed the general behavior of the density matrix of a particle in a potential \cref{eq:potential}, with different initial conditions, we want to investigate the thermalization of this particle surrounded by a thermal heat bath. 
We know, that the ``thermal" state has to be the same for all initial conditions. Therefore
\begin{align}
	\frac{\rho_i(x,y,t_{\text{eq}})}{\rho_k(x,y,t_{\text{eq}})} =1\, ,
\end{align} 
where $i,k$ indicate  initial conditions with $i \neq k$. 
Here, $t_{\text{eq}}$ is some large time, where the system is stationary and therefore  $\partial_t \rho(x,y,t_{\text{eq}}) = 0$.
Making sure, that this condition is fulfilled, we can tackle the question if the stationary  system at large times is really following a thermal statistical distribution, which for high temperatures and bosons should be the Boltzmann distribution.
In general, we have two possibilities to investigate thermalization. 
In \reff \cite{rais2024} we have shown, that the distribution of the matrix entries which are orthogonal to the diagonal follow 
\begin{align}\label{eq:cross_diag}
	\rho(x,-x,t) \sim \ee^{-2mTx^2}\, ,
\end{align}
for a free particle. 
Here, we argue, that the width of the potential, as well as its binding strength are small in comparison to the total size of the computational domain, and therefore the final distribution should be approximately the one of the free particle.

The reason, why we do not comment on the amplitude of this distribution is, that we can not obtain an analytical result for the given potential. 
Fitting \cref{eq:cross_diag} into the numerical results and considering $\rho(x,-x,t)$ allows us to extract a temperature for different values of $\gamma$ and $T$ in different limits.

Before we inspect the fitted temperatures, which we obtain, using \cref{eq:cross_diag}, let us have another look at \cref{fig:3d_init10}, where we can see, that the off-diagonal parts, which are populated at this initial condition seem to be depopulated already after the first $0.25$ fm/c.
To analyze the behavior of this off-diagonal part at the very beginning of the simulation, we plot $\rho(x,-x,t)$ also for early times, right after the bath and the system are brought into contact with each other, cf. \cref{fig:early_times}.
Due to the special choice of the initial condition, one can see, that  $\rho(x,-x,0)$ is purely negative and then starts to develop its final form very fast.

Turning to thermalization, in \cref{fig:T_fit} we show the fit temperatures, using \cref{eq:cross_diag} normalized to the bath temperature for the different choices of $D_{px}$ and the comparison for larger and smaller cut-off. 
The reason, why we do not care about the cut-off frequency in the case, where we consider the pure Caldeira-Leggett model is, that $\Omega$ does not appear in the Lindblad equation at all, if $D_{px} = 0$.

At this point, the intermediate result of the analysis of the off-diagonal  $\rho(x,-x,0)$ in terms of thermalization is that $T_{\text{fit}} \approx T$, the bath temperature, with an maximum deviation of $\pm 1 \%$, even though we are using the solution of the free particle, \cref{eq:cross_diag}. 

The second way to analyze, whether the system is thermalized in some stationary state at some large time, is to evaluate the matrix elements $\rho_{nn}$ calculated via \cref{eq:coeff}
at some large time $t=t_\text{eq}$, where the system is stationary  and therefore should possibly follow the Boltzmann distribution, given by
\begin{align}\label{eq:Boltzmann}
	\rho_{nn} = \exp\left[-\frac{1}{T}(E_n - \mu) \right] \, .
\end{align}
Then one can use $T$ as a fit parameter, allowing for  a free parameter $\mu$, the chemical potential.
 The results for the considered values of $\gamma, T$ in the high temperature limit, where $D_{p x} = -\frac{\gamma T}{\Omega}$ and in the medium temperature limit, where $D_{p x} = \Omega \gamma/6\pi T$ are summarized in \cref{tab:fit_temperatures} and exemplary depicted for one parameter set in \cref{fig:Boltzmann_fit}.
 
\begin{table}[h!]
	\begin{center}
		\begin{tabular}{c|c|c|c} 
			\textbf{$D_{px} = 0$}&  &&\\
			$T$ [MeV] & $\gamma = 0.01$ & $\gamma = 0.1$ & $\gamma = 0.5$ \\
			\hline
			100 &96.74 & 98.79 &100.51\\
			150 & 145.88&149.48 & 152.14\\
			200 & 194.72&199.85  &203.01\\
			250 &243.41 &250.09 &253.66 \\
			300 &292.03 & 300.30 &304.26\\
			\hline
			\textbf{$D_{px} =-\frac{\gamma T}{\Omega} $} &    & & \\
			$\Omega = 4T$ &    & & \\
			$T$ [MeV] & &  &  \\
			\hline
			100 & 96.74 &99.08&102.31\\
			150 &145.89 & 149.71&152.90\\
			200 & 194.71& 199.99 &203.35\\
			250 & 243.363 & 250.13& 253.72\\
			300 &297.92 & 300.23 &304.10\\
				\hline
			 \textbf{$D_{px} =\frac{ \Omega\gamma}{6\pi T}$}&  & &  \\
			 $\Omega = 4T$ &    & & \\
			$T$ [MeV] &  &  & \\
			\hline
			100 &96.70 & 98.46 &96.12\\
			150 & 145.85&149.19&150.71 \\
			200 &194.67& 199.56 &202.22\\
			250 &243.34& 249.78& 253.13\\
			300 & 291.91& 299.95 &303.82\\
				\hline
			\textbf{$D_{px} =-\frac{\gamma T}{\Omega} $}&  & &  \\
			$\Omega = T$ &    & & \\
			$T$ [MeV] & & & \\
			\hline
			100 & 96.62 & 99.66 & \\
			150 &145.83 &150.00& \\
			200 &194.45& 199.81 &\\
			250 &242.78&249.25 & \\
			300 & 290.73& 298.55 &\\
				\hline
			\textbf{$D_{px} =\frac{ \Omega\gamma}{6\pi T}$}&  & &  \\
			$\Omega = T$ &    & & \\
			$T$ [MeV] &  & &  \\
			\hline
			100 &96.73 &98.71 &99.82\\
			150 & 145.88&149.41& 151.88\\
			200 &194.71& 199.79 &202.86\\
			250 &243.40&250.03 & 253.57\\
			300 &292.02 & 300.24 &304.20\\
	\end{tabular}
	\end{center}
	\caption{Fit temperatures of the Boltzmann distribution, \cref{eq:Boltzmann}, for different heat bath temperatures $T$ and different diffusion coefficients $D_{px}$ and different cutoff-frequencies $\Omega$ and for initial condition $n=8$.}
	\label{tab:fit_temperatures}
\end{table}

Comparing \cref{fig:T_fit} with \cref{tab:fit_temperatures} and its corresponding illustration, \cref{fig:Boltzmann_fit}, one observes very good agreement between both ways to fit the system temperature $T$, which shows, that in the stationary case, the system (particle in a P\"oschl-Teller potential) takes the temperature of its thermal environment up to a violation of about 1\%.
The fact that the temperature is not perfectly matching the bath temperature,  has to the best of our knowledge mostly three reasons: (1.) In the case, where we use $\rho(x,-x, t)$ to fit the distribution of a free particle, \cref{eq:cross_diag} and not the one of the Poeschl-Teller potential.
(2.) As was shown in \reffs \cite{Homa2019,Bernad2018}, during the time evolution, the modes of the wave functions can get shifted due to the coupling to the bath, and therefore also the energy eigenvalues of the system. 
Since we are projecting $\rho(x,y,t)$ on the set of unperturbed energy eigenfunctions functions, this can lead to a small deviation.
And (3.), we want to remind the reader, that we are using only $300\times300$ cells on the computational domain, which limits the quality of the norm conservation and was thoroughly discussed in \reff \cite{rais2024}.
However, these are only suggestions, because perfect thermalization of the Lindblad equation was, to the best of our knowledge, never proved mathematically, and small deviations of the heat bath temperature were also observed in other works, as for example \reff \cite{Delorme:2024rdo}.  

Disregarding the deviations, one can see, that for the temperature regime, under consideration, we obtain slightly better results, if the cutoff frequency $\Omega$ is chosen to be large in comparison to the temperature.
This corresponds to the case $\Omega = 4T$.
For setups with $\Omega = T$ the choice of $D_{px}$ has seemingly no impact on the final distribution, but shows the largest deviation. 
One can see, that for all computations, where $\Omega = 4T$ and a damping coefficient $D_{px} = \Omega \gamma/6\pi T$, the coefficient where medium temperatures are assumed, \reff \cite{DIOSI1993517}, the fitted results for the temperature deviate slightly more from the bath temperature, than the ones, where $D_{px} = -\gamma T/\Omega$ (high temperatures \cite{BRE02}), which is closest to the bath temperature, in the cases, where $D_{px} \neq 0$, which satisfies a Lindblad structure. 
The cases, where we solve the pure Caldeira-Leggett equation ($D_{px} = 0$) are usually slightly above the bath temperature. 
For $D_{px} = - \frac{\gamma T}{\Omega}$ and $\Omega = T$ with a comparatively high damping coefficient of $\gamma=0.5$ c/fm we do not provide any results in \cref{tab:fit_temperatures}, because the norm of the density matrix for this parameters is violated significantly.
This is reasonable, because we have already discussed in \cref{fig:cross_4T_0Dpx}, that for $\gamma = 0.5$ c/fm the system is not stationary yet, even though for lower damping coefficients it is. 
As we argued above, this reminds us the over-damped case of the harmonic oscillator, and possibly corresponds to a parameter setup, where fundamental assumptions in the Lindblad equation are violated, and therefore reliable results of the evolution cannot be provided any more.

\subsection{Wave function of the effective Hamiltonian}\label{sec:wafefkt}

\begin{figure}
	\begin{center}
		\hspace*{\fill}%
		\includegraphics[width=1.0\columnwidth,clip=true]{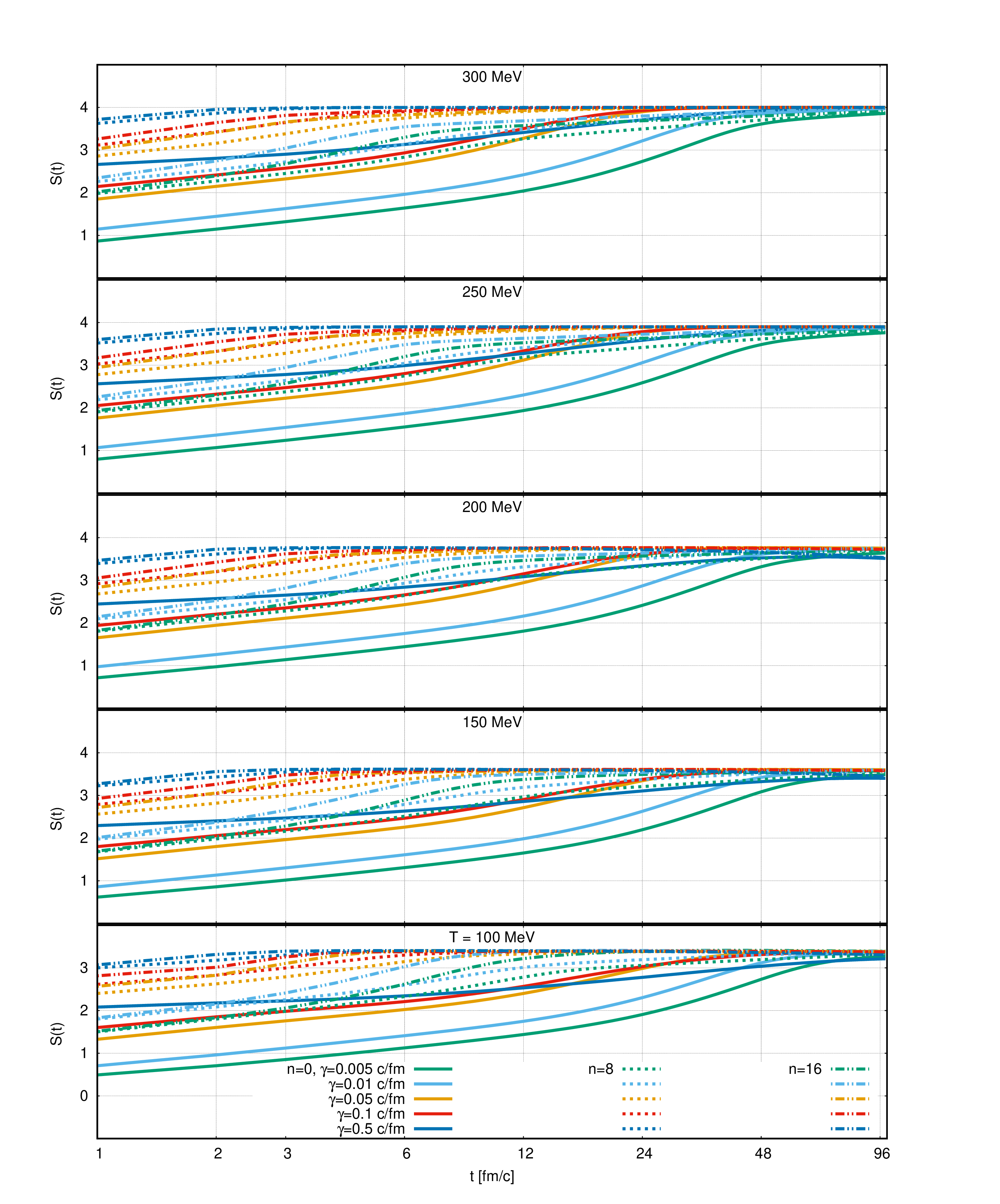}
		\hspace*{\fill}%
		
		\caption{
			Entropy $S(t)$ for different bath temperatures $T$ with $D_{px}=-\gamma T/\Omega$ and $\Omega = 4 T$. The different line colors correspond to different values of the  damping $\gamma$, while the different line-styles correspond to different initial conditions $n=0,8,16$.
		}
		\label{fig:entropy_4T_nDpx}
	\end{center}
\end{figure}
\begin{figure}
	\begin{center}
		\hspace*{\fill}%
		\includegraphics[width=1.0\columnwidth,clip=true]{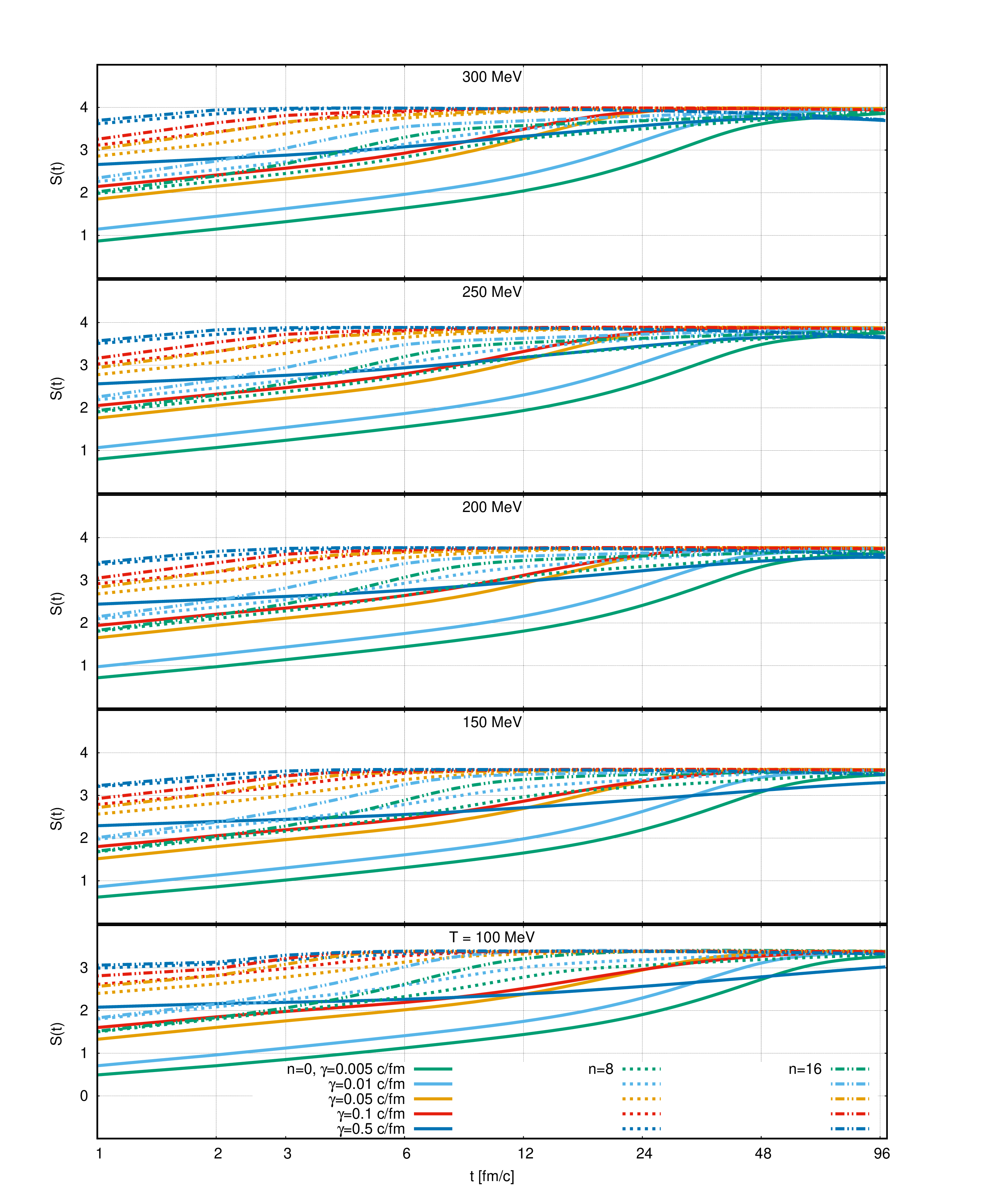}
		\hspace*{\fill}%
		
		\caption{
		Entropy $S(t)$ for different bath temperatures $T$ with $D_{px}=\Omega\gamma/6\pi T$ and $\Omega = 4 T$. The different line colors correspond to different values of the  damping $\gamma$, while the different line-styles correspond to different initial conditions $n=0,8,16$.
		}
		\label{fig:entropy_4T_pDpx}
	\end{center}
\end{figure}
\begin{figure}
	\begin{center}
		\hspace*{\fill}%
		\includegraphics[width=1.0\columnwidth,clip=true]{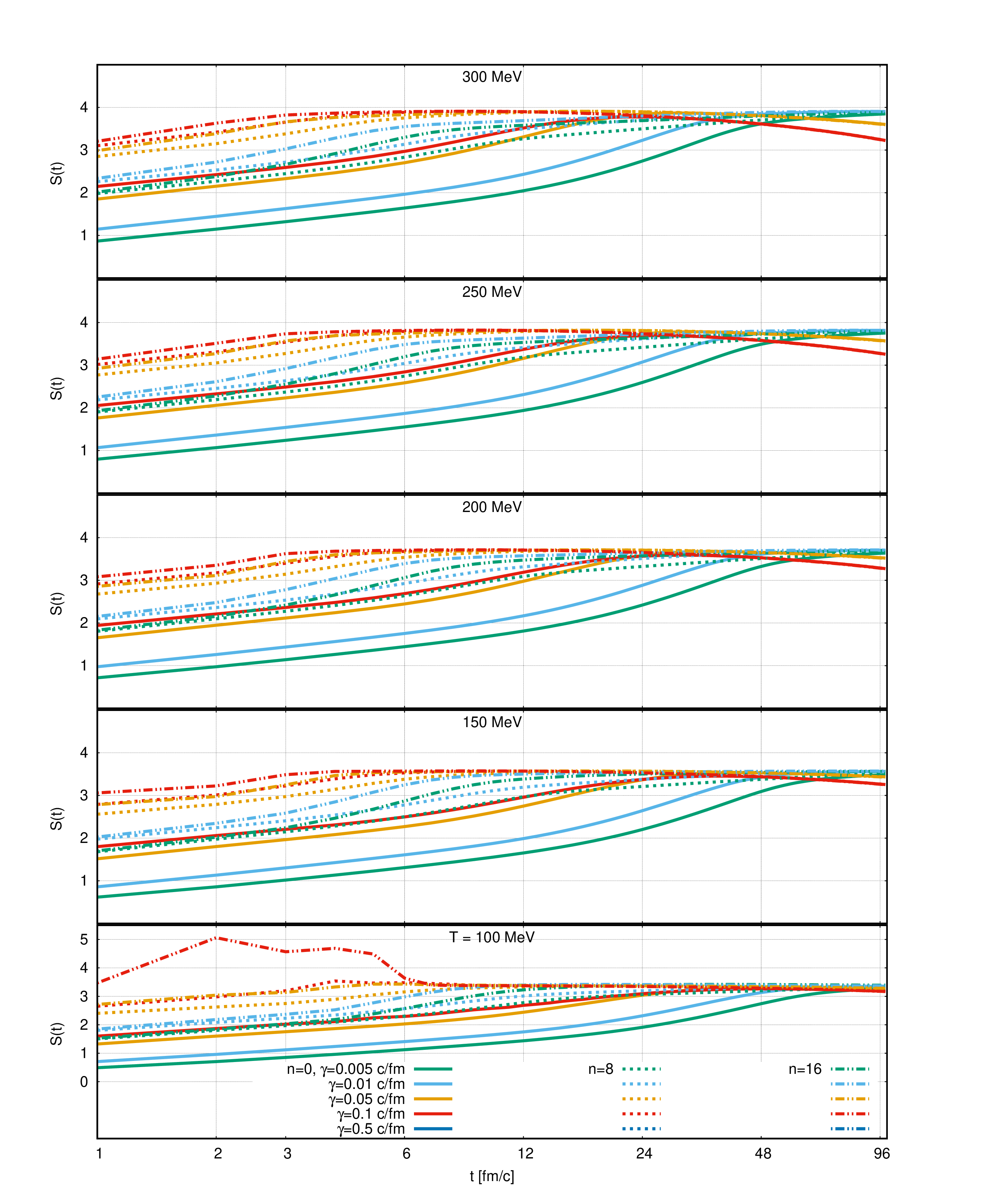}
		\hspace*{\fill}%
		
		\caption{
			Entropy $S(t)$ for different bath temperatures $T$ with $D_{px}=-\gamma T/\Omega$ and $\Omega = T$. The different line colors correspond to different values of the damping $\gamma$, while the different line-styles correspond to different initial conditions $n=0,8,16$.
		}
		\label{fig:entropy_T_nDpx}
	\end{center}
\end{figure}
\begin{figure}
	\begin{center}
		\hspace*{\fill}%
		\includegraphics[width=1.0\columnwidth,clip=true]{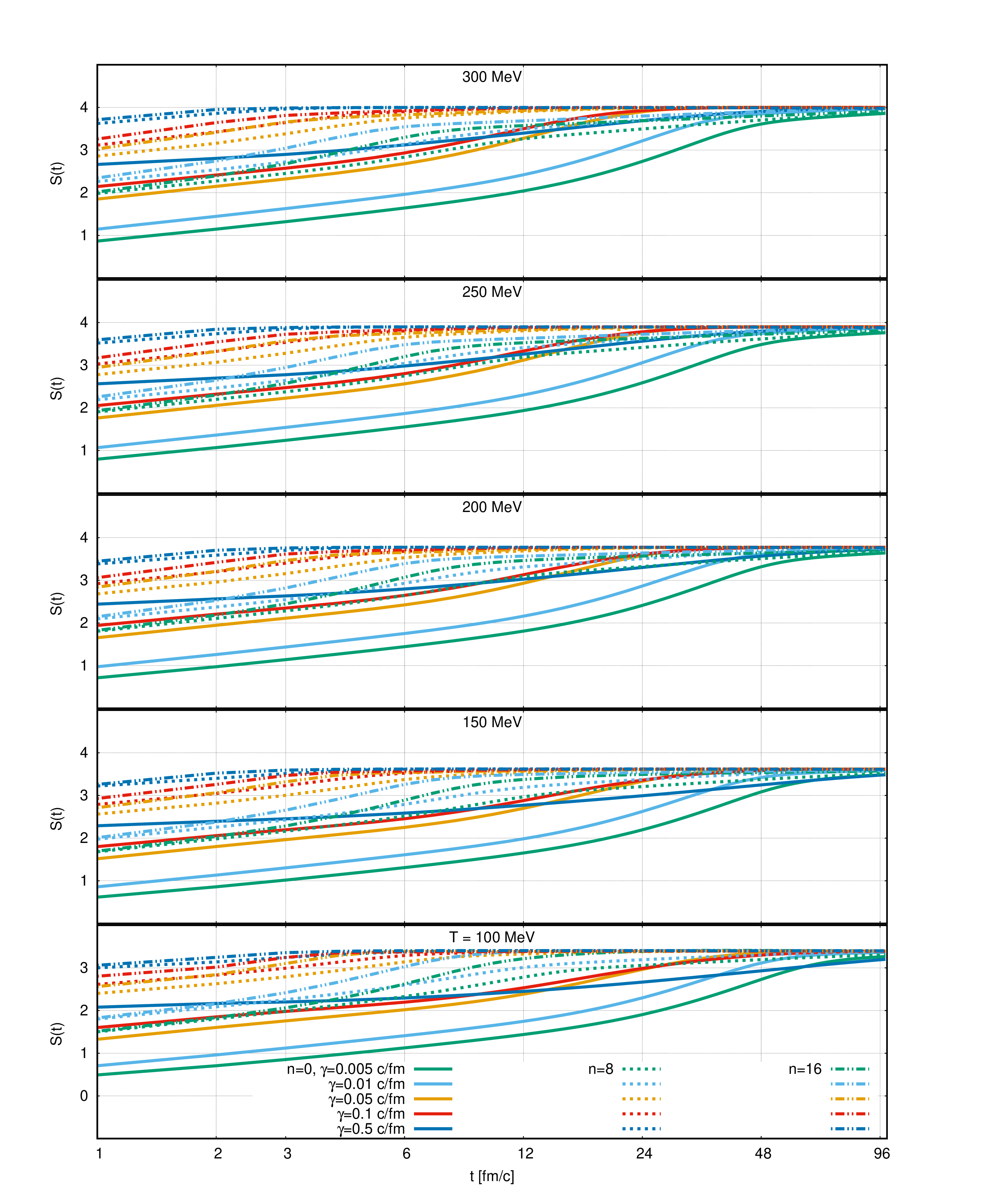}
		\hspace*{\fill}%
		
		\caption{
			Entropy $S(t)$ for different bath temperatures $T$ for the pure Caldeira-Leggett master equation, where $D_{px}=\Omega\gamma/6\pi T$ and $\Omega = T$. The different line colors correspond to different values of the damping $\gamma$, while the different line-styles correspond to different initial conditions $n=0,8,16$.
		}
		\label{fig:entropy_T_pDpx}
	\end{center}
\end{figure}

As we have seen in \cref{sec:thermalization}, the density matrix in the stationary case,  is distributed along the diagonal and symmetric to the left and right off the diagonal.

At this point we want to remind the reader, that, what we call the ``density matrix", actually is the reduced density matrix of the ``system", i.e. the partial trace of the density matrix of the full (closed) quantum system (system-plus-environment).
Generally, a density matrix  in equilibrium for a closed quantum system without interactions has only diagonal entries, because it is composed of orthogonal wave functions. Furthermore, the density matrix is hermitian. 
Therefore, we can construct the full quantum system, which may be of interest in order to calculate the entropy but also to study the interactions, by diagonalizing the reduced density matrix in position space representation.
The corresponding eigenvalues $\lambda_i^\rho$ describe the system-plus-environment entries of the full density matrix.
If we assume, that the final state is Boltzmann-distributed, which has to be the case because we consider a ``bosonic" system at high temperature, the logarithm of the eigenvalues $\lambda_i^\rho$ has to fulfil

\begin{align}\label{eq:effectiveE}
\ln\left(\lambda_i^\rho\right)  = -\frac{\tilde{E}_i}{T}\, ,
\end{align}

where $\tilde{E}_i$ are the energy eigenvalues of the ``effective" Hamiltonian of the system, with the additional interaction with the bath, $\hat{H} = \hat{H}_\text{S} + \hat{H}_{\text{SB}}$.
At this point, comparing to the eigenvalues of the density matrix of the system, we cannot distinguish any more, if there is still a bound state or if the impact of the heat bath has shifted the wave functions of the system, such, that there are only unbound states, because $\tilde{E}_i$ contains also the chemical potential, which leads to an overall shift of the eigenvalues. 
In the case, where the system is in thermal equilibrium, the factor $T$ is the temperature of the full quantum system and the eigenvalues $\lambda_i^\rho$ are real valued.

However, we can compare the effective energies of the Boltzmann distributed equilibrium case, cf. \cref{eq:Boltzmann},
\begin{align}\label{eq:argument}
	\epsilon_i = -\frac{1}{T} (E_i + \mu)\, ,
\end{align} 
satisfying  $\rho_{nn} = \ee^{\epsilon_i}$.

In \cref{fig:eigenvalues} we illustrate the eigenvalues $\lambda_i^\rho$ for different temperatures $T$ and the fitted values from \cref{eq:argument}, from the numerically evaluated values of $\rho_{nn}$, cf. \cref{eq:coeff}, computed with the original initial wave functions. 
Here, we clearly see, that there is no difference between both curves for each temperature, and therefore we cautiously conclude, that the wave functions get only negligibly shifted during the process.
Also for the parameter setups, where $D_{px} = -\gamma T / \Omega$ and  $D_{px} = \gamma \Omega / 6\pi T$, we do not find any difference, even though this might be expected comparing to what is found in \reffs \cite{Homa2019,Bernad2018}.
Also compare to \cref{fig:Boltzmann_fit}, where $\rho_{nn}(t)$ is illustrated depending on the energy eigenvalues of the initial wave functions.
  
Let us mention here, that we illustrate only the first 50 out of 300 eigenvalues, which is also the number of included eigenvalues of the initially prepared system.
Another reason is as follows: even though we have the same amount of eigenvalues as discretization points from the grid, the numerical procedure we are using to extract the eigenvalues, the \textit{armadillo} library \cite{Sanderson2016,Sanderson_2019} to diagonalize the matrix, is more accurate for the largest eigenvalues and less accurate for the numerically smaller values. 
However, for our purposes, it is sufficient to consider only the first few ($\approx$ 50) eigenvalues: 
The energy spectrum of the system rises quadratically due to the square well potential for sufficiently large $i$, cf. \cref{fig:eigenvalues} and comparing to \cref{fig:energies} shows, that already for twice the amount of states, that are illustrated we reach energies that are around the non-relativistic limit.
Besides that, we are comparing the effective energy eigenvalues to the calculated values $\epsilon_i$, which are using the energy eigenvalues of the initial system, and which are 50 in total, too.

Let us again return to \cref{fig:eigenvalues}. 
As already mentioned we illustrate the population of the energy-eigenvalues of the effective system in thermal equilibrium, which, as is expected, rises quadratically with respect to $i$, for large $i$.\footnote{The reason, that the eigenvalues rise quadratically only for high temperatures is, that for small $i$ the impact of the P\"oschl-Teller potential is still larger.}
This has to be the case, because the system's energy spectrum rises approximately quadratically for sufficiently large energies and the contributions of the bath modify only the prefactor, which might also correspond to a potentially modification of the mass.

Another aspect, which can be studied in \cref{fig:eigenvalues} is that for different temperatures one can see a dependency also  proportional to $T$, cf. \cref{eq:effectiveE}.
This, however, cannot be further investigated, because the chemical potential $\mu$ is different for each heat bath temperature. 

However, not only the eigenvalues can be used to obtain a better insight into the full quantum system, also the eigenvectors, which are the corresponding energy eigenfunctions of the full system. 
In \cref{fig:wave_eff_init1} we have illustrated the first eigenfunctions of the effective Hamiltonian.
In \cref{fig:wave_eff_init1} the bound state is originally populated, which means, that for $t=0$ the system is initialized by the wave function which corresponds to $E=-2.3$ MeV. 
One can see, that the interaction with the bath leads to a broadening of the wave functions for the case, where $n=0$ initially.
In the end, when thermal equilibrium is reached, the wave functions of all initial conditions are expected to be equal, which can be seen in the lower right illustration of \cref{fig:wave_eff_init1}. 
But it should be pointed out, that only the eigenfunctions at the end are also eigenfunctions of the ``effective" Hamiltonian.
For the intermediate wave functions, which we calculate during the process of thermalization, we cannot provide a meaningful physical interpretation.

To interpret the final state, the eigenvectors of a thermal Gibbs state have to be the same as the eigenvectors of the initial (non-interacting) system. 
This is, what we obtain in \cref{fig:wave_eff_init1}, in the lower right figure, exemplarily depicted for the bound state.

In \cref{fig:wave_eff_init16} we illustrate the effective eigenvectors for initial conditions, where $n=16$.
Again, we plot the intermediate eigenvectors for different times $t$ and the final distribution.
On the lower right plot one also can see a comparison between $\psi_{0, \text{eff}}(x)$, calculated from the effective eigenvector and  $\psi_{0}(x)$, which we obtain from the initial wave function.
One can see a small deviation between those functions, which is due to boundary effects, and can be also seen on the final distributions for $\rho(x,x,t=100 \text{fm/c})$ in \cref{fig:rho_xx_init16}, where the regions close to the boundaries, before $\rho(x,x,t)$ starts to vanish, are slightly higher than in the case, where the bound state is originally populated.
This can also be interpreted as a Gibbs phenomenon and has seemingly a relevant impact on the numerically computed eigenvectors.
Whereas it is sufficiently good to fit the right temperatures up to a deviation of around 1\%, and to investigate the stationary cases independently of the initial condition, the small boundary effect, which is caused by interferences due to the reflective boundary conditions, lead to a slightly different shaping of the effective wave functions.

Admittedly, this calls for further investigation.
However, we do not consider this result to be of physical origin and therefore accept it at this point.

To summarize this section, we can conclude, that diagonalizing the reduced density matrix leads to energy eigenvalues and eigenvectors, which help to analyse the open quantum system including the bath interaction. 
We have seen, that the energy spectrum also satisfies quadratic behaviour for large energies, and is temperature dependent but is in general under-determined up to free parameters, which are the temperature and chemical potential.
Furthermore, the eigenvectors of the system can be considered as the wave functions of the system-plus-interaction Hamiltonian and therefore give insight into the localization and binding probability of each quantum state.

\subsection{Entropy and thermalization time}\label{sec:entropy}

In this section we use the diagonalization of the density matrix, to calculate and study the von-Neumann entropy,
\begin{align}
 S(t) = -\text{Tr}\left[\rho(t)\ln\rho(t)\right]\, .
\end{align}
The diagonalization is performed on the numeric solution at discrete timesteps from $t=0$, where $S=0$ to $t_{\text{eq}}$, where $\partial_t S(t_{\text{eq}}) = 0$.
We can use the entropy to investigate, at which time the system fully  thermalizes, according to the principle of maximum entropy \cite{Jaynes:1957zza,Jaynes:1957zz}.
Furthermore, we analyze how the thermalization time depends on $\gamma$ and $T$, the damping, and if there are further dependencies, for example on the cutoff $\Omega$ and whether we are in high- or medium temperature, which is mainly described by the parameter $D_{px}$ regime.


Let us start the discussion with the pure Caldeira-Leggett master equation ($D_{px} = 0$), \cref{fig:entropy_4T_0Dpx}. 
We can see, that all cases tend towards the same final entropy in the equilibrated state, which is expected,  for identical temperatures. 
The values for $S(t)$ are higher, if the temperature is higher.
We also find, that for a higher occupied initial state, the equilibration takes place faster. 
The same holds true for the damping $\gamma$. If the damping is small, it takes  longer to equilibrate the system, than for a larger damping $\gamma$.
Let us also mention, that the temperature itself has a negligible impact on the thermalization time. 

Next, we turn to the cases, where $D_{px}\neq 0$, where we furthermore compare $\Omega = T$ and $\Omega = 4T$.
The corresponding (numerical) results are depicted in \cref{fig:entropy_4T_nDpx,fig:entropy_4T_pDpx,fig:entropy_T_nDpx,fig:entropy_T_pDpx}.
One observes from these figures, that the impact of the choice of $D_{px}$ and $\Omega$ on the thermalization time is in fact are very small.
However, we should mention, that we only consider energy and temperature regimes which are of interest for our purpose, namely to study the formation of a bound state in a heavy-ion collision. 
Also the choice of $\gamma$ is motivated from this perspective: if we consider, the typical freeze out time is less then 10 fm, and $\gamma$ is related to the friction coefficient, this means, that $\gamma \approx \frac{1}{\tau_R}$, the relaxation time, which further implies, that $\gamma$ should be of order $0.1$ c/fm.
As already mentioned, and discussed here and in \reff \cite{rais2024} the coefficients in combination among each other, have to fulfil certain conditions, which can be derived, for the case where $V(x)$ is the harmonic potential, purely analytically.
This can not be done for general potentials. 
However, regarding \cref{fig:entropy_T_nDpx}, for late times in contrast to \cref{fig:entropy_4T_0Dpx}, the entropy decreases significantly or even shows some strange behaviour during the temporal evolution. 
This is caused by the choice of $D_{px}$, which in this case is $D_{px} = - \gamma T /\Omega$, the one for high temperatures.
This explains, why we see this behaviour for ``lower" temperatures, considering high damping (0.1 and 0.5 c/fm).
We conclude, that these damping coefficients, for the given choice of $D_{px}$ violate some analytically non-derivable constraints, and they are not suitable for this type of scenario.
This also becomes visible in the norm of the density matrix, which, while solving the Lindblad equation get violated significantly.
Except for this, we cannot state any remarkable discrepancies concerning the choices of parameters comparing the pure Caldeira-Leggett master equation and the Lindblad form.

From a numerical point of view let us point out, that considering \cref{fig:entropy_T_nDpx}, for the setups, where $\gamma = 0.1$ the entropy shows some spurious behaviour. 
Not only, that it increases in one case, to then decrease again, but also that towards the final distribution, the entropy decreases in all cases.
This corresponds to the norm, which is in the setups, where we observe this behaviour violated by about 10\%.
This indicates, that these parameter setups are incompatible with a proper Lindblad description, which can be easily understood: the diffusion coefficient $D_{px}$ corresponds to the high temperature limit, while the cut-off frequency is the same as the heat bath temperature. 
This is contradicting the first assumption. 

As a final remark we want to mention, that for the initial condition, where there is a mixture of states initially, as illustrated in \cref{fig:inhom_xx}, also the entropy and the purity is illustrated.  The entropy $S(t)$ gets constant after approximately 16 fm/c, and therefore somewhere in between the cases, where we have initialized a pure state.

\subsection{Formation time of the bound state}\label{sec:bound}
\begin{figure*}
	\begin{center}
		\hspace*{\fill}%
		\includegraphics[width=2.0\columnwidth,clip=true]{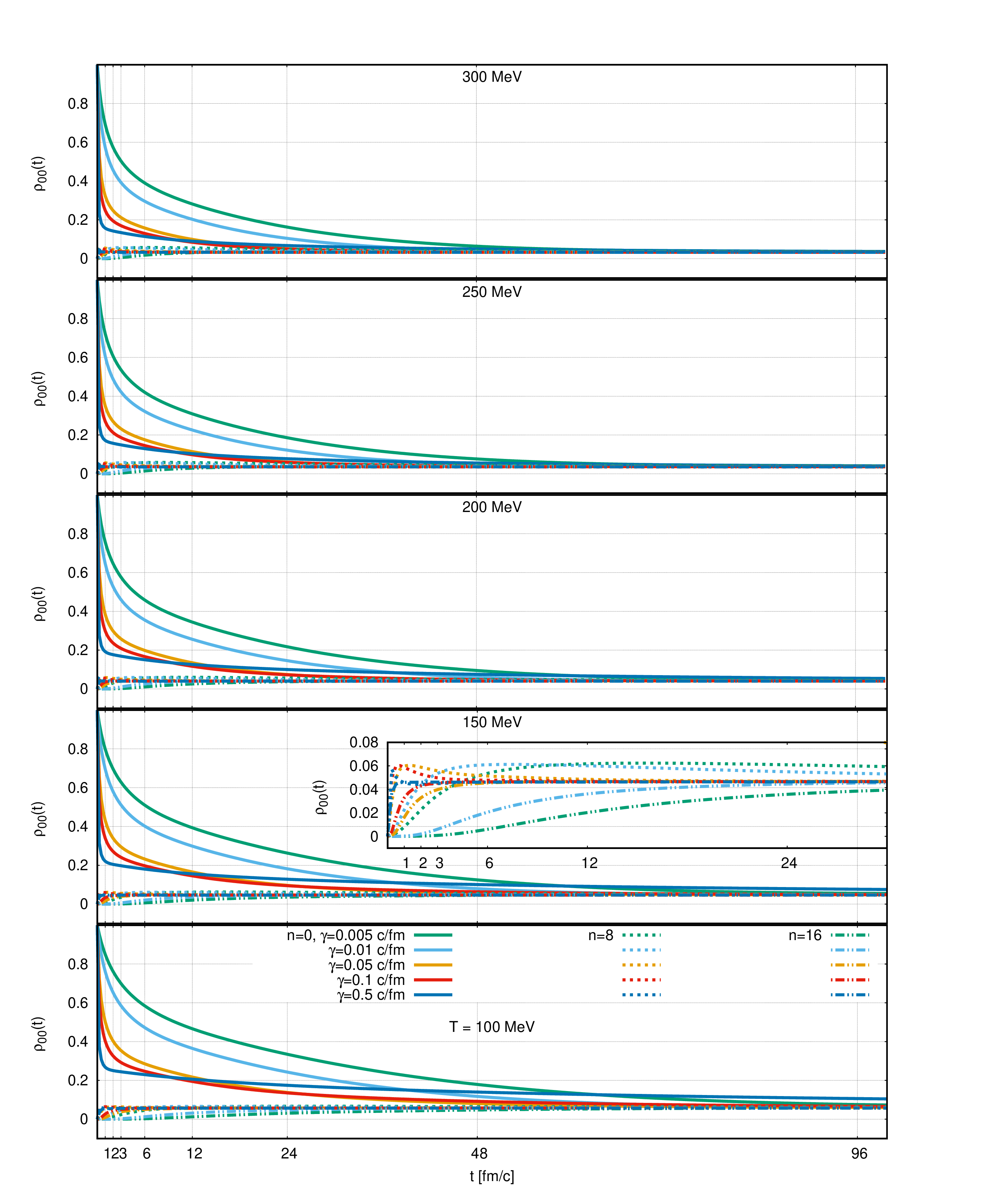}
		\hspace*{\fill}%
		
		\caption{
			$\rho_{00}(t)$ for different bath temperatures $T$ for the pure Caldeira-Leggett master equation, where $D_{px}=0$ and $\Omega = 4 T$. The different line colors correspond to different damping $\gamma$, while the different line-styles correspond to different initial conditions $n=0,8,16$.
		}
		\label{fig:ground_state}
	\end{center}
\end{figure*}
\begin{figure*}
	\begin{center}
		\hspace*{\fill}%
		\includegraphics[width=2.0\columnwidth,clip=true]{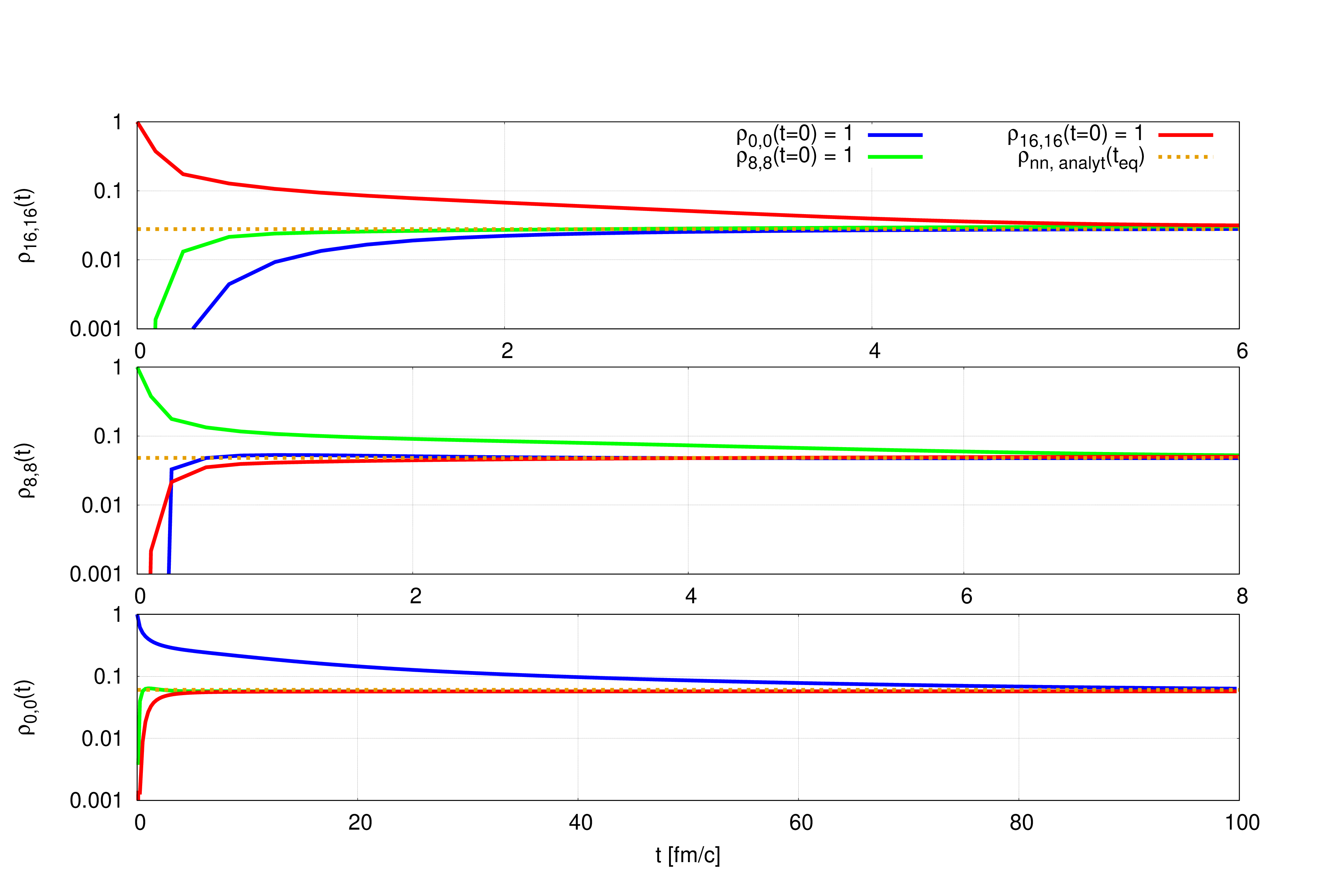}
		\hspace*{\fill}%
		
		\caption{
			$\rho_{0\,0}(t)$, $\rho_{8\, 8}(t)$ and $\rho_{16\,16}(t)$ depending on different initial conditions, $\rho_{nn} = 1$ with $n=0,8$, and $16$ for bath temperature $T = 100$ MeV for the pure Caldeira-Leggett master equation, where $D_{px}=0$ and $\Omega = 4 T$ and damping $\gamma=0.1$ c/fm. The dashed line illustrates the equilibrium result, which corresponds to the fit of the Boltzmann distribution, \cref{eq:Boltzmann}, as indicated in \cref{fig:Boltzmann_fit}, where $T_{\text{fit}} = 97.43$ and $\mu_{\text{fit}} = 275.39$ MeV.
		}
		\label{fig:rho_nn}
	\end{center}
\end{figure*}
Besides the equilibration time of the system, which we are determining considering the entropy $S(t)$, another interesting time scale is the formation and/or destruction time of a bound state, respectively of an arbitrary other state of the given system.
In the following, we are going to consider those states, that serve as a setup in terms of different initial conditions, i.e. the bound state, and the states where $n = 8$ and $n=16$.

In \cref{fig:ground_state} we show the probability of the bound state, i.e. the $\rho_{00}$ component, cf. \cref{eq:coeff}, of the density matrix with the before-mentioned parameter choices for $\gamma$ and $T$ and for different initial conditions, where the states $n=0, n=8$ and $n=16$ are initially populated.
Firstly, we see, that higher values of $n$ for the initial condition lead to a faster population of the bound state, than if the bound state is already populated, and then  depopulates again during the temporal evolution.
This appears plausible, because the binding potential is most attractive as it is the lowest energy in the system.
On the other hand, it takes the system much longer to depopulate an initially bound state.

It is not surprising, that the relaxation time of the bound state depends on the damping $\gamma$ and shows reciprocal behaviour. 
Rather more exciting it seems to be, that the shape of the temporal evolution of $\rho_{00}$ is not purely exponential. 
This can be especially seen in the cases, where $\gamma = 0.5$ c/fm, which shows, that during the temporal evolution various time scales have to be incorporated.
In the case, where $\gamma = 0.5$ c/fm, regarding $\rho_{00}(t)$ one can see, that the main dynamics, that leads to thermalization happens very fast and follows the expectation,  that the system should thermalize faster for larger values of the damping coefficient. 
However, another timescale, $t\geq 6 $ fm/c shows, that for originally fully populated bound state, the bound state decreases asymptotically but very slowly towards its final distribution.
This also explains, why in \cref{fig:cross_4T_0Dpx}, the cases, where $\gamma = 0.5$ c/fm do also not yet show the final distribution. 
This is a case, where final thermalization takes longer, even though the damping is higher, and seem to mark a limit of a satisfactory applicability of Lindblad dynamics. 
 
This high value for $\gamma$ and the fact, that for some choices of $D_{px}$ it was not possible to calculate $S(t)$ properly, cf. \cref{sec:entropy}, indicates, that a certain choice of $\gamma$ can lead to a potential behaviour of a (over-)damped system, which can also be seen in \cref{tab:fit_temperatures}, where for one parameter setup, it was not possible to calculate $\rho_{nn}(t)$ properly.
This however is only a hand-wavy interpretation, since we cannot derive the over-damped solution for a potential given by \cref{eq:potential} exactly, which is why we use the numeric approach in the first place.

Furthermore, we want to mention the impact of the  temperature on the (de-)population time of the bound state is rather weak. 
However one can see a faster (de-)population towards full thermal equilibrium for higher temperatures.

In the following, we discuss \cref{fig:rho_nn}, where we have illustrated the states $\rho_{1\, 1}$, $\rho_{8\, 8}$ and $\rho_{16\, 16}$ in the different panels, depending on the different initial conditions $\rho_{nn}(t=0)=1$ for $n=0,8,$ and $16$.
In the stationary case, all identical states reach the same value, independent of the initial condition. 
As already seen, only for the case, where the bound state is originally populated, the full thermalization time for $\rho_{00}(t)$ is recognizably larger than for all other cases. 
This is reasonable, because for an already fully populated bound state it is less attractive to depopulate, since it is confined in the potential.
The thermalization time $\tau_R$ is smaller for higher states: Despite having used the same  damping coefficient $\gamma = 0.1$ c/fm, the individual state-dependent relaxation time  $\tau_R= \tau_R(n)$ is different for each state. 

As a final remark, we want to emphasize, that the (de-)population time of the bound state obeys a different timescale than the thermalization timescale of the full system and therefore the time at which the entropy gets constant (``pre-thermalization"). 
Especially towards higher initialized states and higher values of the damping coefficient one can see, comparing for example \cref{fig:entropy_4T_0Dpx} with \cref{fig:ground_state}, that the bound-state formation takes place much earlier than the full equilibration of the system.

\subsection{Purity of the density matrix}\label{sec:purity}
\begin{figure*}
	\begin{center}
		\hspace*{\fill}%
		\includegraphics[width=2.0\columnwidth,clip=true]{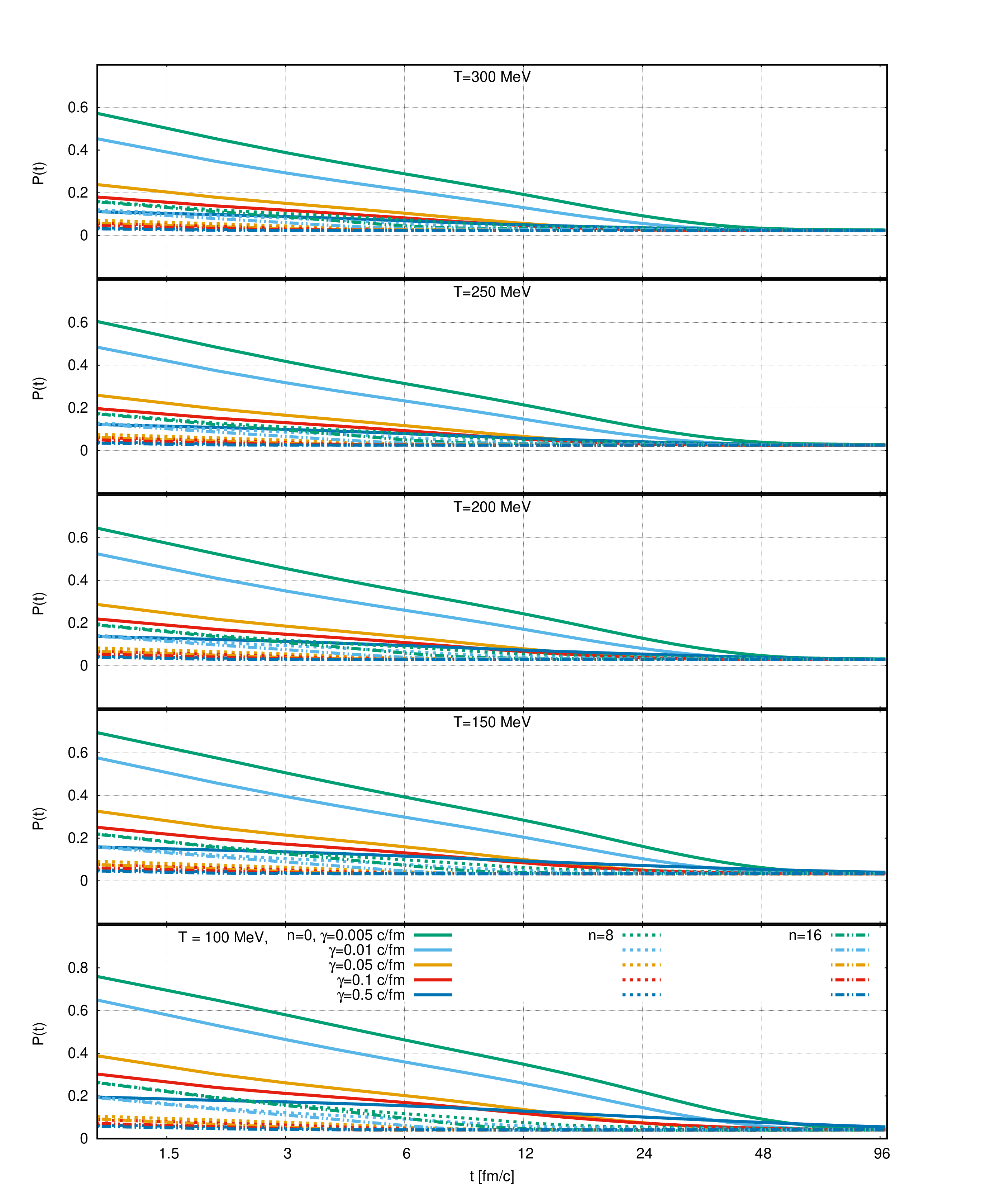}
		\hspace*{\fill}%
		
		\caption{
		Purity $P(t)$ for different bath temperatures $T$ for the pure Caldeira-Leggett master equation, where $D_{px}=0$ and $\Omega = 4 T$. The different line colors correspond to different dampings $\gamma$, while the different line-styles correspond to different initial conditions $n=0,8,16$.
		}
		\label{fig:purity_4T}
	\end{center}
\end{figure*}

\begin{figure}
	\begin{center}
		\hspace*{\fill}%
		\includegraphics[width=1.0\columnwidth,clip=true]{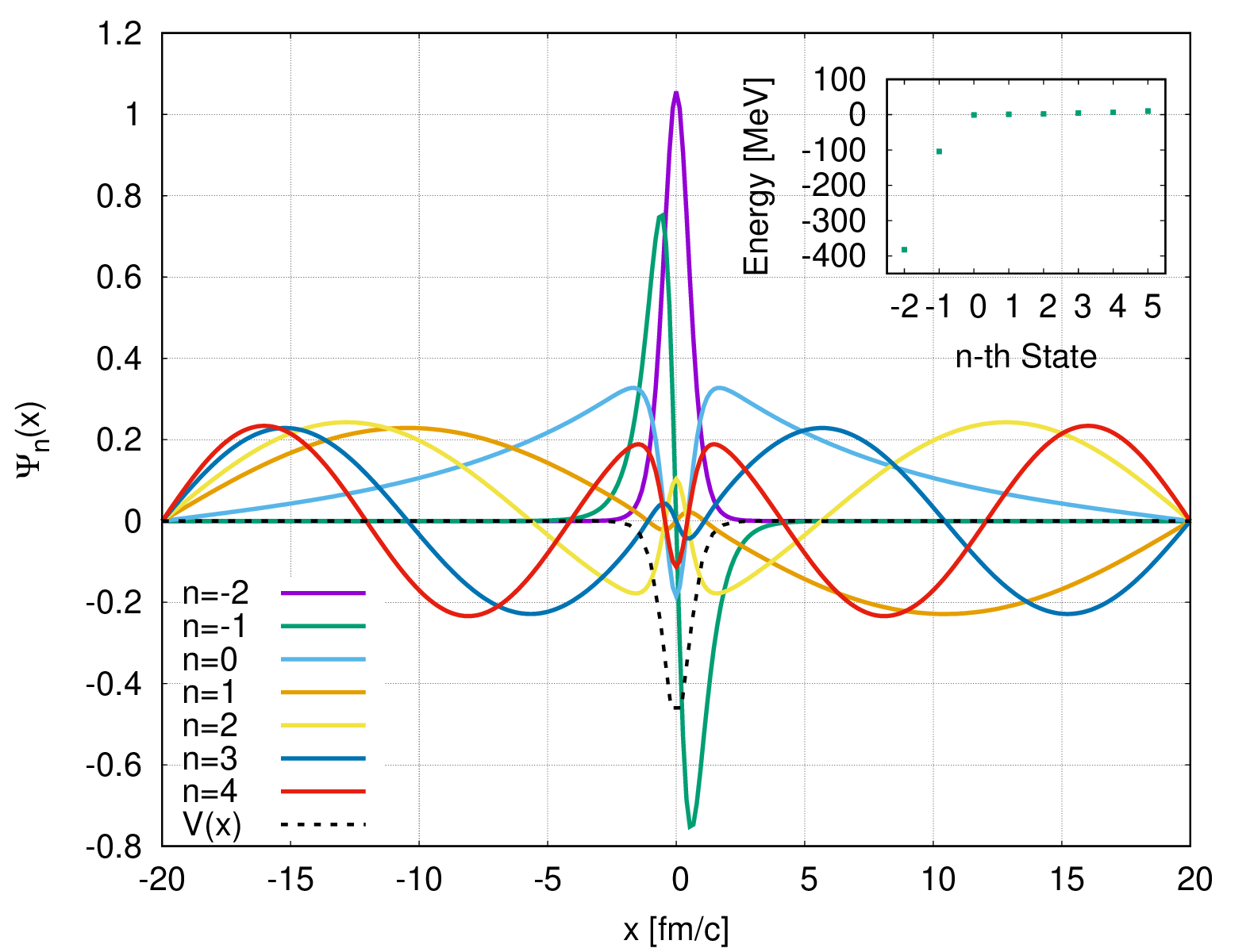}
		\hspace*{\fill}%
		
		\caption{
			The wave function for the potential given in \cref{eq:potential} with parameters $\alpha = 1.45$ fm$^{-1}$ and $V_{0} = 565$ MeV, in order to generate three bound states $n=-2$, $n=-1$ and $n=0$. Furthermore, we illustrate the first four eigenstates with positive energy eigenvalues and the rescaled potential (dashed black line). In the figure in the upper right corner we see the first 8 energy eigenvalues of the given potential.
		}
		\label{fig:3bound_wave_and_energy}
	\end{center}
\end{figure}
\begin{figure}
	\begin{center}
		\hspace*{\fill}%
		\includegraphics[width=1.0\columnwidth]{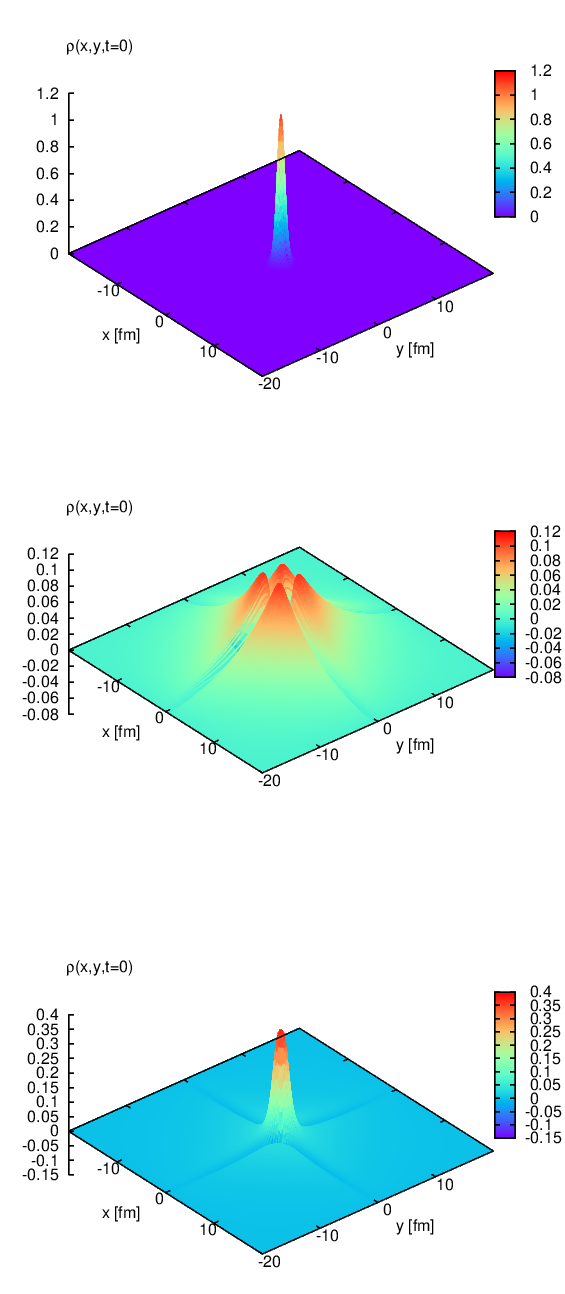}
		\hspace*{\fill}%
		
		\caption{
			$\rho(x,y,0)$ for three different initial conditions of the setup given in \cref{fig:3bound_wave_and_energy}. The upper plot shows the initial condition, where the $n=-2$ state, the state with the largest negative energy is originally populated, the middle plot shows the case, where the $n=0$ state, the state with the highest negative energy is originally populated and the lower plot shows the case, where all three bound states are populated equally, with purity 1/3.
		}
		\label{fig:3bound_3d_init}
	\end{center}
\end{figure}
\begin{figure}
	\begin{center}
		\hspace*{\fill}%
		\includegraphics[width=\linewidth]{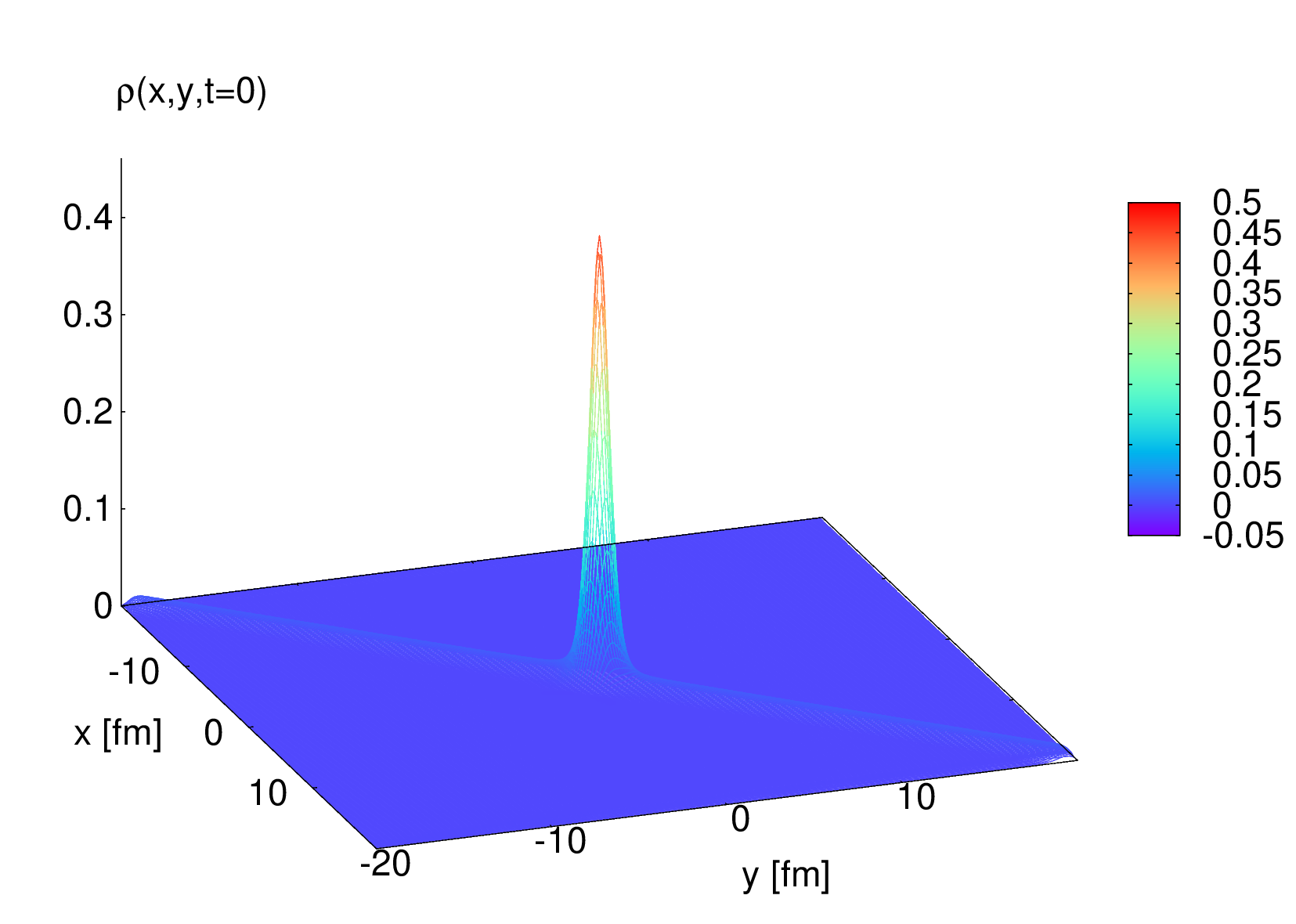}
		\hspace*{\fill}%
		
		\caption{
			$\rho(x,y,t)$ at $t= 2000$ fm/c, as the final distribution of the setup given in \cref{fig:3bound_wave_and_energy} of one of the initial conditions introduced in \cref{fig:3bound_3d_init} following Lindblad dynamics. 
			Here the bath temperature is given by $T=200$ MeV, the cut-off frequency $\Omega = 4T$, $\gamma = 0.1$ c/fm and $D_{px} = - \gamma T /\Omega$.
		}
		\label{fig:3bound_3d_final}
	\end{center}
\end{figure}

\begin{figure*}
	\begin{center}
		\hspace*{\fill}%
		\includegraphics[width=2.0\columnwidth,clip=true]{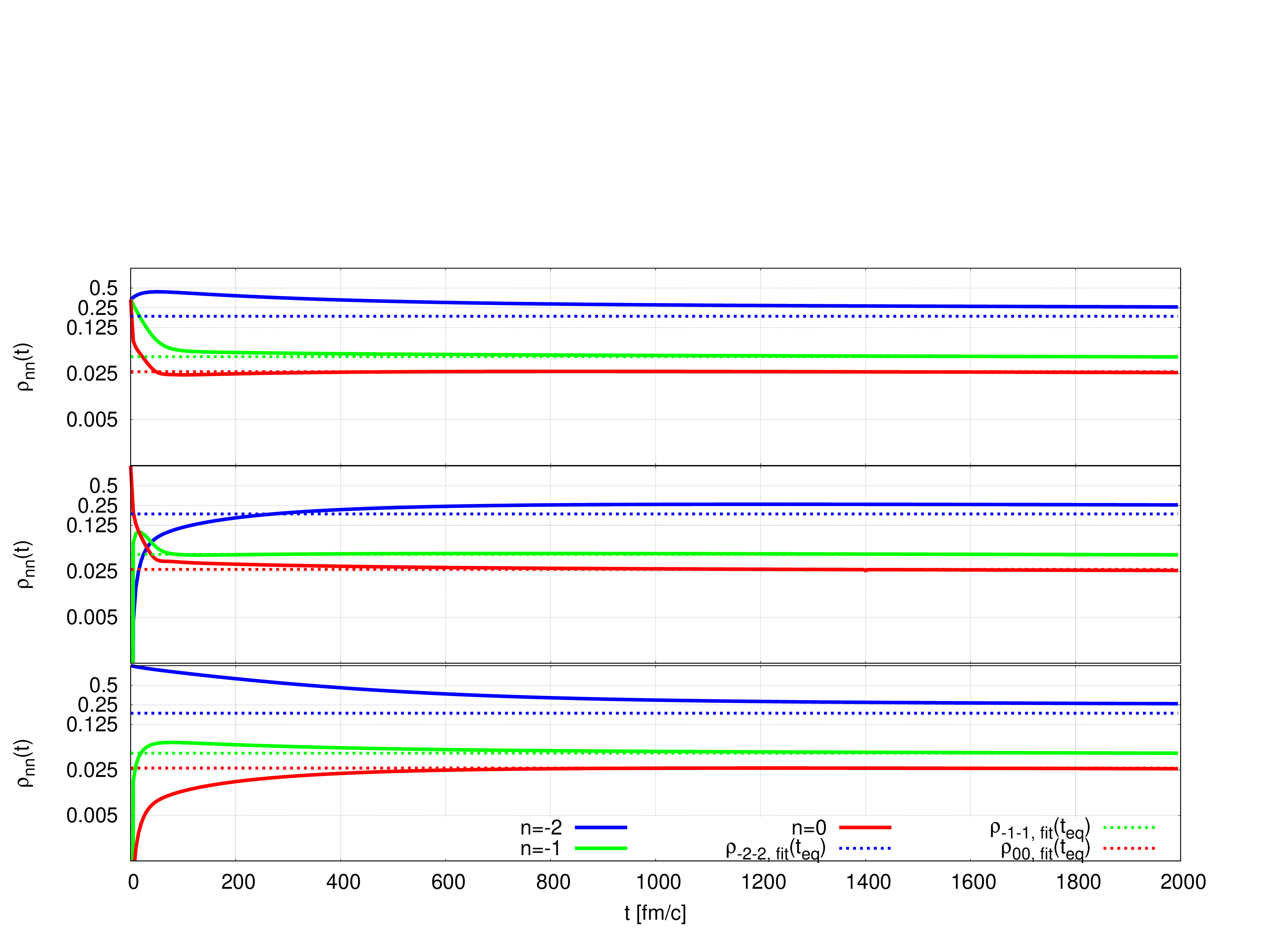}
		\hspace*{\fill}%
		
		\caption{
			The matrix coefficients $\rho_{-2,-2}$, $\rho_{-1,-1}$, $\rho_{0,0}$, corresponding to the three bound states of the setup introduced in \cref{fig:3bound_wave_and_energy} for the three initial conditions given in \cref{fig:3bound_3d_init}. Here,  $T=200$ MeV, $\Omega = 4T$, $\gamma = 0.1$ c/fm, and $D_{px} = - \gamma T /\Omega$. The dashed lines correspond to the fitted Boltzmann distribution, \cref{eq:Boltzmann} (the expected equilibrium), where $T_{\text{fit}} = 196.41$ MeV and $\mu_{\text{fit}} = 712.31$ MeV.
		}
		\label{fig:3bound_states}
	\end{center}
\end{figure*}

\begin{figure}
	\begin{center}
		\hspace*{\fill}%
		\includegraphics[width=1.0\columnwidth,clip=true]{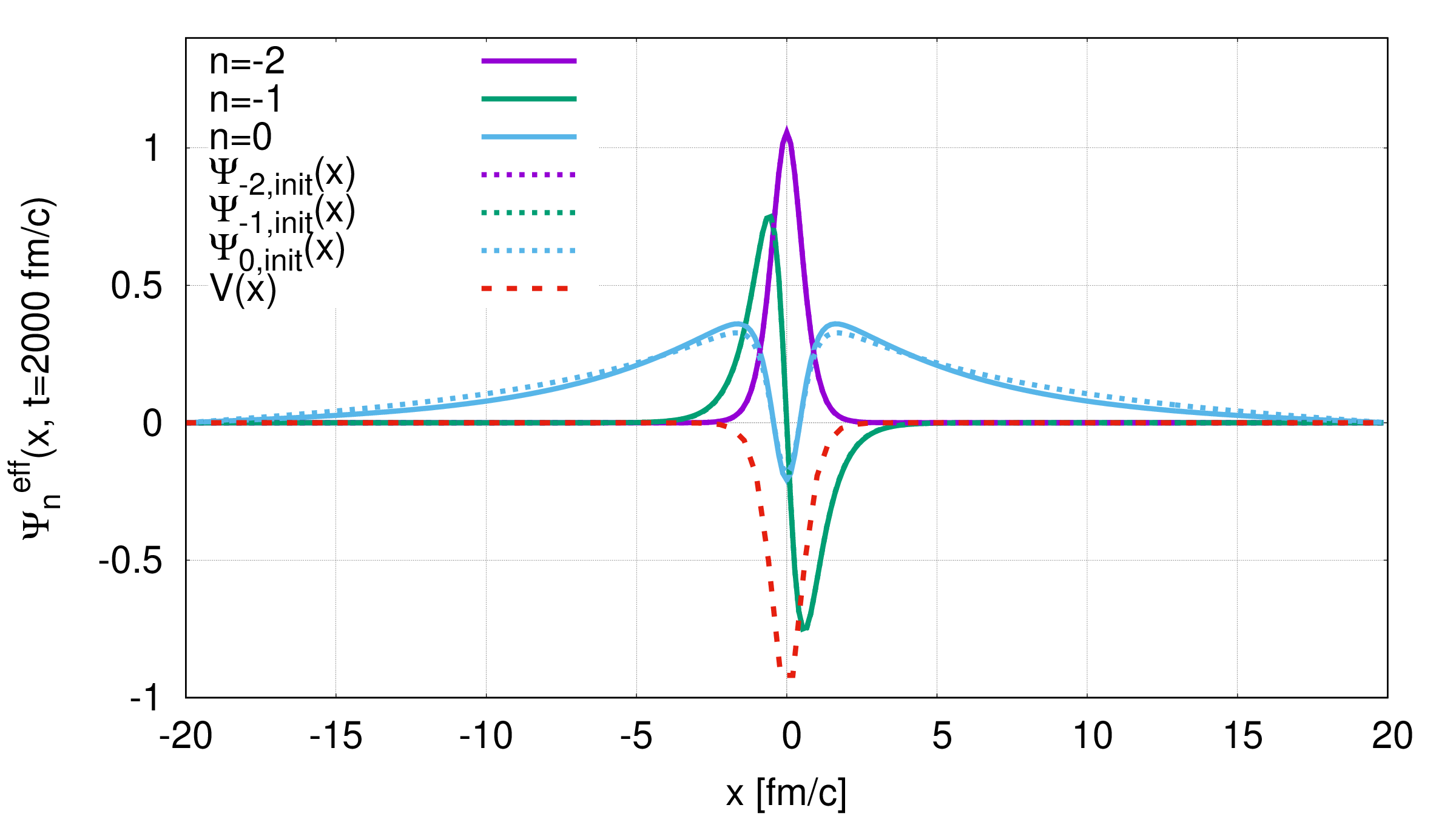}
		\hspace*{\fill}%
		
		\caption{ Eigenvectors of the diagonalized density matrix at time $t=2000$ fm/c, where the system is equilibrated, for the case, where $\rho_{-2\, -2}$ are populated originally. The parameters are the ones given in \cref{fig:3bound_wave_and_energy,fig:3bound_entropy}. The dashed lines are the initial wave functions at $t=0$ and therefore provide a comparison between the effective wave function in equilibrium and the initial wave functions.
		}
		\label{fig:3bound_waves}
	\end{center}
\end{figure}

\begin{figure}
	\begin{center}
		\hspace*{\fill}%
		\includegraphics[width=1.0\columnwidth,clip=true]{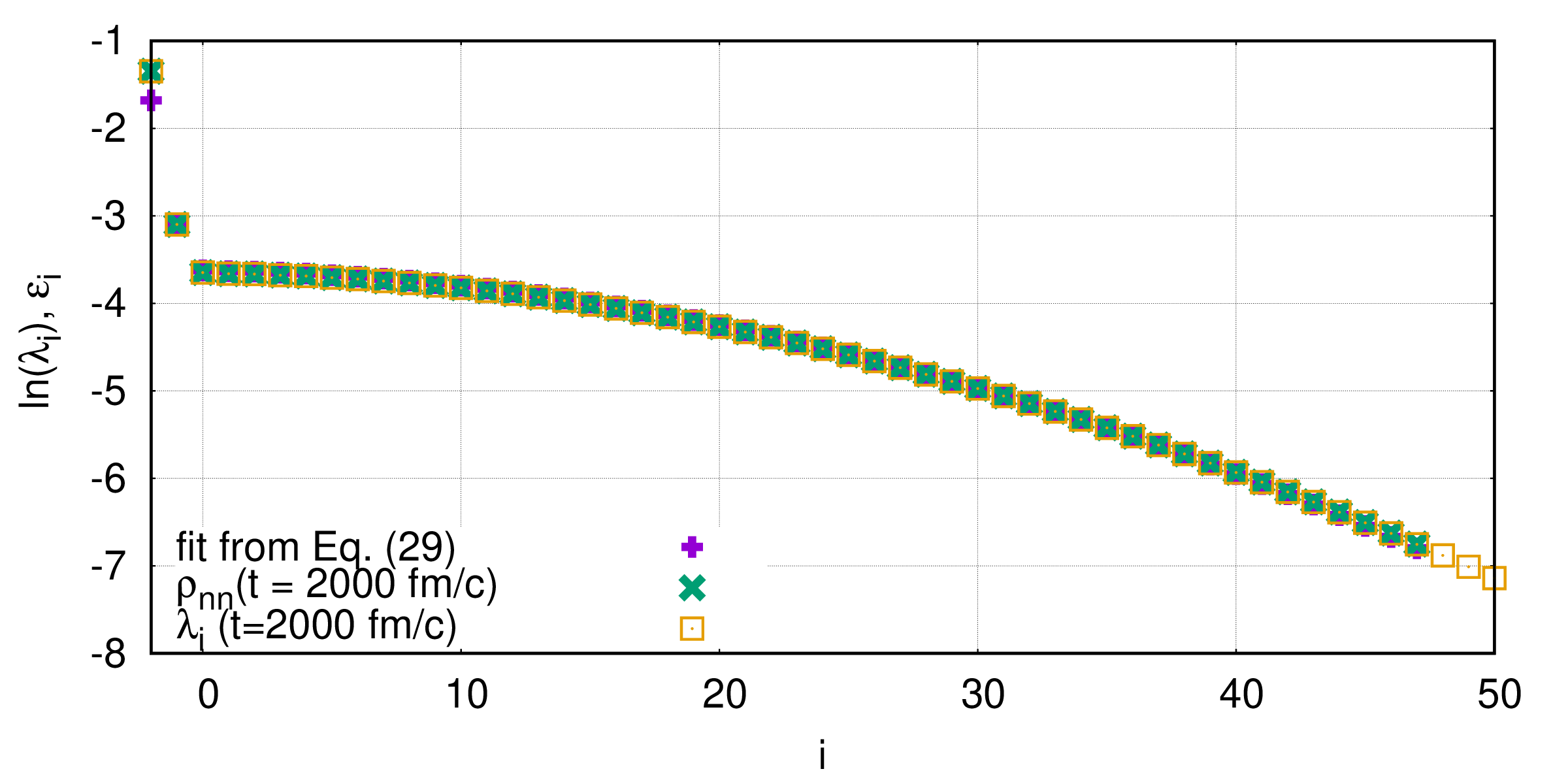}
		\hspace*{\fill}%
		
		\caption{
			Logarithm of the eigenvalues $\lambda_i$, \cref{eq:effectiveE}, compared to the argument of \cref{eq:Boltzmann}, \cref{eq:argument} and the values for  $\rho_{nn}$, calculated via \cref{eq:coeff}, using the initial wave functions, for different heat bath temperatures $T$ and a damping $\gamma = 0.1$ c/fm, cutoff frequency $\Omega= 4T$ for the pure Caldeira-Leggett model, $D_{px} = 0$. The parameters $T$ and $\mu$ of $\epsilon_i$ are the fitted values obtained also \cref{fig:Boltzmann_fit}.				
		}
		\label{fig:eigenvalues_3bound}
	\end{center}
\end{figure}

\begin{figure}
	\begin{center}
		\hspace*{\fill}%
		\includegraphics[width=1.0\columnwidth,clip=true]{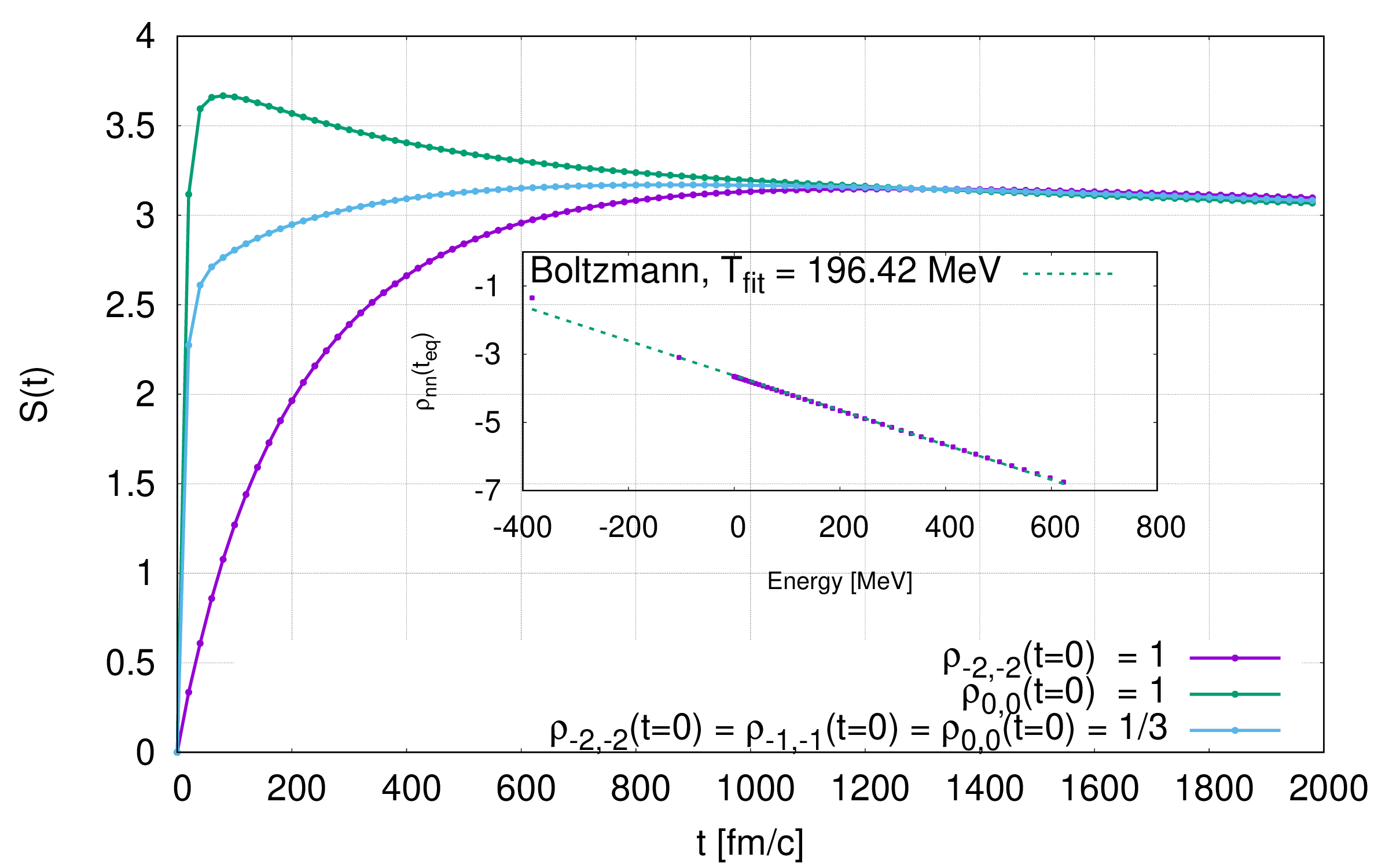}
		\hspace*{\fill}%
		
		\caption{
			Entropy $S(t)$ of the system described by \cref{fig:3bound_wave_and_energy} for the initial conditions introduced in \cref{fig:3bound_3d_init} for $T=200$ MeV, $\Omega = 4T$, $\gamma = 0.1$ c/fm, and $D_{px} = - \gamma T /\Omega$. 
			The inner illustration shows $\rho_{nn} (t=2000)$, calculated with \cref{eq:coeff}, where the dashed green lines follows a Boltzmann distribution and serves as a fit function.
		}
		\label{fig:3bound_entropy}
	\end{center}
\end{figure}

\begin{figure}
	\begin{center}
		\hspace*{\fill}%
		\includegraphics[width=1.0\columnwidth,clip=true]{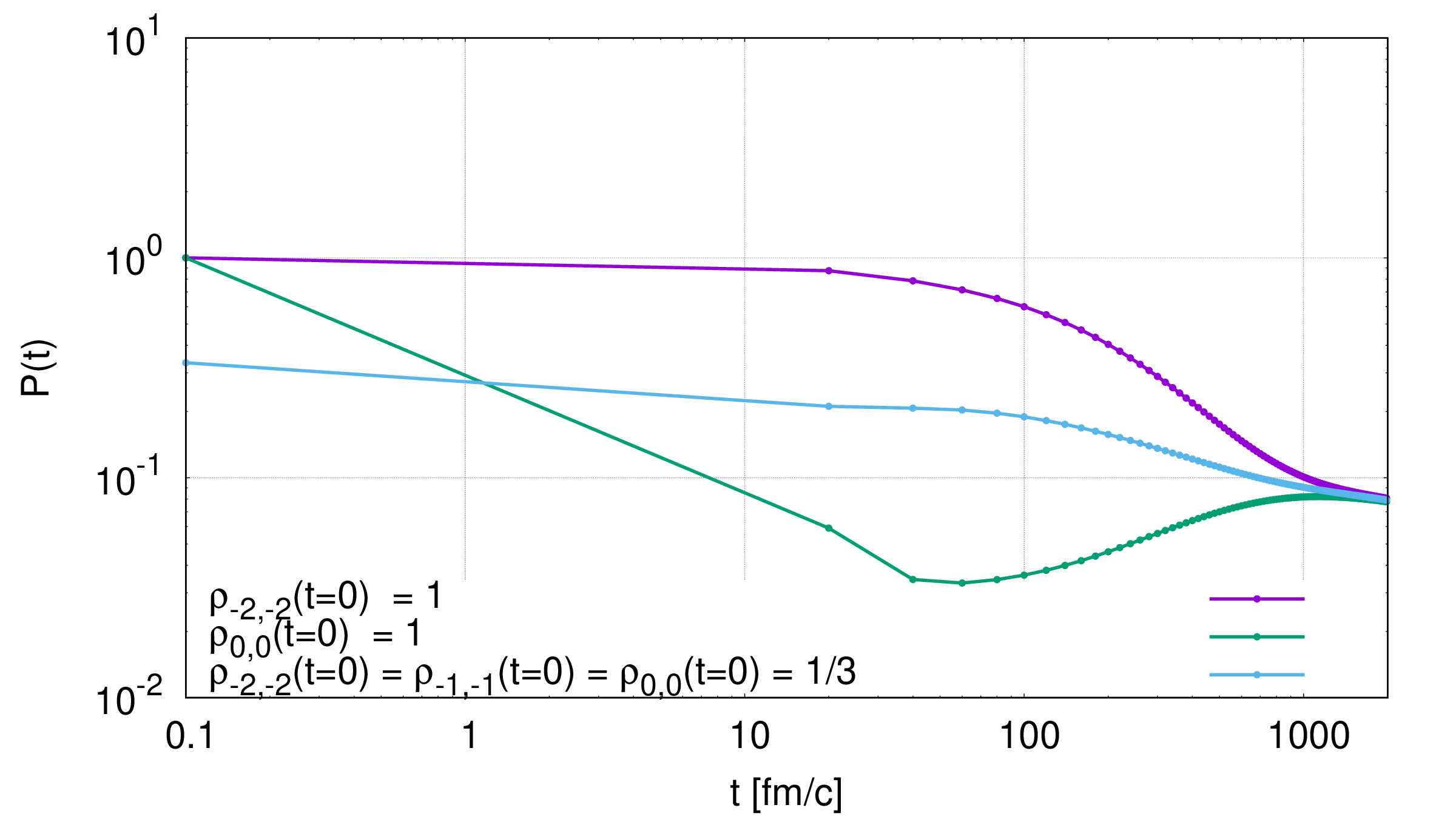}
		\hspace*{\fill}%
		
		\caption{
			Purity $P(t)$ of the system described by \cref{fig:3bound_wave_and_energy} for the initial conditions introduced in \cref{fig:3bound_3d_init} with $T=200$ MeV,  $\Omega = 4T$, $\gamma = 0.1$ c/fm, and $D_{px} = - \gamma T /\Omega$. 
		}
		\label{fig:3bound_purity}
	\end{center}
\end{figure}

 Investigating the purity, 
 \begin{align}\label{eq:purity}
 	P(t) = \text{Tr}(\rho^2) \, ,
 \end{align}
helps us to measure the ``pureness" of the considered system. Therefore, if  $P = 1$, the system is called pure, i.e. for $P=1$, $\hat{\rho} = \ket{\psi}\bra{\psi}$ is representing a pure state. 
Except for the setup presented in \cref{fig:inhom_xx}, we initialize our calculation as a pure state.
We are always expecting a Boltzmann distribution as the final thermal state, whose density matrix has a low purity, since each state is populated. 
However, the purity should be constant in the end, which also can be taken as a measure for thermalization.
In order to calculate the purity, applying \cref{eq:purity}, we again diagonalize the density matrix for different times.
We again discuss the purity of different parameters, initial conditions, bath temperatures, and damping $\gamma$.
In \cref{fig:purity_4T} we illustrate this comparison for the pure Caldeira-Leggett master equation.
We do not show other choices of parameters, because here again, we do not find any remarkable differences. 
To discuss the result we again find, that the impact of the temperature on the purity $P(t)$ is not dominant. 
However, for lower temperatures, the time when $\partial_t P(t) \approx 0$ is slightly larger. 
As we have already seen in \cref{fig:entropy_4T_0Dpx}, the purity gets constant fastest for initial conditions, where $n$ is large and for larger damping, which is not surprising in the context of the previous discussions.

\subsection{Increasing the number of bound states}\label{sec:morebound}

In this section we want to modify the system we have considered until now such, that we strongly increase the value for $V_0$ of \cref{eq:potential}. Here, $V_0 = 565$ MeV. 
This allows us to bind more states, which in our case leads to three bound states.
In \cref{fig:3bound_wave_and_energy} we illustrate the energy eigenfunctions of these states and  the first few states for $E>0$ as well as the corresponding energy eigenvalues.

Let us prepare the following three initial conditions: (1.) only the lowest bound state, with eigenvalue
\begin{align}
 E_{-2} = -382 \text{ MeV}\, ,
\end{align}
(2.) only the highest bound state, which is quite on the border to be unbound, 
 \begin{align}
 	E_{0} = -0.74 \text{ MeV}\, ,
 \end{align}
and (3.) mixed state, where all three bound states are populated equally at $t=0$. 
The second bound state has an eigenvalue 
 \begin{align}
	E_{-1} = -104.3 \text{ MeV}\, .
\end{align}
The density matrices for these initial conditions are illustrated in \cref{fig:3bound_3d_init}.
After evolving the system until it reaches its ``final" distribution, which we illustrate in \cref{fig:3bound_3d_final}, we can tackle the same questions as already done in the case, where we have considered only one (loosely) bound state: Does the system thermalize? What time does it need to thermalize? How do the entropy and purity behave? 
We compute the Lindblad dynamical evolution of this setup for a parameter set, which appears most suitable in terms of time scales and energy regimes from our above discussions, applying the findings of the case with one bound state.
As we can directly see in \cref{fig:3bound_3d_final} the impact of the potential which is located in the centre of the diagonal is  more visible, because it is much more attractive than in the case with smaller $V_0$ and only one bound state.
As a following step, we compute the coefficients $\rho_{nn}(t)$ using \cref{eq:coeff} to project on the very state of interest, and illustrate the bound states for the above mentioned initial conditions, cf. \cref{fig:3bound_states}.
Here we can see, that independent of the initial condition, all three cases reach the same final distribution, which does not mean, that they follow a monotonic behaviour towards this distribution. 
This was already found in \reff \cite{Rais:2022gfg}, where each state adjusted to the thermal distribution during the temporal evolution as well.
In \cref{fig:3bound_states} one can also see the final distribution of the bound states, following the procedure of fitting a Boltzmann distribution to obtain the fit parameters $\mu_{\text{fit}}$ and $T_{\text{fit}}$.
One can see, that the lowest bound state does not reach its final value for $\rho_{nn}(t)$, which is predicted analytically by \cref{eq:coeff,eq:Boltzmann}  after $2000$ fm/c! 
However, the system is stationary after approximately 1500 fm/c, which seems to correspond to a strengthening of the effective potential, and allows the speculation, if for strongly bound states the binding energy gets enforced due to bath interactions.
To investigate this further, we apply the procedure of calculating the eigenvectors (effective wave functions of the stationary system) to compare the initial wave functions to the effective ones.

Furthermore, recapitulating what was shown in \reffs \cite{Homa2019,Bernad2018}, the wave functions and therefore the energy eigenvalues of the system can get shifted during the temporal evolution.
In consequence, it is theoretically possible to make a bound state unbound by bringing it into contact with a thermal heat bath. 
This is the reason, why we tuned the potential such, that the bound state with the highest energy is bound only very weakly.
To discuss these questions, we illustrate the three bound states, cf. \cref{fig:3bound_waves} and the initial wave function. 
Having in mind, that the lowest bound state is above its analytically predicted value, we would expect, that the wave functions of this state appear different for the effective Hamiltonian. 
Surprisingly, the two lowest bound states are exactly the same as initially and only the most weakly bound state differs slightly from the initial value.  
Also comparing the eigenvalues $\lambda_i^\rho$, \cref{eq:effectiveE} with   \cref{eq:argument}, cf. \cref{fig:eigenvalues_3bound} shows, that all values are the same, except that the lowest bound state analytically, following \cref{eq:Boltzmann}, is expected to be slightly above the numerical result, which was already mentioned, cf. \cref{fig:3bound_states}.
One possible explanation for this is, that due to the interaction with the heat bath, the effective potential is stronger than the background potential \cref{eq:potential}.

Diagonalizing the density matrix allows us to  calculate the entropy and purity of the system.
This we have illustrated in \cref{fig:3bound_entropy}, where we also show the final distribution of $\rho_{nn}$ at time $t= 2000$ fm/c.
Fitting a Boltzmann distribution, to the distribution given by $\rho_{nn}$ we can extract a temperature $T_\text{fit} = 196.42$ MeV, which is slightly lower than the bath temperature of $200$ MeV.
 Let us point out here, that the system achieves equilibrium after $\approx 1200$ fm/c, which can be seen considering the entropy in \cref{fig:3bound_entropy}. 
 For times $t>1200$ fm/c, the entropy seems to decrease, which has a purely numerical reason: the norm in all three setups decreases linearly and is violated at $t=2000$ fm/c by up to 10\%. 
 This of course affects also the entropy but can be controlled easily by increasing the number of cells in the computational domain. 
 For our phenomenological interests, we schedule this problem for an upcoming work about strongly bound particles \cite{Rais2025}.
 One other interesting remark is deduced from the case where $\rho_{0,0}$ is originally populated. This corresponds to an energy, which is $\approx 0$, in comparison to the heat bath temperature of $200$ MeV and the lowest energy in the system, which is $-382$ MeV.

 The total entropy decreases for the initial condition, where $\rho_{0,0}(t=0) = 1$ after rapidly increasing for early times, cf. \cref{fig:3bound_entropy}. 
 One can also observe, considering the purity in \cref{fig:3bound_purity}, that at the same time, where the entropy reaches its maximum value, the purity is minimal, and decreases afterwards.
 This should be the case, because purity and entropy are related to each other.\footnote{In general, it would be also possible to construct a mixed state as an initial condition, where the purity is lower than the purity of the equilibrated thermal state. For this condition, the purity should increase, while the entropy should decrease.}
 In the Lindblad approach the heat bath is assumed to be a thermostat, and therefore is constant in temperature. 
 Since the Lindblad approach do not provide any information about the back reaction of the heat bath to the system, but energy is exchanged between the bath and the system, it is generally not impossible to observe a decreasing entropy. 
 However, to get full information about work and heat, one should introduce a more sophisticated definition of an entropy, which takes also the heat bath into account \cite{Aurell:2023sng}. 
 
 The thermalization time of 1200 fm/c is of course way over the thermalization times, we found in the cases, where only one state was (slightly) bound, and corresponds to the large energies which are needed to bind three states (the lowest energy is close to -400 MeV and the bath temperature is 200 MeV).
 However, this can be easily understood. 
 Phenomenologically, the system is perturbed by environmental particles during the process and therefore with energy. 
 Due to the large energy gaps among the bound states, it takes the system much longer to absorb enough energy to transition between different states.  
 Interestingly, for the case, where the highest of the three states is populated initially, which has an energy of only $E_0 = -0.72$ MeV, the entropy decreases after increasing rapidly.
 
Next, let us discuss the purity, \cref{fig:3bound_purity}.
 We initialize two cases as pure states and one as a mixed state, where the purity is 1/3 at $t=0$.
Similarly to the entropy, also the purity decreases slightly for larger times, which we neglect here in order to visualize and focus on the behaviour for times $t<1000$ fm/c.
For all three conditions we can see, that the purity reaches the same value at a time $t\approx 1000$ fm/c which is slightly earlier than $S(t)$.

\section{Summary and Outlook}
\subsection{Summary}
In the last decades open quantum system turned out to be a major topic in physics. 
Despite providing a deeper fundamental understanding of the quantum nature of a physical system and its connection to the classical world, they are practically applied also to physical systems in a broader sense, such as biological systems or quantum-information devices. 
In particular, understanding the non-linearity and interactions/impacts of thermal environments on a system of interest is a necessity when it comes to highly sensitive measurements.

A special physical system, which itself is part of a many-body system is a deuteron in nuclear matter with a binding energy much lower than the energy of the surrounding system.
Therefore, the question of how the deuteron is formed in such a many-body system, and what its formation time is, is an interesting case for studying the properties of open quantum systems within the context of heavy-ion collisions. 
 
To turn to a physically motivated model, which was constructed to mimic bound states and therefore investigate the formation or destruction time of such a bound state and to treat thermalization, we found, that for various regimes and various parameters thermalization is reached up to some (small) numerical discrepancy, and the equilibration time is usually dependent on the temperature of the heat bath and the damping coefficient, as well as the cut-off frequency of the Ohmic heat bath spectrum. 
We found, that the solution of the pure CLME leads to a thermal state as well as the various modifications, which are incorporated in order to obtain Lindblad form.

We have used different measures to approach the questions of  thermalization and the thermalization time. 
Therefore, we calculated the reduced density matrix $\rho(x,y,t)$ at arbitrary times towards equilibration. 
We used $\rho(x,-x,t=t_\text{eq})$ and the coefficients $\rho_{nn}(t=t_\text{eq})$ to clarify the question of thermalization and calculated the entropy and purity of the system to extract the thermalization time. 
This we have done for various heat bath temperatures $T$, damping coefficients $\gamma$, and initial conditions, as well as for different types and formulations of the Lindblad equation.

Generally, the damping parameter $\gamma$ seems not to satisfy the simple relation $\gamma = \frac{1}{\tau_R}$ in a rigorous way.
Moreover, the thermalization time depends on the initial condition, such, that for higher populated states, the thermalization time is much smaller.
Furthermore, we found strong dependencies on the damping, which is  phenomenologically understandable, and the binding energies set by the system potential. 
The heat bath temperature and the choice of the diffusion coefficients $D_{px}$ seems to be of subleading relevance for the above questions.

Therefore, we cautiously claim, to have shown, that the considered system thermalizes.
The weak-coupling limit and the Markovianity lead to thermalization times of the typical order of the formation times, which are observed in heavy-ion collisions and which are typically $<10$ fm/c, if states with higher energies are originally populated.

We have shown, that a bound state can be stably bound due to environmental effects.

All numerical computations were done with a finite volume method, which is successfully used in hydrodynamics and is applicable for systems, which preserve conserved quantities, such as in our case the norm of the density matrix.
Therefore, we have recapitulated the conservative form of the Lindblad equation in terms of a diffusion-advection equation, which was derived in a previous work \cite{rais2024}. 
This additionally opened up a new viewpoint of dissipative quantum systems from a fluid dynamical perspective.

\subsection{Outlook}
Lindblad master equations are demanding in two ways: firstly, they are numerically demanding, and even if we hope to have convinced the reader to have a numerically powerful tool, there is still room for improvements.
One aspect, that has been already discussed in \cite{rais2024} is norm conservation on the numerical level. 
Especially, when we want to implement spatial diffusion on a finite spatial grid, which is necessary to reproduce the full Lindblad equation, small violations of the norm conservation turns out to be a mathematical fact \cite{rais2024} and is not totally under control yet.

On the numerical side, it would be desirable to extend the model to two or three dimensions and to implement the diffusion coefficient $D_{xx}$. Therefore, it would be necessary to increase the computational domain such, that the boundary effect, which is caused by the spatial diffusion turn to be negligible.
This of course leads to numerically more expensive calculations.

Furthermore, it is desirable to understand more of the general structure of the Lindblad approach framework. 
Therefore, as a ``work in progress" we perform a comparison between a non-Markovian approach from a fully solvable path integral to the Lindblad master equation, and a comparison to another popular approach for open quantum systems, the Kadanoff-Baym approach, applying Keldysh-Schwinger methods \cite{Neidig:2023kid,Rais2025}.

In future works, we also want to implement time-dependent damping coefficients, which makes it necessary to calculate the momentum and coordinate space expectation values, which can be obtained by a Wigner transformation.
Additionally, this allows us to study decoherence and gain a deeper understanding of the physical interpretation of the real and imaginary part of the density matrix in the hydrodynamical picture, which automatically brings us to the fundamentals of quantum nature.
We also plan to investigate other systems, described by different potentials to mimic for example charmonia-like potentials.

\begin{acknowledgements}	
	J.~R.\  acknowledges support by the \textit{Helmholtz Graduate School for Hadron and Ion Research for the Facility for Antiproton and Ion Research} (HGS-HIRe for FAIR) and the \textit{Deutsche Forschungsgemeinschaft} (DFG, German Research Foundation) through the CRC-TR 211 ``Strong-interaction matter under extreme conditions'' -- project number 315477589 -- TRR 211.
	
	J.~R.\ thanks J.P.~Blaizot for extensive and inspiring discussions, creative ideas and collaboration during his visit in Saclay.
	
	J.~R.\ thanks D.~Schuch for initializing the need of a deeper understanding of the fundamentals of quantum nature, and educational discussions.
	
	J.~R.\ thanks N.~Zorbach  for providing a highly reliable code, which was used to perform all numerical simulations and collaborating on a reliable and professional way in the preceding project.
	
	J.~R.\ thanks A.~Koenigstein  for creative discussions and impulses, motivations, and careful proofreading.
	
	J.~R.\ thanks T.~Neidig  for creative and useful discussions and impulses.
\end{acknowledgements}


\bibliography{bibliography.bib}

\end{document}